\documentclass[a4paper,11pt]{article}
\usepackage{commands}
\pdfoutput=1 

\usepackage{jheppub} 

\usepackage[T1]{fontenc} 
\usepackage{amsmath}
\usepackage{graphicx}
\usepackage{booktabs}
\usepackage{multirow}
\renewcommand{\figurename}{Fig.}

\usepackage[colorinlistoftodos]{todonotes}

\title{\boldmath Heavy resonances at energy-frontier hadron colliders}

\author[a]{Clement~Helsens}
\author[a,c]{\!\!, David~Jamin}
\author[a]{\!\!, Michelangelo~L.~Mangano}
\author[b]{\!\!, Thomas~G.~Rizzo}
\author[a]{and Michele~Selvaggi}

\affiliation[a]{CERN, CH-1211 Geneva 23, Switzerland}
\affiliation[b]{SLAC National Accelerator Laboratory 2575 Sand Hill Rd., Menlo Park, CA, 94025 USA}
\affiliation[c]{Academia Sinica, Institute of  Physics, Taipei, Taiwan}

\emailAdd{clement.helsens@cern.ch}
\emailAdd{david.jamin@cern.ch}
\emailAdd{michelangelo.mangano@cern.ch}
\emailAdd{rizzo@slac.stanford.edu}
\emailAdd{michele.selvaggi@cern.ch}

\abstract{This paper explores the physics reach of the proton-proton Future Circular Collider (FCC-hh) and of the High-Energy LHC (HE-LHC) for searches of new particles produced in the $s$-channel and decaying to two high-energy leptons, jets (non-tops), tops or W/Z bosons. We discuss the expected discovery potential and exclusion limits for benchmark models predicting new massive particles that result in resonant structures in the invariant mass spectrum. We also present a detailed study of the HE-LHC potential to discriminate among different models, for a $\Zp$ that could be discovered by the end of High-Luminosity LHC (HL-LHC).}

\begin{document}
\maketitle
\flushbottom

\section{Introduction}
\label{sec:intro}
Extensive searches at the LHC, addressing a broad range of final states, have set stringent limits on the existence of new physics beyond the Standard Model (SM). While the LHC has still a long way to go in exposing new phenomena~\cite{CidVidal:2018eel}, several projects are being proposed for future colliders, whose principal goal is to push even further the sensitivity to the theoretical scenarios beyond the SM (BSM) that have been proposed to address the shortcomings of the SM. These range from the theoretical puzzle of hierarchy problem, to the limitations of the SM to account for observed phenomena such as dark matter, neutrino masses or the matter-antimatter asymmetry of the universe. These future projects are designed to address the possible reasons for the current lack of BSM evidence: either the BSM phenomena appear at energy scales consistent with the reach of the LHC, but elude the LHC searches due to the stealthy nature of their final states or to very weak couplings leading to very low rates, or the BSM phenomena are tied to physics living at mass scales beyond the LHC reach. Future $e^+e^-$ colliders have a limited energy reach for direct observation, but their clean experimental environment and high precision have the potential to expose the most elusive manifestations of TeV-scale BSM models. Future pp colliders, on the other hand, allow us to extend the direct discovery reach to masses well beyond those probed by the LHC, and their high luminosity may give visibility to the rarest of the processes. The HE-LHC~\cite{Zimmermann:2651305} is designed to increase the LHC energy to 27~TeV, thus almost doubling the LHC mass reach. The FCC-hh~\cite{Benedikt:2651300,Mangano:2651294} and the SPPC~\cite{CEPCStudyGroup:2018rmc,CEPCStudyGroup:2018ghi} are designed to reach a pp centre of mass energy of 100~TeV, relying on new facilities built around a 100~km accelerator ring, with a potential increase in mass reach of a factor of 5-7, depending on their integrated luminosity and on the production channel. In addition to greatly extending the discovery reach, future pp colliders would enhance the production rate of new particles that the LHC could still discover during its forthcoming operations. This would enable a much more precise study of the properties of these new particles, and help to pin down the underlying physics models that will extend the SM. 

General overviews of the BSM discovery potential of HE-LHC and FCC-hh, spanning across a broad range of models, can be found in Refs.~\cite{Golling:2016gvc,CidVidal:2018eel}.
This paper focuses on a quantitative study of the discovery potential at the highest masses, using as benchmarks $s$-channel resonances. 
Section~\ref{sec:bsmmodels} presents the expectations of some relevant BSM scenarios, and defines the models that we shall discuss.  Details on the generators, detector performance, statistical method and other analysis techniques are presented in Section~\ref{sec:simulation}. The leptonic ($ee$, $\mu\mu$, $\tau\tau$) and the hadronic (WW, $t\bar{t}$ and jj) analyses are detailed in Sections~\ref{sec:lep} and~\ref{sec:hadronic} respectively. The results of the analyses are discussed in the context of the anticipated FCC-hh accelerator and detector performance; the sensitivity of the HE-LHC to the same benchmark models is summarised in Section~\ref{sec:ana27tev}. Finally, a study of \Zp models' discrimination at the HE-LHC, in case of a discovery at the HL-LHC, is presented in Section~\ref{sec:zprimedisc}.

\section{BSM \texorpdfstring{\Zp}{zp} models and their possible probes}
\label{sec:bsmmodels}
In order to explore and contrast the capabilities of future colliders to discover and examine the properties of possible new physics, a broad set of benchmark models needs to be studied. In the
case of new heavy resonances, this benchmark set should be sufficiently complete that all of the major discovery channels of relevance are represented. Here we
are particularly interested in the 2-body final states of these resonances, 
consisting of opposite sign dilepton
pairs ($e^+e^-, \mu^+\mu^-$ and $\tau^+\tau^-$), dijets, $ t\bar t$ and $W^+W^-$.  We note that typically at least one or possibly more of these 2-body channels will possess
a significant branching fraction, particularly into jets, given the partonic production mechanism. Note that decays into pairs of secondary objects that
then themselves decay hadronically can often populate the dijet channel if the final state jets are sufficiently boosted. The dijet channel can thus represent many different final states, unless
substructure studies are performed.  When there are 2 or more of these channels available for simultaneous study we have an increased chance to learn more about
the underlying physics model. The most important properties of a newly discovered resonance that need to be determined (other than the mass) are its
production cross section, which, especially for a broad resonance, will sometimes require a good understanding of the underlying background shape, and its spin (as was the case of the Higgs boson). These properties alone can provide important information about the BSM model from which the signal originated. The spin measurement usually requires
the reconstruction of the angular distribution of the resonance decay products and, hence, a respectable amount of statistics, although the observation of certain final states can
immediately exclude some spin possibilities as was the case with the observation of $H\rightarrow \gamma \gamma$.

A new, neutral, spin-1 gauge boson, $\Zp$, which is usually a color-singlet object produced in the $q\bar q$ channel, is a ubiquitous feature of many BSM models~\cite{Langacker:2008yv,Rizzo:2006nw,Carena:2004xs,Salvioni:2009mt}.  While falling into several distinct classes, $\Zp$ are most commonly associated with the
extension of the SM EW gauge group by an additional U(1) or SU(2) factor, although more significant additions are possible. When the additional factor is non-abelian,
as in the case of SU(2), a new $W^\pm$' gauge boson  also appears in the spectrum together with the $\Zp$, and with a comparable mass.  Of this subset of models, those
that arise from Grand Unified Theory frameworks are the ones most commonly encountered in the literature, and include familiar examples such as the Left-Right Symmetric
Model (LRM)~\cite{Senjanovic:1975rk,Mohapatra:1980yp} which results from SO(10)
(or larger GUT groups) and where the SM is augmented by an SU(2)$_R$ factor. For example, the LRM can arise from SO(10) breaking at the GUT scale directly to
SU(2)$_L \times$SU(2)$_R \times$U(1)$_{B-L}$, which then breaks to the SM at a few to multi-TeV scale. A second set of GUT-based $\Zp$ models arise from
$E_6$~\cite{Robinett:1982tq,London:1986dk,Hewett:1988xc,Joglekar:2016yap}, where, e.g., $E_6 \rightarrow$ SM$\times$U(1)$_\psi \times$U(1)$_\chi \rightarrow $ SM$
\times$U(1)$_\theta$, giving rise to an additional U(1)$_\theta$ gauge group. Note that here $\theta$ labels the linear combination of U(1)$_\psi$-U(1)$_\chi$ that remains
unbroken to energies below the GUT scale.  A common set of features of this GUT-based class of models include their generation-universal couplings
of the $\Zp$ to the SM fermions, their charges commuting with those of the SM, so that, e.g., $u_L$ and $d_L$ have the same $\Zp$ coupling and the resonances themselves are usually
narrow, reflecting EW strength or weaker couplings with width to mass ratios $\Gamma/M < 0.01-0.03$. In particular, the GUT origin of these models
implies that this class of $\Zp$ can be used to simultaneously study all of the dileptonic channels: $e^+e^-,~\mu^+\mu^+$ as well as $\tau^+\tau^-$ together with the dijets, $W^+W^-$  and 
$t\bar t$ channels as well.

With this much information potentially available from the observation of a given $\Zp$ in multiple channels, one may try to distinguish it from another \Zp{} of similar type, given
sufficient statistics and
well-controlled systematics. In addition to relative cross section measurements,  e.g., that of dijets and/or $t\bar t$ compared to dileptons, the cleanliness of the dilepton channel itself
provides additional information.  Since the lepton charges can be determined, their angular distribution allows us to probe the forward-backward asymmetry, $A_{FB}$, whose sensitivity to the quark and lepton couplings is complementary to that of the dilepton production cross section\footnote{The scattering angle is defined by the direction of the outgoing charged lepton w.r.t. the incoming quark direction. For the heaviest resonances, where the quark is dominantly a valence parton inside the proton, the latter is {\it usually} also the direction of the boost of the \Zp{} in the lab frame. We adopt this convention in our analysis, using the Monte Carlo to correct on average for the dilution of $A_{FB}$ induced by antiquark contributions.}. A second handle~\cite{delAguila:1993ym} to probe the couplings to the various quark flavours is the different rapidity distributions of the $u\bar u$ and $d\bar d$ initial states. Since the various $\Zp$ will generally couple differently to the $u$ and $d$ quarks, the
rapidity distributions of the dilepton final state will probe these coupling variations. The relevant observable in this case is the rapidity ratio, $r_y$, defined by the ratio of the number of
dilepton pairs produced at central versus forward rapidities (see below for details).

Returning to our discussion of these specific GUT-inspired models, we note that in the LRM with the assumption of left- and right-handed gauge couplings, i.e., $\kappa=g_R/g_L=1$,
all of the various interactions of the $\Zp$ with the SM fields are completely fixed. However, in the $E_6$ model case, the
single new mixing parameter, $\theta$, controls the couplings of the $\Zp$ to the various SM particles; four particular choices for the value of this parameter correspond to the more
specific model cases discussed here and are denoted as $\psi$, $\chi$, $\eta$ and I. As in the SM, the $\Zp$ in GUT models generally couple to all the familiar quarks and leptons and
thus can easily populate simultaneously the various fermionic 2-body final states listed above at various predictable rates. The measurement of these rates (as well as other associated
observables) can be then used to discriminate among the different  $\Zp$ possibilities after discovery, as will be discussed further below. Note that the decay rate for $\Zp$ into the
$W^+W^-$ final state in GUT frameworks is highly dependent on the details of the model building assumptions within a specific scenario and especially upon the detailed nature
of spontaneous symmetry breaking as manifested by the amount of mixing (if it occurs at all) between the $\Zp$ and SM $\Z$; the $\Zp$ coupling to $W^+W^-$ in U(1) extensions is
always controlled solely by the amount of this gauge boson mixing, i.e., this coupling in such extensions in the absence of mixing is zero. 

The $\Zp$ of the Sequential Standard Model~\cite{Altarelli:1989ff} (SSM) has been used very frequently for many years as a standard candle by experimenters, since it conveniently posits the existence of heavier copies of the SM gauge bosons,
with heavier masses but identical couplings; this provides a useful yardstick for easier performance comparisons.

Alternative models of EW symmetry breaking, including the topcolor assisted technicolor scenarios~\cite{Hill:1994hp}, also lead frequently to $\Zp$-like states
with resonance signatures. The greatest difference of such theories from the GUT-type $\Zp$ models lies in their generation-dependent couplings,
potentially of QCD strength. (The color-octet versions of such states in this model class are called colorons.) This implies that the corresponding resonances will likely not be narrow and
will preferentially couple, by construction, to the
third generation, leading to highly boosted $t\bar t$ final states, and proving a useful benchmark model for this channel. Similar new $\Zp$ states can also arise in Little Higgs
models~\cite{ArkaniHamed:2001nc}, which also have preferential decays to third generation states.

Occasionally, the expected properties of a new $\Zp$ models are suggested by the attempt to interpret and model anomalies observed in the data. For example, a $\Zp$ with an unusual flavor-dependent coupling structure has recently been suggested as
a (partially complete) UV model to explain the apparent anomaly seen in semileptonic $b\rightarrow sl^+l^-$ decays~\cite{Aaij:2014ora,Aaij:2017vbb}. In effective field theory language,
a new interaction of the form $\sim \bar b\gamma_\mu P_Ls \bar \mu \gamma^\mu P_L \mu$ of proper strength can provide a reasonable fit to these experimental observations~\cite{Bifani:2018zmi}. This operator can
be induced by the exchange of a
heavy $\Zp$ potentially accessible to high energy colliders~\cite{Allanach:2017bta,Allanach:2018odd}. This $\Zp$, in the weak basis, couples only to the third generation quark doublet and to the
muon lepton doublet, so that it will have a suppressed production cross section at hadron colliders. Such a $\Zp$ could be observed in both the dimuon and ditop channels.

Models of composite quarks and leptons offer another path wherein new resonances are predicted. Excited quarks~\cite{Baur:1987ga,Baur:1989kv}, $Q^*$, are spin-1/2, color triplet
states with the same SM quantum numbers of quarks. 
There is, as of yet, no fundamental, UV-complete
model encompassing this idea so that this framework is purely phenomenological. The SM quarks couple to these excited states via a magnetic-dipole-like interaction together with an associated SM gauge boson ($g$, $\gamma$ or $W/Z$). This dimension-5 interaction is suppressed by a large `compositeness scale', $\Lambda$, and the relative coupling strengths to the different gauge bosons are partially controlled by a set of essentially free parameters, $f_i$. Excited quarks can be singly produced in
the $gq$ channel, to which they will also dominantly decay due to the presence of the strong coupling constant, yielding the dijet signature of interest to us here (the $q\gamma$ channel is also possible, but will not be considered here). It is useful to have a benchmark model with dijet decays which take place in the $gq$ channel (as opposed to a $\Zp$ that can
only populate the $q\bar q$ dijet channel) with which to compare and contrast. The angular distributions of the 2 jets in the dijet decay, which will require significant statistics to
determine, can provide
us information about the spin of the original resonance and the nature of its couplings to the decay products~\cite{Harris:2011bh,Boelaert:2009jm,Chivukula:2014pma,Chivukula:2017nvl}.

Spin-2 graviton resonances occur in extra-dimensional scenarios that attempt to address the hierarchy problem, as in the case of the warped extra dimensional model of
Randall and Sundrum (RS)~\cite{Randall:1999ee}. In such setups, the SM gauge fields and fermions are generally allowed to propagate in the 5-dimensional
bulk~\cite{Pomarol:1999ad,Davoudiasl:1999tf,Grossman:1999ra,Davoudiasl:2000wi,Gherghetta:2000qt} whereas EW symmetry breaking occurs at or near the TeV/SM brane
via the usual Higgs mechanism. 
Due to the shape of their 5-D wavefunctions, the Kaluza-Klein excitations of the familiar graviton, $G_{RS}$~\cite{Davoudiasl:1999jd} will dominantly decay into
objects localized near to where SM symmetry breaking occurs, i.e., Higgs boson pairs and $t\bar t$, as well as to the longitudinal components of the massive SM gauge bosons, e.g.,
$W^+_L W^-_L$, all with relatively fixed branching factions with only some small allowed variations. Thus $G_{RS}\rightarrow
W^+W^-$ in the RS framework provides an excellent benchmark model for the study of resonant and highly boosted $W$-pairs. For hadronic $W$ decays, given the high boost, this final state may also (appear to) populate the resonant dijet channel. One notes that apart from the $G_{RS}$ mass scale itself, essentially the only other
free parameter in this RS model setup (wherein the lighter fermions are essentially decoupled from the graviton resonances), is frequently denoted by $c=k/\bar M_{Pl}$, which simply
controls the overall coupling strength to all of the various SM particles.

\section{Simulation setup}
\label{sec:simulation}

\subsection{Monte Carlo production}
\label{subsec:mcprod}

Monte Carlo~(MC) event samples are used to simulate the response of the detector to signals and backgrounds processes. Signal events are generated with \py~\cite{Sjostrand:2014zea} version 8.201 with the \pdf{NNPDF2\!.\!3NNLO} PDF set~\cite{Ball:2014uwa} using the leading order cross-section from the generator with no K-factor. The SM backgrounds considered are Drell-Yan, di-jet (QCD), top pairs (\ttbar), $VV$ and \vj\ where $V=W/Z$ and were generated using \MGAMC~\cite{Alwall:2014hca} version 2.5.5 at leading order only with the \pdf{NNPDF3\!.\!0NLO}~\cite{Ball:2014uwa} PDF set in bins of $\hht$. A K-factor of 2 is applied to all the background processes to account for higher order corrections and is considered to be very conservative.

\subsection{Simulation of the detector response}
\label{subsec:detparam}

This study discusses the discovery potential of heavy resonances decaying to multi-TeV final states. The ability to accurately reconstruct highly boosted final states is largely dependent on the nature of the object and on the detector assumptions. Generally speaking, the energy-momentum resolution of calorimetric objects such as electrons, photons and jets improves as a function of the energy. Conversely, the momentum resolution of charged particles reconstructed as tracks decreases with the momentum as the curvature of the trajectory vanishes. In addition, at high energies, composite objects such as jets, or hadronically decaying $\tau$'s and heavy bosons are highly collimated. This results in an effectively coarser granularity of the detector, which can potentially limit the ability to resolve and identify the decay products inside the jets, thereby limiting the identification and QCD background rejection capabilities. 

The detector response has been simulated via the \delphes{}~software package~\cite{deFavereau:2013fsa}. For the \sqrtsfcc\ collider, the reference FCC-hh detector configuration has been used as a baseline~\cite{Benedikt:2651300, delphes_card_fcc}. For the HE-LHC study with \sqrtshelhc\, a generic detector configuration~\cite{delphes_card_helhc} has been designed to reproduce an average response of the HL-LHC general purpose ATLAS~\cite{Aad:2008zzm,Capeans:2010jnh} and CMS~\cite{Chatrchyan:2008aa} detectors. Hereafeter, we will simply refer to the HE-LHC and the FCC-hh detectors and will only discuss detector specifications that are relevant for high \pt\ objects. The overall contribution of pile-up is neglected altogether, as it is expected to have a  negligible impact of multi-TeV objects. Even though only one specific configuration of the detector response has been studied in details for FCC-hh, the effect of the degradation of some key parameters have been examined and is documented in Appendix~\ref{sec:app:detperf}.

\subsubsection{Tracking}
\label{appsub:tracking}

After collision, parton showering, hadronisation, and decays, the first step of \delphes{}~is the propagation of long-lived particles inside the tracking volume within a uniform axial magnetic field parallel to the beam direction. The magnetic field strength $B$, the size of the tracking radius, $L$, and the single hit spatial resolution, $\sigma_{r\phi}$, are the main parameters that determine the resolution on the track transverse momentum:
\begin{equation}
\frac{\sigma(\pt)}{\pt} \approx \frac{\sigma_{r\phi}~ \pt}{B\cdot L^2}\,.
\end{equation}
The radius of the FCC-hh inner tracking detector is $3/2$ that of the HE-LHC detector with a similar magnetic field of 4 Tesla.  The spatial resolution $\sigma_{r\phi}$ is 3 times smaller than at HE-LHC, which is possible thanks to a more granular pixel detector~\cite{CMS:2012sda}. These specifications of the FCC-hh detector would allow measurements of $\pt=$~1 TeV charged hadrons with a precision of $\sigma(\pt)/\pt  \simeq 2\%$, compared to $\sigma(\pt)/\pt \simeq 10\%$ for the HE-LHC detector. 

\begin{figure}[htb!]
  \centering
  \includegraphics[width=0.56\columnwidth]{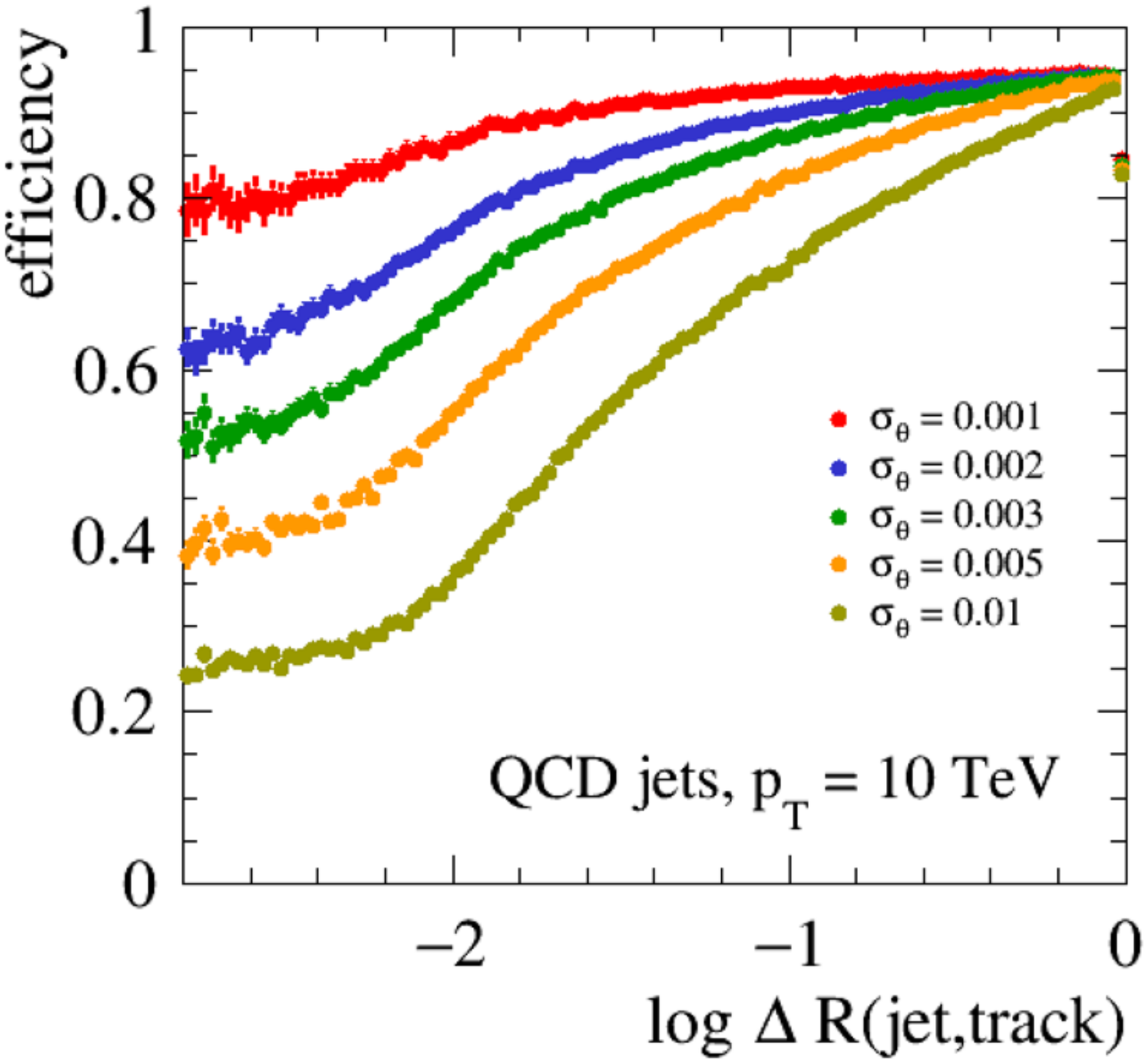}
  \includegraphics[width=0.42\columnwidth]{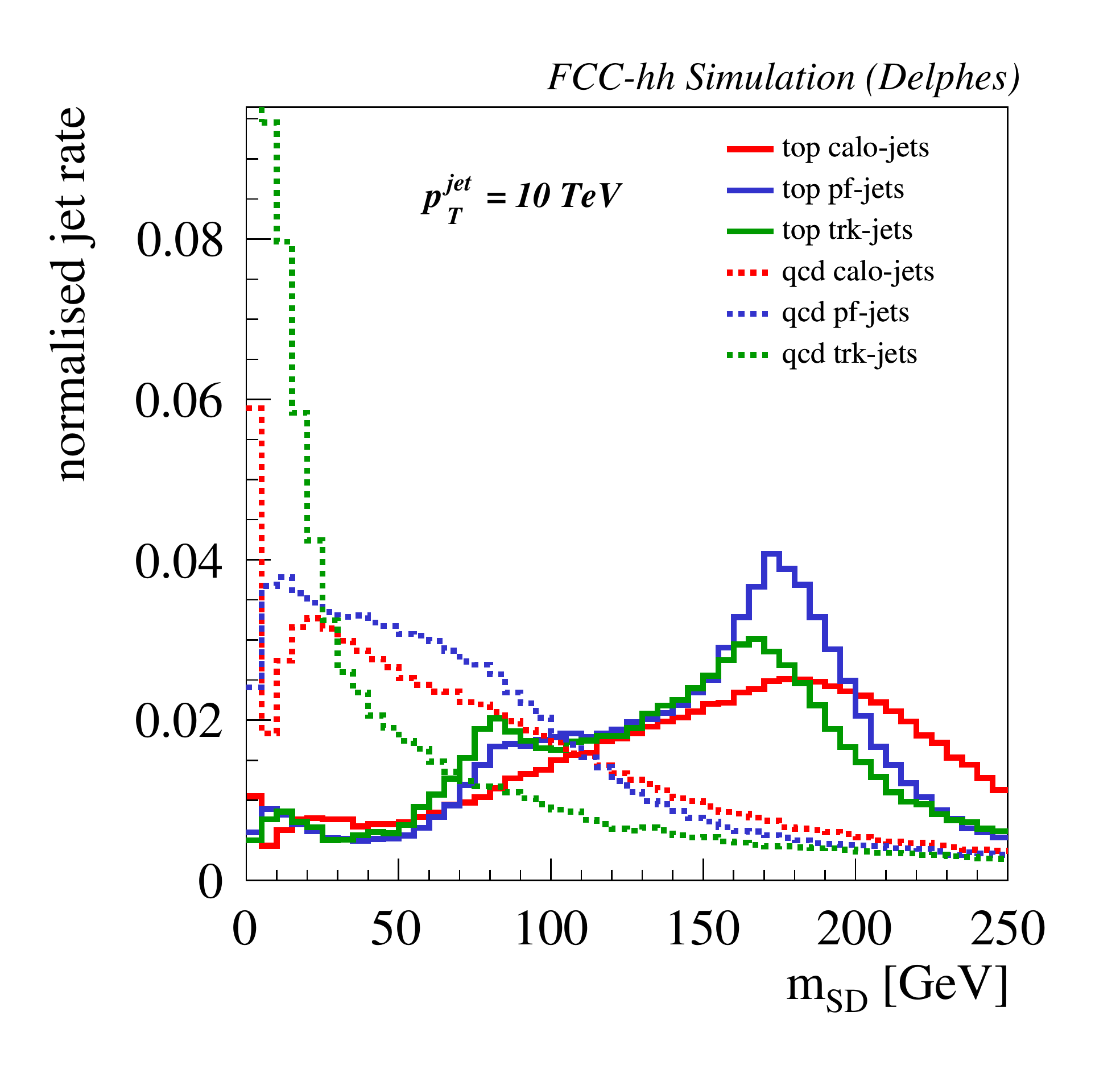}
  \caption{Left: Track reconstruction efficiency inside highly boosted QCD jets as function of the angular distance $\Delta R$ between the track and the center of the jet for different assumptions on the tracker spatial resolution. Right: Reconstructed 
  "soft-dropped" jet mass of highly boosted top and QCD jets with various sets of input to the jet clustering algorithm: tracks only, calorimeters towers only and particle-flow candidates. }
  \label{fig:substructure}
\end{figure}

Central and isolated high momentum charged hadron tracks are assumed to be reconstructed with an efficiency $\epsilon = 95\%$. However, charged particles confined inside a highly boosted jet can be extremely collimated, resulting in unresolvable tracker hits, especially in the innermost tracking layers. Although an accurate description of this feature would require a full event reconstruction by means of a GEANT4-based simulation~\cite{Agostinelli:2002hh,Allison:2006ve,Allison:2016lfl}, a specific \delphes{}~ module aiming at reproducing this effect has been designed. Whenever two or more tracks fall within an angular separation $\sigma(\eta,\phi)$, only the highest momentum track is reconstructed. This effect can result in an additional inefficiency to that shown in Table~\ref{tab:trk_param}, and can affect the ability to reconstruct tracks in the core of highly boosted jets, as shown in Fig.~\ref{fig:substructure} (left).

\begin {table}[htb!]
\begin{center}
\begin{tabular}{l||c|c}
& FCC-hh & HE-LHC \\
  \hline
  \hline
$B_z$ $(T)$ &  4 & 4 \\
  \hline
Length $(m)$ & 10 & 6 \\
 \hline
Radius $(m)$ & 1.5 &  1.1 \\
 \hline
$\epsilon$ & 0.95 & 0.95\\
\hline
$\sigma(\eta,\phi) (mrad) $ & 1  & 3  \\
\hline
$\sigma(\pt)/\pt$ (tracks) &  $0.02\cdot \pt$ (TeV/c) &  $0.1\cdot \pt$ (TeV/c)\\
\hline
$\sigma(\pt)/\pt=5\%$ (muons) & $\pt=15$~TeV & $\pt=2$~TeV\\

\end{tabular}
\caption{Tracking-related parameters for the FCC-hh and HE-LHC detectors in Delphes.}
\label{tab:trk_param}
\end{center}
\end{table}

Muons are also reconstructed using tracking. However, an additional stand-alone muon measurement is provided by the angular difference between track angle in the muon system and the radial line connection to the beam axis, giving a large improvement on the resolution at high \pt~\cite{Benedikt:2651300}. Assuming a 2 times better position resolution of the muon system for the FCC-hh detector, a combined muon momentum resolution of $\sigma(\pt)/\pt  \simeq 5\%$ can be achieved for momenta as high as $\pt=$~15 TeV, as opposed to $\pt=2~$TeV for the HE-LHC detector.

\subsubsection{Calorimetry and Particle-Flow}
\label{appsub:calorimetry}

After propagating within the magnetic field, long-lived particles reach the electromagnetic (ECAL) and hadronic (HCAL) calorimeters. Since these are modeled in \delphes{}~by two-dimensional grids of variable spacing, the calorimeter deposits natively include finite angular resolution effects. Separate grids for ECAL and HCAL have been designed for both the FCC-hh and the HE-LHC detectors in order to accurately model the angular resolution on reconstructed jets. The FCC-hh detector features an improved angular resolution by a factor 2 in the ECAL and a factor 4 in the HCAL compared to the HE-LHC detector. The energy resolution of the calorimeters is assumed to be the same for both detectors and the calorimeter parameters are summarised in Table~\ref{tab:cal_param}.

In \delphes{}~the information provided by the tracker and calorimeters is combined within the particle-flow algorithm for an optimal event reconstruction. If the momentum resolution of the tracking system is better than the energy resolution of calorimeters (typically for momenta below some threshold) the charged particles momenta are measured mainly through tracking. Vice-versa at high energy, calorimeters provide a better momentum measurement. The particle-flow algorithm exploits this complementarity to provide the best possible single charged particle measurement --- the \emph{particle-flow tracks}; these contain electron, muons and charged hadrons. Jet collections are then formed using several different input objects such as tracks (\emph{Track-jets}), calorimeter (\emph{Calo-jets}) and particle-flow candidates (\emph{PF-jets}). The \delphes{}~framework integrates the FastJet package~\cite{Cacciari:2011ma}, allowing for jet reconstruction with the most popular jet clustering algorithms. In the present study the anti-$k_T$ algorithm~\cite{Cacciari:2008gp} is used with several jet clustering R parameters ($R=0.2, 0.4, 0.8, 1.5$).

Common jet shape observables used for jet substructure analysis such as N-subjettiness~\cite{Thaler:2010tr} and the soft-dropped mass~\cite{Larkoski:2014wba} are computed on-the-fly and stored in the output jet collections. As an illustration, the reconstructed soft-dropped mass in the FCC-hh detector for top and QCD jets with $\pt=10$~TeV and cone size $R=0.2$ is shown in Fig.~\ref{fig:substructure}(right). Thanks to the superior tracker segmentation, we find \emph{Track-jets} to perform better in terms of QCD background rejection despite the slightly worse jet mass resolution. A recent study shows that the reconstruction of jet substructure variables for highly boosted objects will benefit from small cell sizes of the hadronic calorimeter which confirms the FCC baseline design~\cite{Yeh:2019xbj}.

\begin {table}[htb!]
\begin{center}
\begin{tabular}{l||c|c}
& FCC-hh & HE-LHC  \\
  \hline
  \hline
$\sigma(E)/E $ (ECAL)& 10\%/$\sqrt{E} \oplus 1\%$ &  10\%/$\sqrt{E} \oplus 1\%$\\
   \hline
$\sigma(E)/E $ (HCAL)& 50\%/$\sqrt{E} \oplus 3\%$  &  50\%/$\sqrt{E} \oplus 3\%$\\
  \hline
  \hline
$\eta \times \phi$ cell size (ECAL)& $(0.01\times0.01)$ &  $(0.02\times0.02)$\\
  \hline
 $\eta \times \phi$ cell size  (HCAL)& $(0.025\times0.025)$ & $(0.1\times0.1)$
\end{tabular}
\caption{Calorimeter parameters for the FCC-hh and HE-LHC detectors in Delphes.}
\label{tab:cal_param}
\end{center}
\end{table}

\subsubsection{Object identification efficiencies}
\label{appsub:objid}

Trigger, reconstruction and identification efficiencies are parametrised as function of the particle momentum in~\delphes{}. Given that these parameterisations depend on the detailed knowledge of the detector, we simply use a global parameterisation for each object.

For electron and muons, the isolation around a cone is computed as the sum of the full list of \emph{particle-flow candidates} within a cone of radius $R$ excluding the particle under consideration. No selection on the isolation variable is applied during \delphes\ processing since the optimal selection working point is analysis and object dependent. Electrons and muons originating from heavy resonances are highly boosted and populate the central rapidity region of the detector. For the purpose of this study, flat reconstruction identification efficiencies are assumed (see Table~\ref{tab:effs}).

The identification of jets that result from $\rm{\tau}$ decays or heavy flavour quarks --- $\rm{b}$ or $\rm{c}$ quarks --- typically involves the input from tracking information, such as vertex displacement or low-level detector input such as hit multiplicity~\cite{PerezCodina:2631478,PerezCodina:2635893}. Such information is not available as a default in~\delphes{}. Instead, a purely parametric approach based on MC generator information is used. The probability to be identified as $\rm{b}$ or $\rm{\tau}$ depends on user-defined parameterizations (see Table~\ref{tab:effs}). The behaviour of usual heavy flavour tagging algorithms in regimes of extreme boosts is yet unknown. We make the conservative assumption of vanishing efficiency as a function of the transverse momentum for both $\rm{b}$ and $\rm{\tau}$-jets, as shown in Table~\ref{tab:effs}. This choice is motivated by the fact that decay products originating from highly boosted $\rm{b}$ and $\rm{\tau}$ decays will be extremely collimated and highly displaced, making their reconstruction difficult. 

\begin {table}[htb!]
\begin{center}
\begin{tabular}{ l | c | c | c | c | c }
  & electrons & muons & photons & b-jets & $\tau$-jets\\
  \hline
  \hline
FCC-hh & 99\% & 95\% & 95\%  & (1 - \pt~[TeV]/15)$\cdot$85\% & (1 - \pt~[TeV]/30)$\cdot$60\% \\
HE-LHC & 95\% & 95\% & 95\% & (1 - \pt~[TeV]/5)$\cdot$75\% & (1 - \pt~[TeV]/5)$\cdot$60\%  \\
\end{tabular}
\caption{Global reconstruction efficiency of high \pt\ central objects for the HE-LHC and FCC-hh detectors in Delphes.}
\label{tab:effs}
\end{center}
\end{table}

A mis-tagging efficiency, that is, the probability that a particle other than $\rm{b}$ or $\rm{\tau}$ will be wrongly identified as a $\rm{b}$ or a $\rm{\tau}$ has been included in the simulation and assumes a similar falling behaviour as a function of the jet momentum. For b-tagging, the mistag efficiency are parameterised separately for light-jets (uds-quarks) and c-jets. For $\tau-$tagging, we consider only mis-identification from QCD jets. Table~\ref{tab:mistag} summarises the main values for the mis-tagging efficiency.

\begin {table}[htb!]
\begin{center}
\begin{tabular}{ l | c | c | c }
  & light (b-tag) & charm (b-tag) & QCD ($\tau$-tag)\\
  \hline
  \hline
FCC-hh & (1 - \pt~[TeV]/15)$\cdot$1\% & (1 - \pt~[TeV]/15)$\cdot$5\% & (8/9 - \pt~[TeV]/30)$\cdot$1\% \\
HE-LHC & (1 - \pt~[TeV]/5)$\cdot$1\%  & (1 - \pt~[TeV]/5)$\cdot$10\% & (1 - \pt~[TeV]/5)$\cdot$1\%  \\
\end{tabular}
\caption{Mis-identification efficiency of high \pt\ central heavy flavour jets for the HE-LHC and FCC detectors in Delphes.}
\label{tab:mistag}
\end{center}
\end{table}

\subsection{Treatment of the Monte Carlo samples}
\label{subsec:mctreat}
The modelling of the backgrounds in the high-$\pt$ tagging regimes is a challenging task. The requirement of $b$ tagging in some MC samples can drastically reduce the available statistics. This shortage of events that pass the $b$-tagging cut in the signal regime, in conjunction with the large cross section of some of the backgrounds can lead to very spiky templates. To overcome this problem the tag rate function (TRF) method is used. By using the TRF method, no event is cut based on its $b$-tagging count, but instead all the events are weighted. This weight can be interpreted as the probability of the given event to contain the desired number of $b$ jets. To achieve this, the tagging efficiency (a function of $\eta$, $\pt$ and true jet flavour) was used to calculate the event weight based on the kinematics and flavour of the jets found in each event. 
Despite the fact that very large amount of Monte Carlo statistics have been simulated in bins of $\hht$ and the usage of TRF to save events, there are still large statistical fluctuations from high weight events. In order to reduce this effect, and when large fluctuations are observed, the background spectrum is fitted. Further details on the TRF and fitting procedure are given in Appendix~\ref{sec:app:trf} and~\ref{sec:app:bgfit}, respectively.

\subsection{Statistical analysis}
Hypothesis testing is performed using a modified frequentist method based on a profile likelihood that takes into account the systematic uncertainties as nuisance parameters that are fitted to the expected Monte Carlo. The full shape information is used, with help from the sidebands to reduce the effect of systematic uncertainties in the signal region. The test statistics $q_\mu$ is defined as the profile log-likelihood ratio: $q_\mu = -2\ln({\cal L}(\mu,\hat{\hat{\theta}}_\mu)/{\cal L}(\hat{\mu},\hat{\theta}))$, where $\hat{\mu}$ and $\hat{\theta}$ are the values of the parameters that maximise the likelihood function (with the constraint $0\leq \hat{\mu} \leq \mu$), and $\hat{\hat{\theta}}_\mu$ are the values of the nuisance parameters that maximise the likelihood function for a given value of $\mu$. In the absence of any significant deviation from the background expectation, $q_\mu$ is used in the CL$_\text{s}$ method~\cite{Junk:1999kv,Read:2002hq} to set an upper limit on the signal production cross-section times branching ratio at the 95\%~CL. For a given signal scenario, values of the production cross section (parameterised by $\mu$) yielding CL$_\text{s} < 0.05$, where CL$_\text{s}$ is computed using the asymptotic approximation~\cite{Cowan:2010js}, are excluded at 95\%~CL. For a $5\sigma$ discovery, the quantity 1-CL$_\text{b}$ must be smaller than $2.87 \cdot 10^{-7}$~\cite{Junk:1999kv} and is also computed using the asymptotic approximation.

\section{Studies at 100 TeV}
\subsection{Leptonic final states}
\label{sec:lep}

The decay products of heavy resonances are in the multi-TeV regime and the capability to reconstruct their momentum imposes stringent requirements on the detector design. 
In particular, reconstructing the track curvature of multi-TeV muons requires excellent position resolution and a large lever arm. In this section, the expected sensitivity is presented for a \Zpll\ (where $\ell=e,\mu$) and \Zptata\ separately.

\subsubsection{The \texorpdfstring{\ee}{ee} and \texorpdfstring{\mumu}{mumu} final states}
\label{sec:lepee}

\begin{figure}[!htb]
  \centering
  \includegraphics[width=0.48\columnwidth]{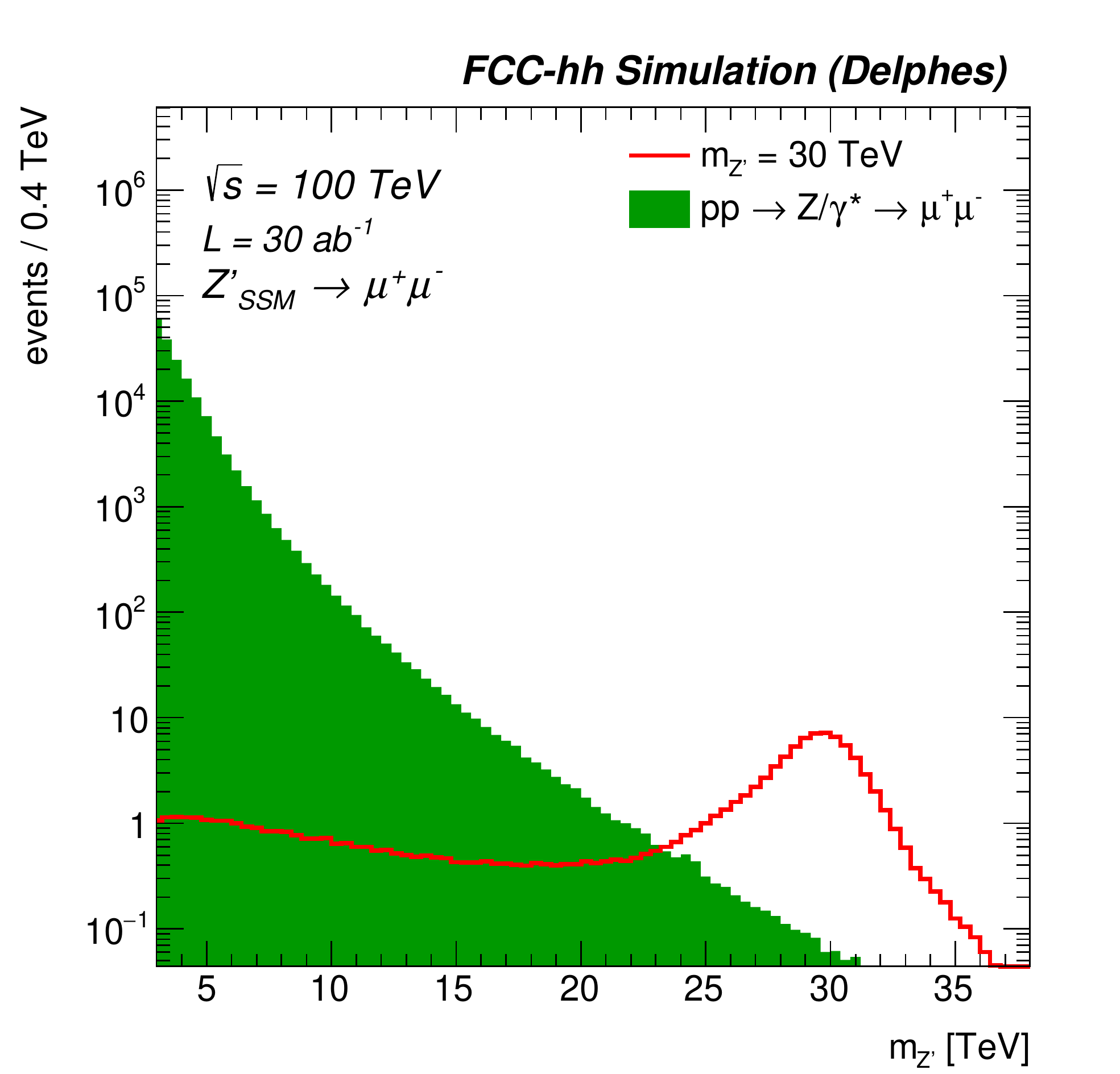}
  \includegraphics[width=0.48\columnwidth]{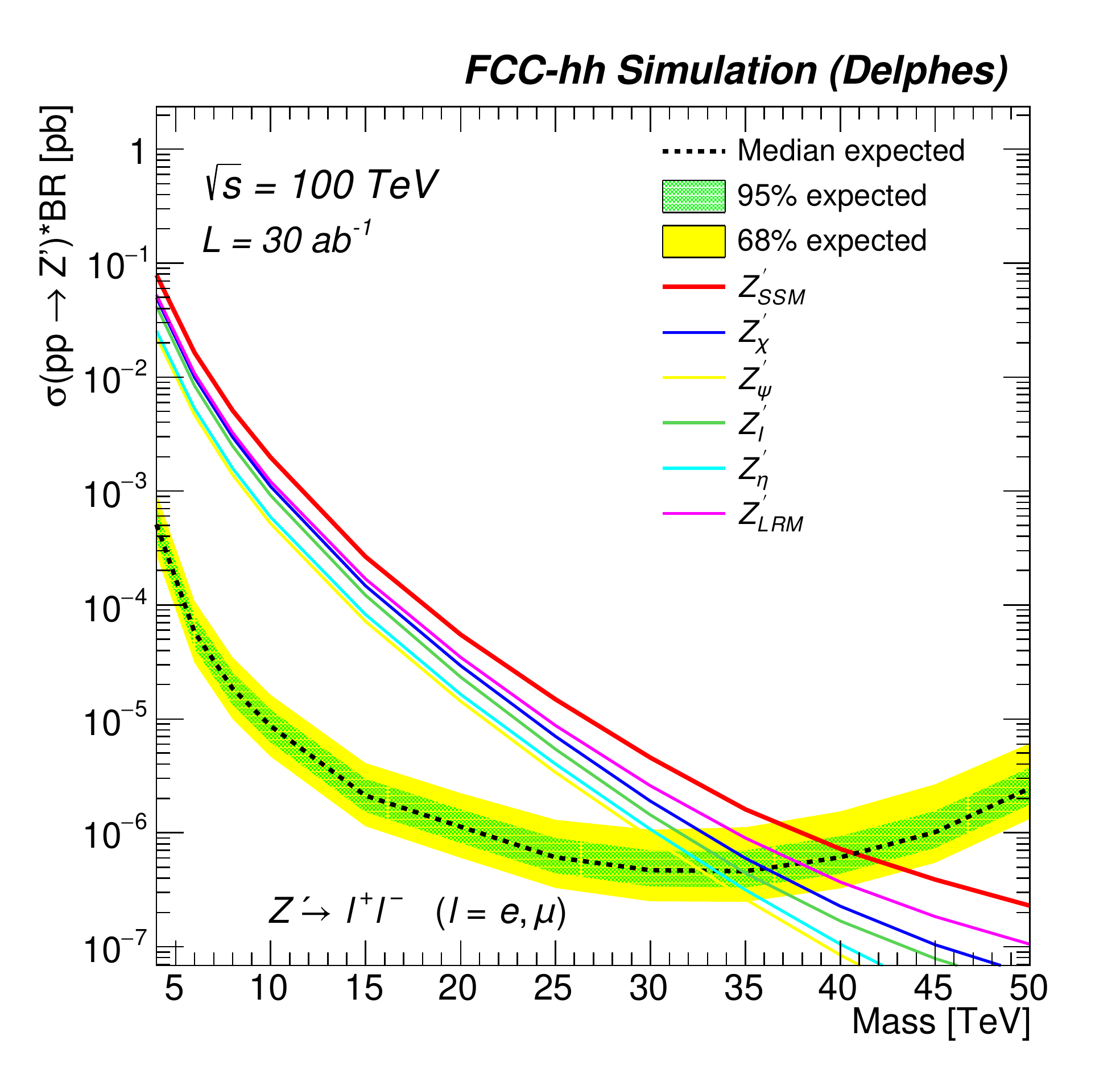}
  \caption{Left: Invariant mass for a 30\,TeV signal after full event selection for the $\mu\mu$ channel. Right: 95\% CL limit versus mass for the combined di-lepton ($ee$, $\mu\mu$) channel.}
  \label{figure:leptonicresonances:ll}
\end{figure}

Events are required to contain two isolated opposite-sign leptons with $\pt > 1$\,TeV, $|\eta|$<4 and an invariant mass \mll\ > 2.5\,TeV.
Figure~\ref{figure:leptonicresonances:ll} left shows the invariant mass for a 30\,TeV \ZpSSM\ signal for the $\mu\mu$ channel for FCC-hh. The di-electron invariant mass spectrum is not shown, but as expected from the calorimeter constant term that dominates the resolution at high \pt, the mass resolution is better for the $ee$ channel.
The di-lepton invariant mass spectrum is used as the discriminant and a 50\% normalisation uncertainty on the background is assumed (this uncertainty is extremely conservative, but does not affect the final results, due to the negligible background in the largest mass regions).
Figure~\ref{figure:leptonicresonances:ll} (right) shows the 95\% CL exclusion limit obtained with 30\,ab$^{-1}$ of data combining ee and $\mu\mu$ channels. Figure~\ref{figure:leptonicresonances:vslumi} (left) shows the integrated luminosity required to reach a $5\sigma$ discovery as a function of the mass of the heavy resonance. The \Zpee\ and \Zpmumu\ channels display very similar performance due to the low background rates. We conclude therefore that the reference detector design features near to optimal performance for searches involving high \pt\ muon final states. Combining $ee$ and $\mu\mu$ channels, masses up to $\sim$ 42\,TeV can be excluded or discovered.

\subsubsection{The \texorpdfstring{\tautau}{tautau} final state}
\label{sec:leptautau}

\begin{figure}[htb!]
  \centering
  \includegraphics[width=0.48\columnwidth]{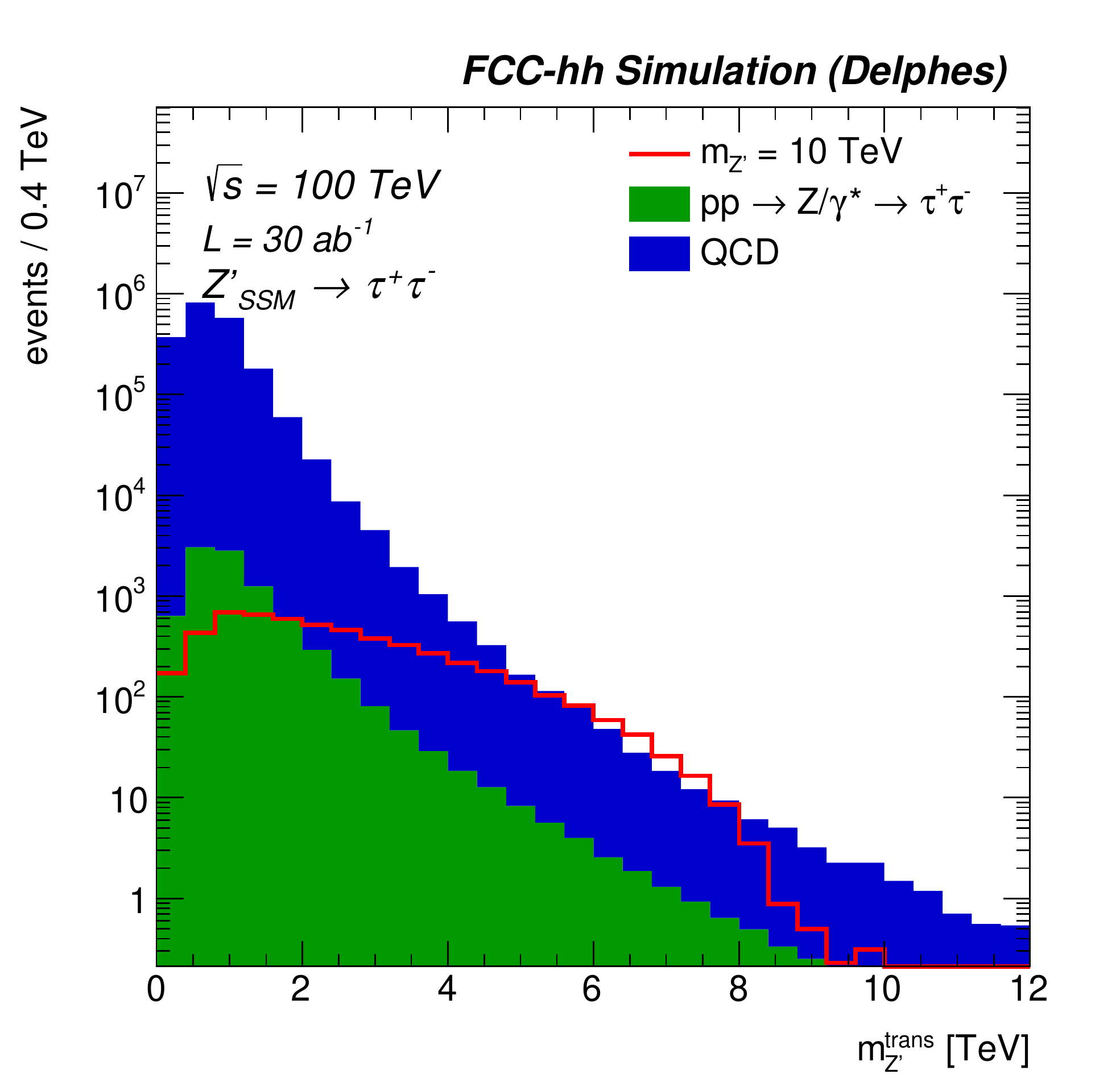}
  \includegraphics[width=0.48\columnwidth]{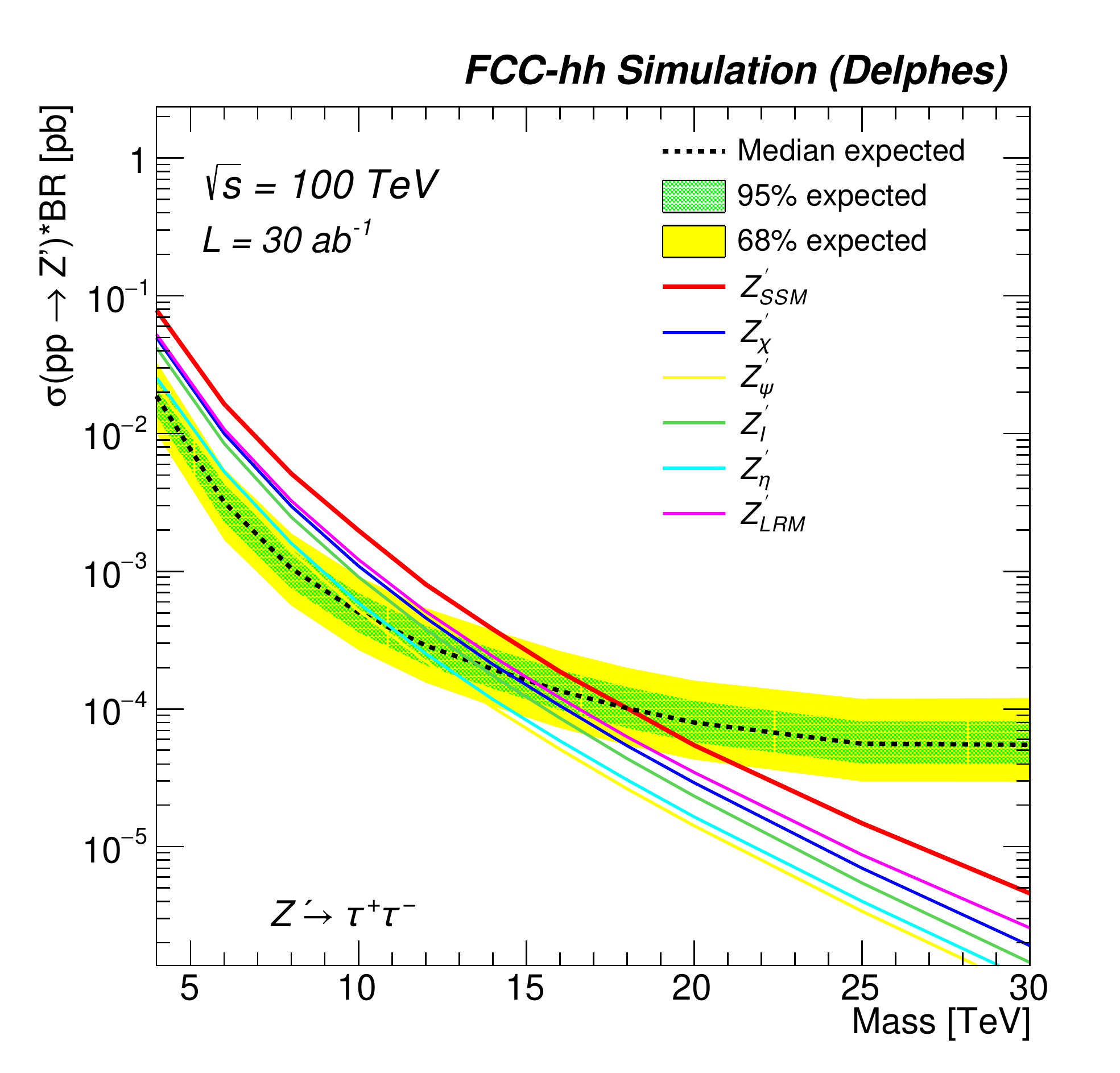}
  \caption{Left: Di-$\tau$ transverse mass for a 10\,TeV signal after full event selection. Right: 95\% CL limit versus mass.}
  \label{figure:leptonicresonances:tautau}
\end{figure}

\begin{figure}[htb!]
  \centering
  \includegraphics[width=0.48\columnwidth]{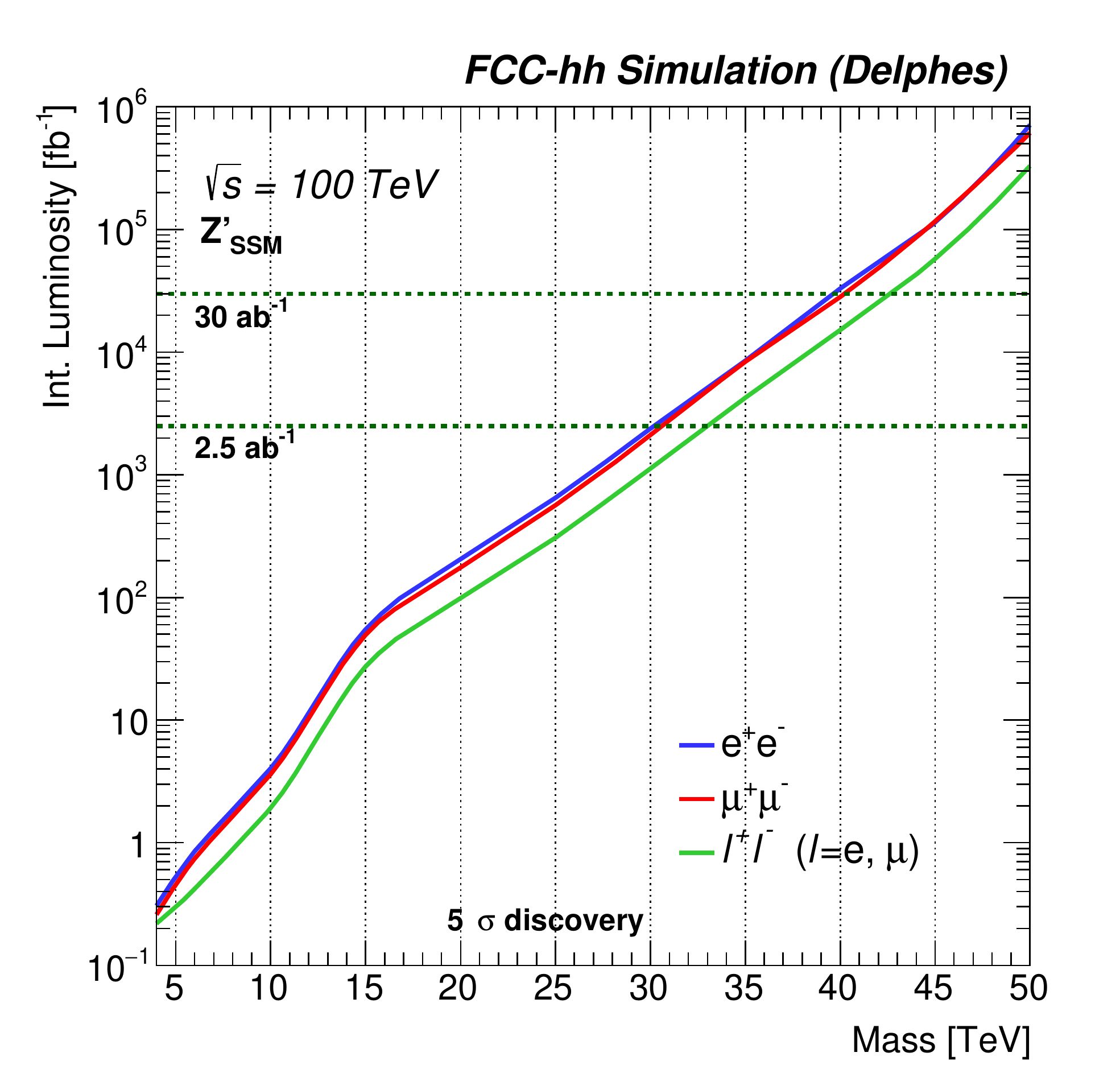}
  \includegraphics[width=0.48\columnwidth]{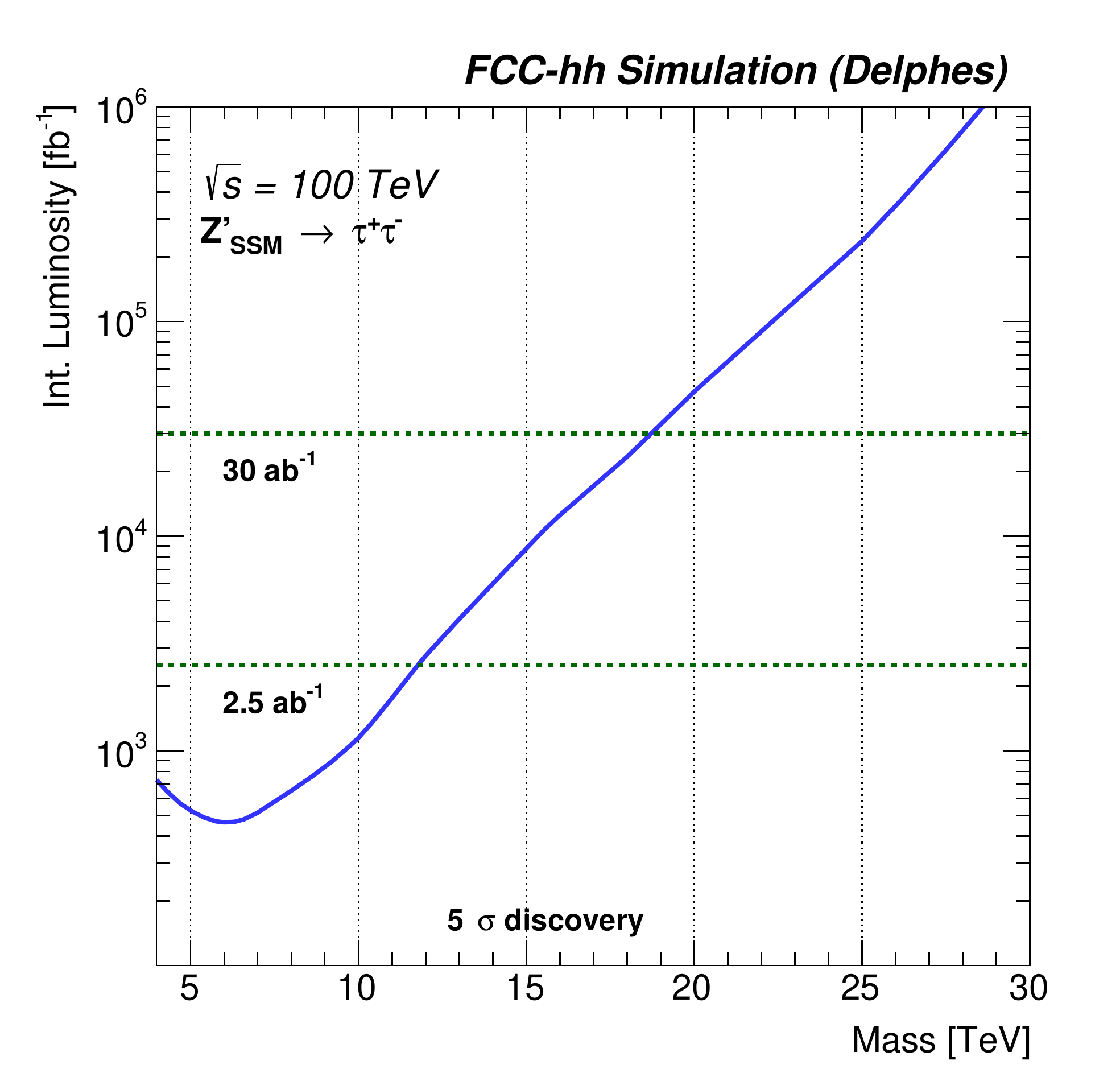}
  \caption{Integrated luminosity versus mass for a $5\sigma$ discovery, for $e/\mu$ final states (left) and for the $\tau$ final states (right).}
  \label{figure:leptonicresonances:vslumi}
\end{figure}

At the LHC, the most sensitive channel to search for high-mass di-$\tau$ resonances is when both $\tau$ leptons decay hadronically~\cite{Khachatryan:2016qkc}. The analysis presented in this section focuses on this decay channel alone. The event selection requires two jets with $p_{T} > 0.5$\,TeV and $|\eta|<2.5$, both identified as $\tau$'s. To ensure no overlap between the $\ell=e,\mu$ and $\tau$ final states, jets containing an electron or a muon with $\pt > 100$\,GeV are vetoed. The requirements of $\Delta \phi(\tau_1, \tau_2)> 2$ and $2.5<\Delta R(\tau_1, \tau_2)<4$ are applied to suppress multi-jet backgrounds. Furthermore, mass dependent cuts are applied to maximise the signal significance and are summarised in Table~\ref{tab:leptonicresonances:tautau}. Several proxies for the true resonance mass have been tested, such as the invariant mass of the two $\tau$'s, with and without correction for the missing energy, however the transverse mass~\footnote{The transverse mass is defined as $m_{T}  =  \sqrt{2\ptZp*\met*(1-cos\Delta\phi(\Zp,\met))} $.} provides the best sensitivity and is therefore used to estimate the sensitivity.
Figure~\ref{figure:leptonicresonances:tautau} shows the di-$\tau$ transverse mass (left) for a 10\,TeV \ZpSSM{} and the 95\% CL exclusion limits for 30\,ab$^{-1}$ of data (right). The required integrated luminosity versus mass of the resonance to reach a $5\sigma$ discovery is shown in Fig.~\ref{figure:leptonicresonances:vslumi} (right). 

Heavy resonances decaying to $\tau$ leptons reconstructed in the hadronic decay mode are more challenging than $ee$ or $\mu\mu$, given the overwhelming multi-jet background. We find that masses up to 18\,TeV can be probed at the FCC-hh. We note that the assumed $\tau$ identification efficiency considered in this analysis is assumed to be conservative (see Section~\ref{appsub:objid}), and only a study based on full detector simulation could provide more realistic numbers. 

\begin{table}[htb!]
   \centering
\begin{tabular}{c|c|c|c}
   $\Zp$ mass [TeV] &  $\Delta \phi(\tau_1, \tau_2)$&  $\Delta R(\tau_1, \tau_2)$ & $\met$\\
  \hline
  \hline
  $4-8$ & > 2.4 & > 2.5 and < 3.5 & > 400 GeV\\
  $10$ & > 2.4 & > 2.7 and < 4 & > 300 GeV\\
  $12-14$ & > 2.6 & > 2.7 and < 4 & > 300 GeV\\
  $16-18$ & > 2.7 & > 2.7 and < 4 & > 300 GeV\\
  $>18$ & > 2.8 & > 3 and < 4 & > 300 GeV\\
  \end{tabular}
  \caption{List of mass dependent cuts optimised to maximise the sensitivity for the \Zptata\ search.}
  \label{tab:leptonicresonances:tautau}
\end{table}

\subsubsection{Sensitivity to flavour-anomaly inspired \texorpdfstring{\Zp}{zp} models}
\label{sec:zprimeflav}
LHCb measurements in $B\rightarrow~K^*\mu^+\mu^-$ decays are somewhat discrepant with SM predictions~\cite{Aaij:2014ora,Aaij:2017vbb}. They may be harbingers of new physics at an energy scale potentially accessible to direct discovery. The reach for the "naive" flavour anomaly $Z'$ model from Ref.~\cite{Allanach:2017bta} is studied in this section. In this model it is assumed that the $Z'$ only couples to $b/s$ quarks ($g_{sb}$) and to muons ($g_{\mu\mu}$). This assumption is maximally conservative in the sense that it assumes a minimal set of non-vanishing couplings of the new resonance to quarks. Additional quark couplings would have the effect of increasing the production cross section thereby increasing the reach. For each $Z'$ mass hypothesis $m_{Z'}$, the product of $g_{sb}$ and $g_{\mu\mu}$ is fixed by the observed $R_{K^{(*)}}$ anomalies:

\begin{equation}
\frac{g_{bs}~g_{\mu\mu}}{m_{Z'}^2} \approx \frac{1}{(30~\mathrm{TeV})^2}.
\label{eq:anomaly_constraint}
\end{equation}

In this study only one direction in the ($g_{sb}$, $g_{\mu\mu}$) plane~\footnote{More precisely, constraints from $B_{s}~-~\bar{B}_{s}$ mixing measurements and from unitarity set an upper limit on the allowed $g_{sb}$ and $g_{\mu\mu}$ parameters.} has been explored in which both couplings are re-scaled by a common mass dependent factor compatible with Equation~\ref{eq:anomaly_constraint}. The scaling of the couplings is defined by: $g_{\mu\mu} = m_{Z'}/(5~\mathrm{TeV})$ and $g_{sb} = 0.023~m_{Z'}/(5~\mathrm{TeV})$. 

\begin{figure}[!htb]
  \centering
  \includegraphics[width=0.48\columnwidth]{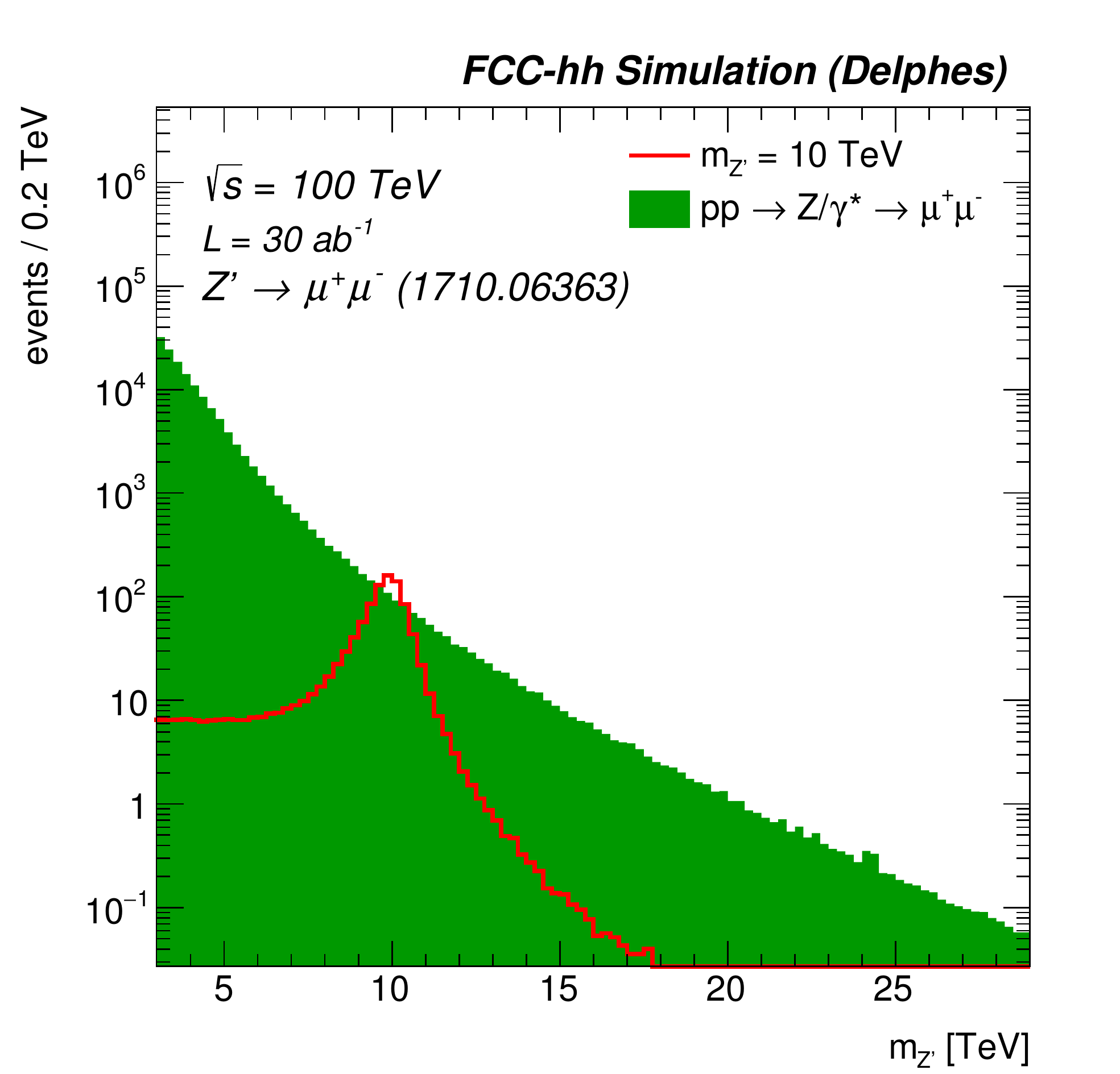}
  \includegraphics[width=0.48\columnwidth]{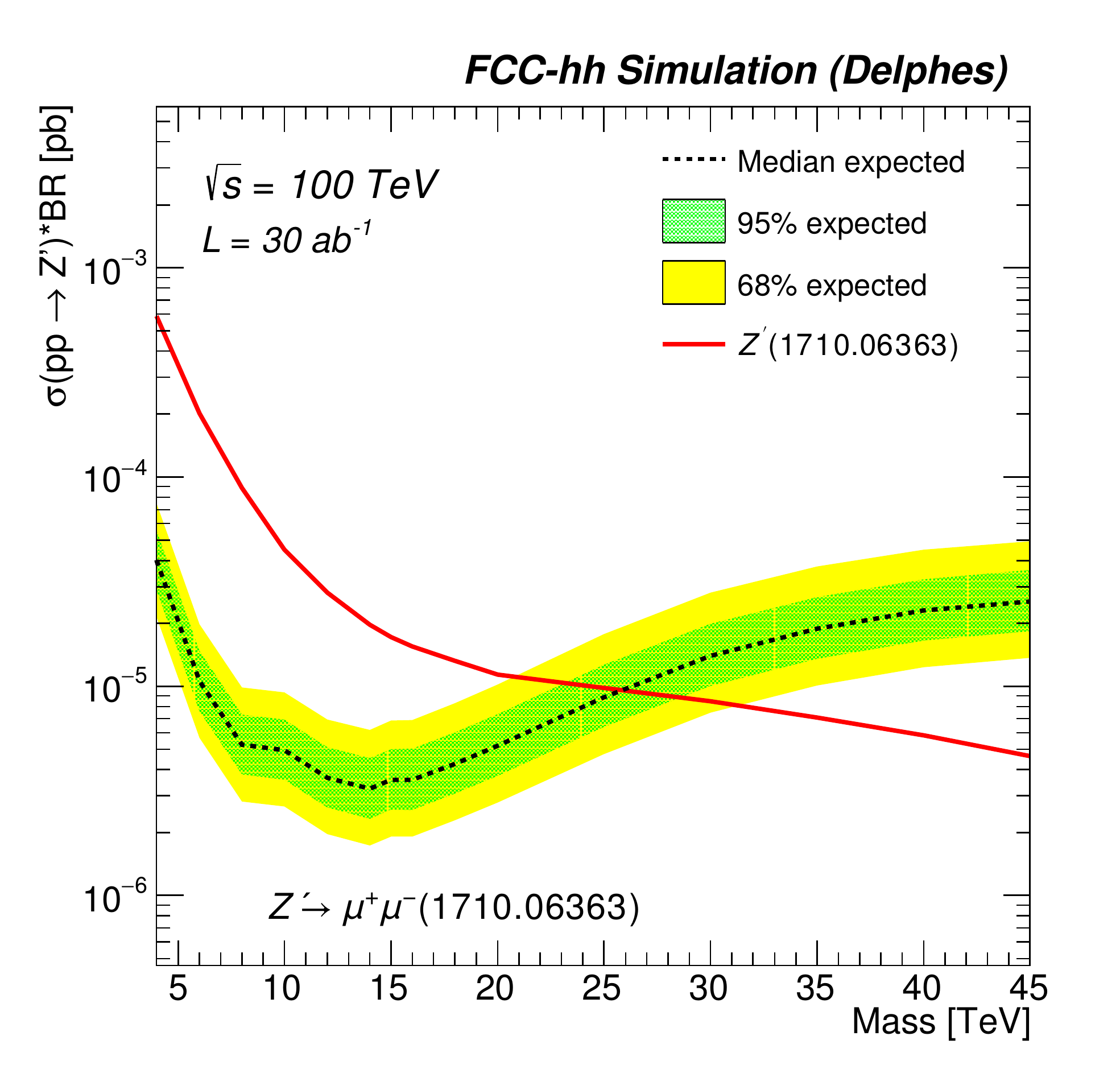}
  \includegraphics[width=0.48\columnwidth]{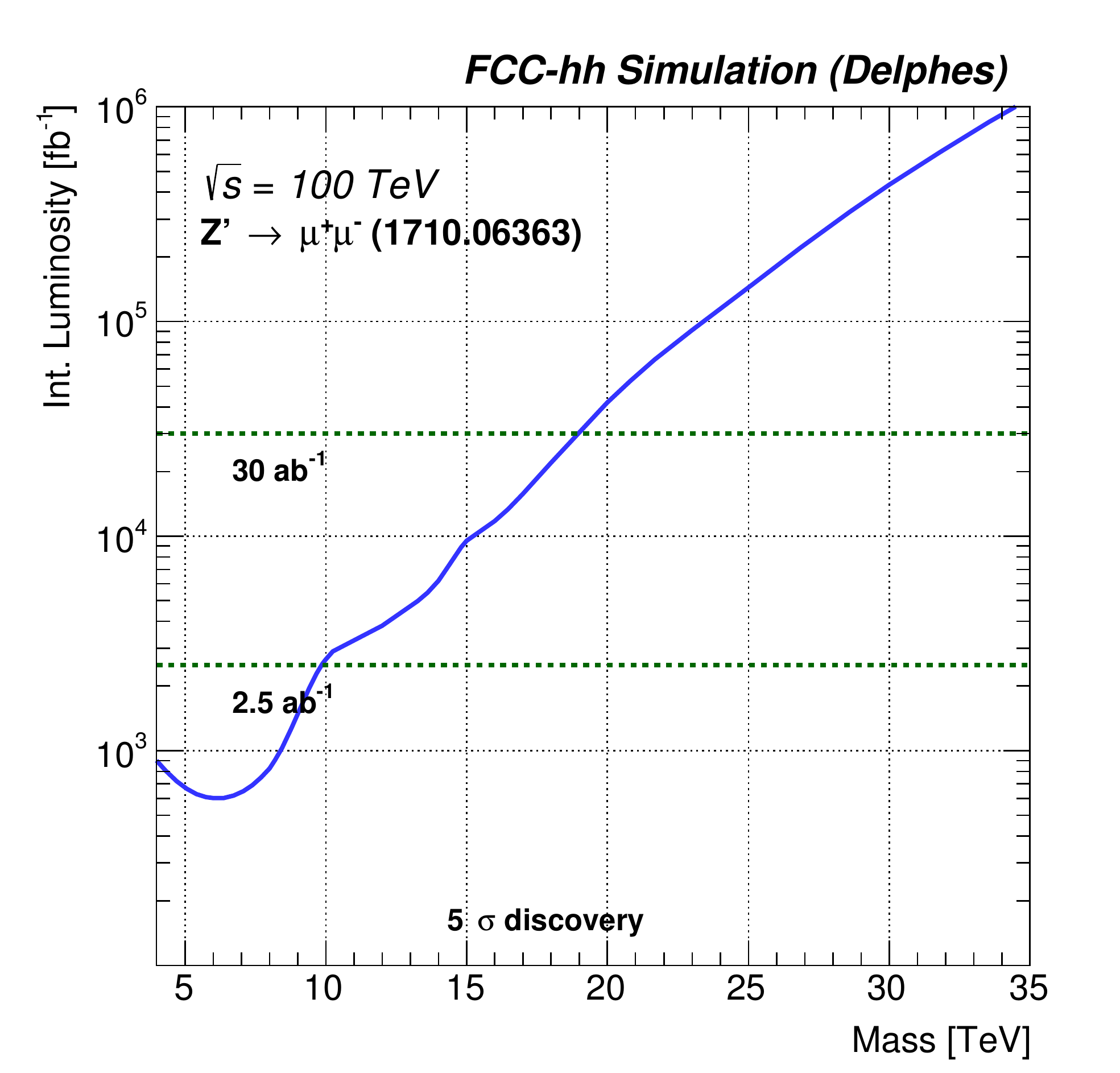}
  \caption{Left : Invariant mass for a 10~TeV signal after full event selection in flavour anomaly scenario. Limit versus mass (right) and luminosity for a $5\sigma$ discovery (bottom). }
  \label{figure:leptonicresonances:resultsmumu_flav}
\end{figure}

The event selection strategy described in Section~\ref{sec:lepee} has been applied here. Figure~\ref{figure:leptonicresonances:resultsmumu_flav} shows the invariant mass of the $\mu^+\mu^-$ system (left) for $m_{Z'} = 10$~TeV, the 95\% CL exclusion limit obtained (right) and the integrated luminosity required to reach a $5\sigma$ discovery as a function of the mass of the $\mu^+\mu^-$ system (bottom). We find that $m_{Z'} < 25$~TeV can be excluded with \intlumifcc. Models with $m_{Z'} > 18$~TeV are not allowed because they either violate unitarity or they are incompatible with $B_{s}~-~\bar{B}_{s}$ mixing measurements~\cite{DAmico:2017mtc}. We conclude therefore that the full allowed mass range can be excluded at the FCC-hh with the detector performance discussed in Section~\ref{subsec:detparam}, in agreement with the findings in Ref.~\cite{Allanach:2017bta}. We note that further studies involving more realistic models can be found in Ref.~\cite{Allanach:2018odd}.

\subsection{Hadronic final states}
\label{sec:hadronic}

Heavy resonances decaying hadronically also impose stringent requirements on the detector design. Precise jet energy resolution requires full longitudinal shower containment. Highly boosted top quarks and $W$ bosons decay into highly collimated jets and differ from standard QCD jets by a characteristic inner jet sub-structure. An excellent granularity both in the tracking detectors and calorimeters is therefore required to resolve the sub-structure of jets since a high discrimination power against QCD backgrounds can result in increased sensitivity in heavy resonance searches. 

\subsubsection{Multi-Variate object tagging}
\label{subsec:mvatagger}

An important ingredient of the hadronic searches is the identification of heavy boosted top quarks and $W$ bosons. Two object-level taggers using Boosted Decision Trees (BDT) were developed to discriminate $W$ and top jets against the light jet flavours treated as background.
Top and $W$ taggers were optimised using jets with a transverse boost of $\pt=$10 TeV. At these extreme energies, $W$ and top jets have a characteristic angular size $R=0.01-0.02$, i.e., smaller than the typical electromagnetic and hadronic calorimeter cells. Following the approach described in~\cite{Larkoski:2015yqa}, we exploit the superior track angular resolution and reconstruct jets from tracks only using the anti-$k_T$ algorithm with a parameter R=0.2, but also larger values are used to increase the discrimination power of the BDT. The missing neutral energy is corrected for by rescaling the track 4-momenta by the factor $\ptSub{,trk}/\ptSub{,PF}$, where $\ptSub{,trk}$ is the track jet \pt\ and $\ptSub{,PF}$ is the Particle-Flow (PF) Jet \pt. In what follows, we will simply refer to ``track jets'' as the jet collection that includes the aforementioned rescaling. The boosted top tagger is built from jet substructure observables: the soft-dropped jet mass~\cite{Larkoski:2014wba} (\mSD) and N-subjettiness~\cite{Thaler:2010tr} variables $\tau_{1,2,3}$ and their ratios $\tau_{2}/\tau_{1}$ ($\tau_{21}$) and $\tau_{3}/\tau_{2}$ ($\tau_{32}$). The $W$-jet tagger also uses ``isolation-like'' variables, first introduced in Ref.~\cite{Mangano:2016jyj} that exploit the absence of high \pt\ final-state radiation (FSR) in the vicinity of the $W$ decay products. We call these variables $E_{F}(n,\alpha)$ and define them as:


\begin{equation}
E_{F}(n,\alpha) =  \sum \limits_{\frac{n-1}{5}\alpha < \Delta R(k,jet)< \frac{n}{5}\alpha} \ptSup{(k)} \Biggm/ \sum \limits_{\Delta R(k,jet)< \alpha} \ptSup{(k)}, 
\end{equation}
with $k$ running over the jet constituents and $\alpha=0.05$. We construct 5 variables $E_{F}(n,\alpha)$ with $n=[1,2,3,4,5]$ and use them as input to the BDT.
The $W$ tagger has significantly better performance than the top tagger thanks to the excellent discrimination power of the energy-flow variables. We choose the working points for the analyses presented later, with a top and $W$ tagging efficiencies of $\epsilon_S^{\text{top}}=60\%$ and $\epsilon_S^{\text{W}}=90\%$, corresponding to a background rejection of $\epsilon_B^{\text{top}}=\epsilon_B^{\text{W}}=90\%$. More details on the multi-Variate object tagging can be found in Appendix~\ref{sec:app:mva}.

\subsubsection{The \jj\ final state}
\label{sec:hadjj}

\begin{figure}[!htb]
  \centering
  \includegraphics[width=0.48\columnwidth]{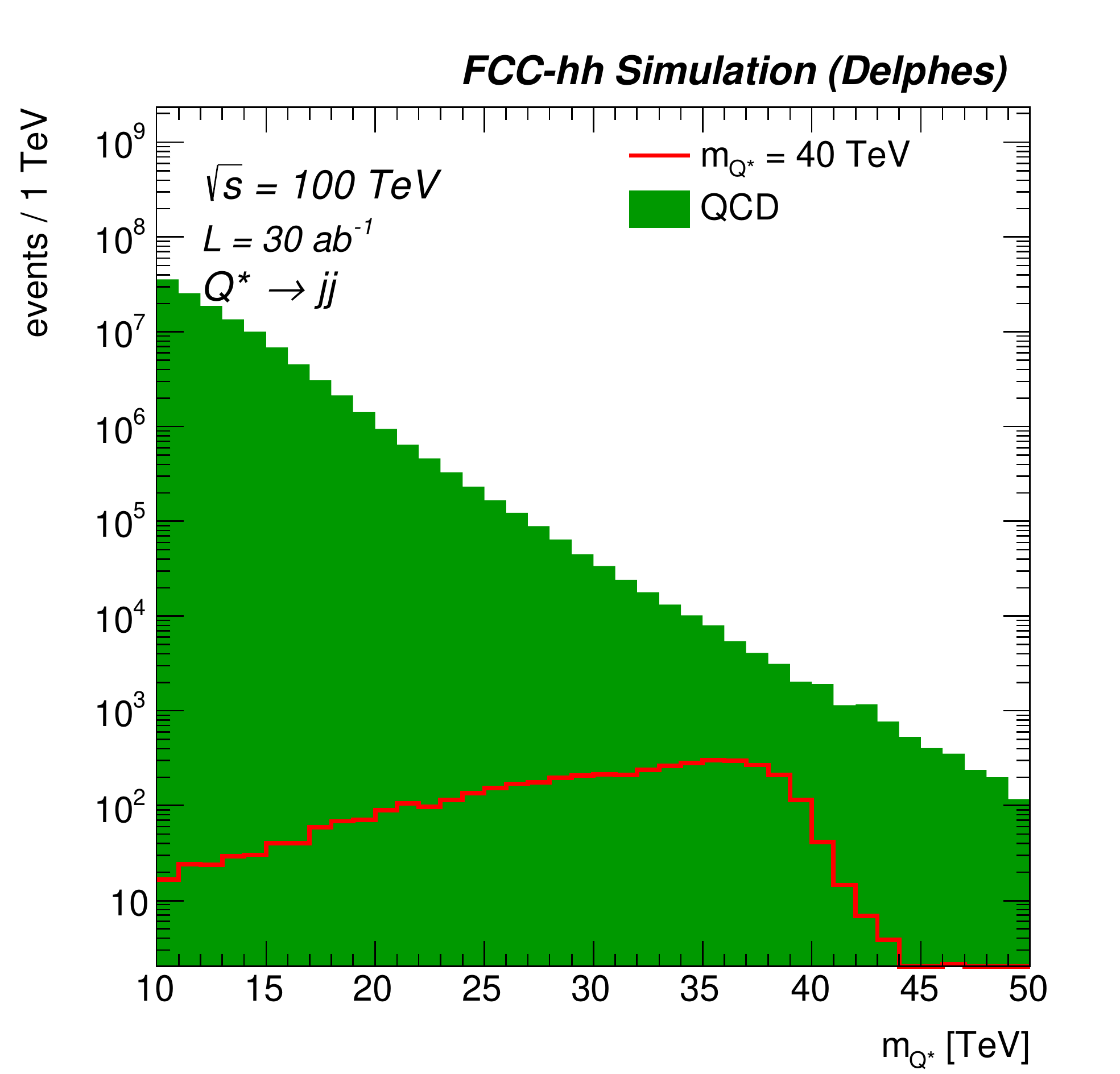}
  \includegraphics[width=0.48\columnwidth]{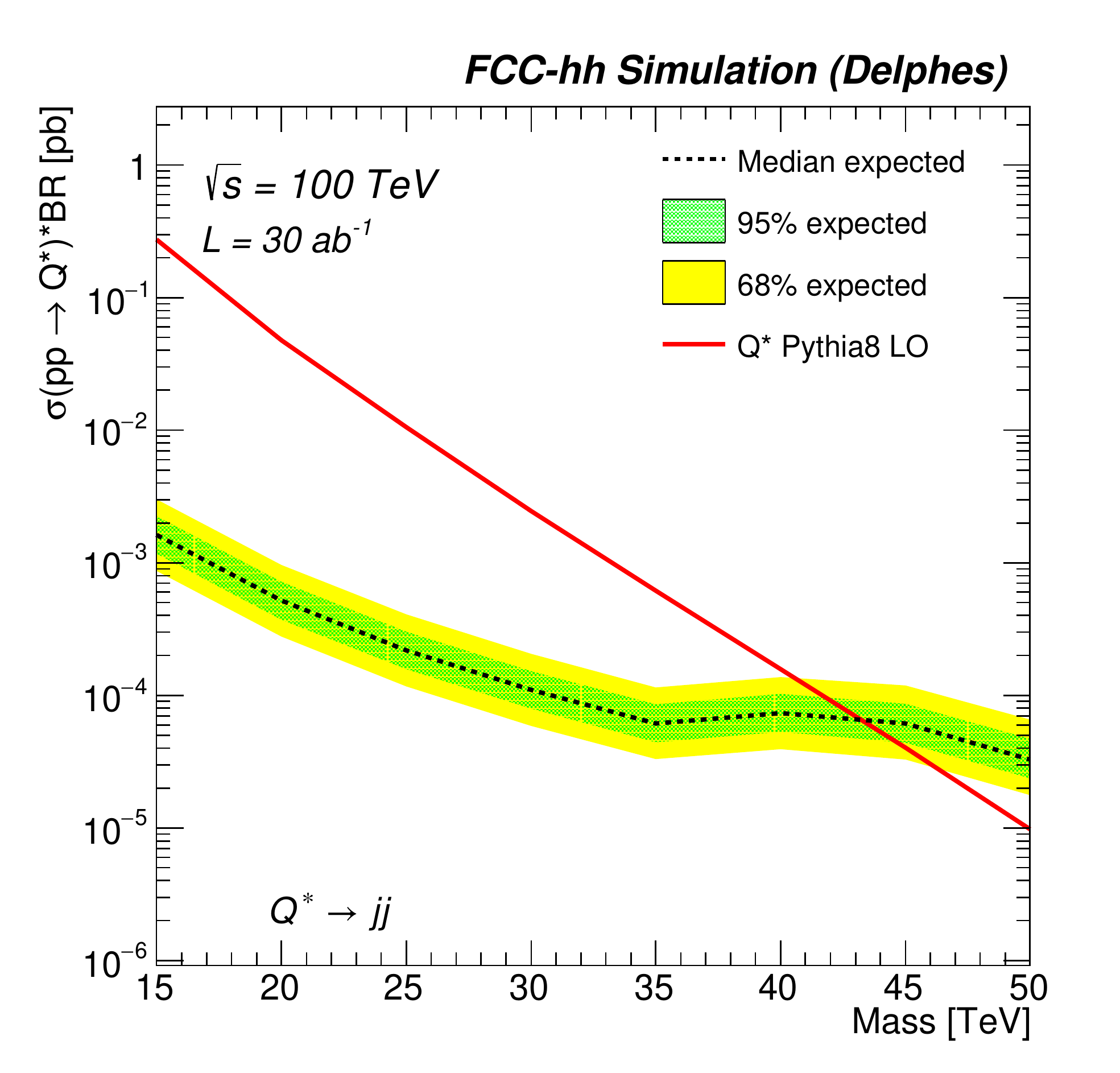}
   \caption{Invariant mass distribution of the two selected jets for a 40\,TeV signal (left), 95\% CL limit versus mass (right).}
  \label{figure:hadronicresonances:jj}
\end{figure}

Jets are clustered using particle-flow candidates with the anti-$k_T$~\cite{Cacciari:2008gp} algorithm and  parameter R=0.4. We require at least two very energetic jets with $\pt$>3\,TeV and $|\eta|<3$. As di-jet events from the signal will tend to be more central than for the background, the rapidity difference between the two leading jets $\Delta(\eta_1, \eta_2)$ is required to be smaller than 1.5. The di-jet invariant mass for the \qjj\ signal with a mass of 40\,TeV together with the QCD contribution after the full event selection, is shown on Fig.~\ref{figure:hadronicresonances:jj} (left). The right figure shows the 95\% CL exclusion limit obtained with 30\,ab$^{-1}$ of data, and the left panel of Fig.~\ref{figure:hadronicresonances:vslumi} shows the integrated luminosity required to reach a $5\sigma$ discovery as a function of the Q$^*$ mass. For this very simple case of a strongly coupled object we reach 95\%~CL exclusion limits of 43\,TeV and 5$\sigma$ discovery reach of 40\,TeV with 30\,ab$^{-1}$ of integrated luminosity. 

\subsubsection{The \texorpdfstring{\ttbar}{tt} final state}
\label{sec:hadtt}

\begin{figure}[!htb]
  \centering
   \includegraphics[width=0.48\columnwidth]{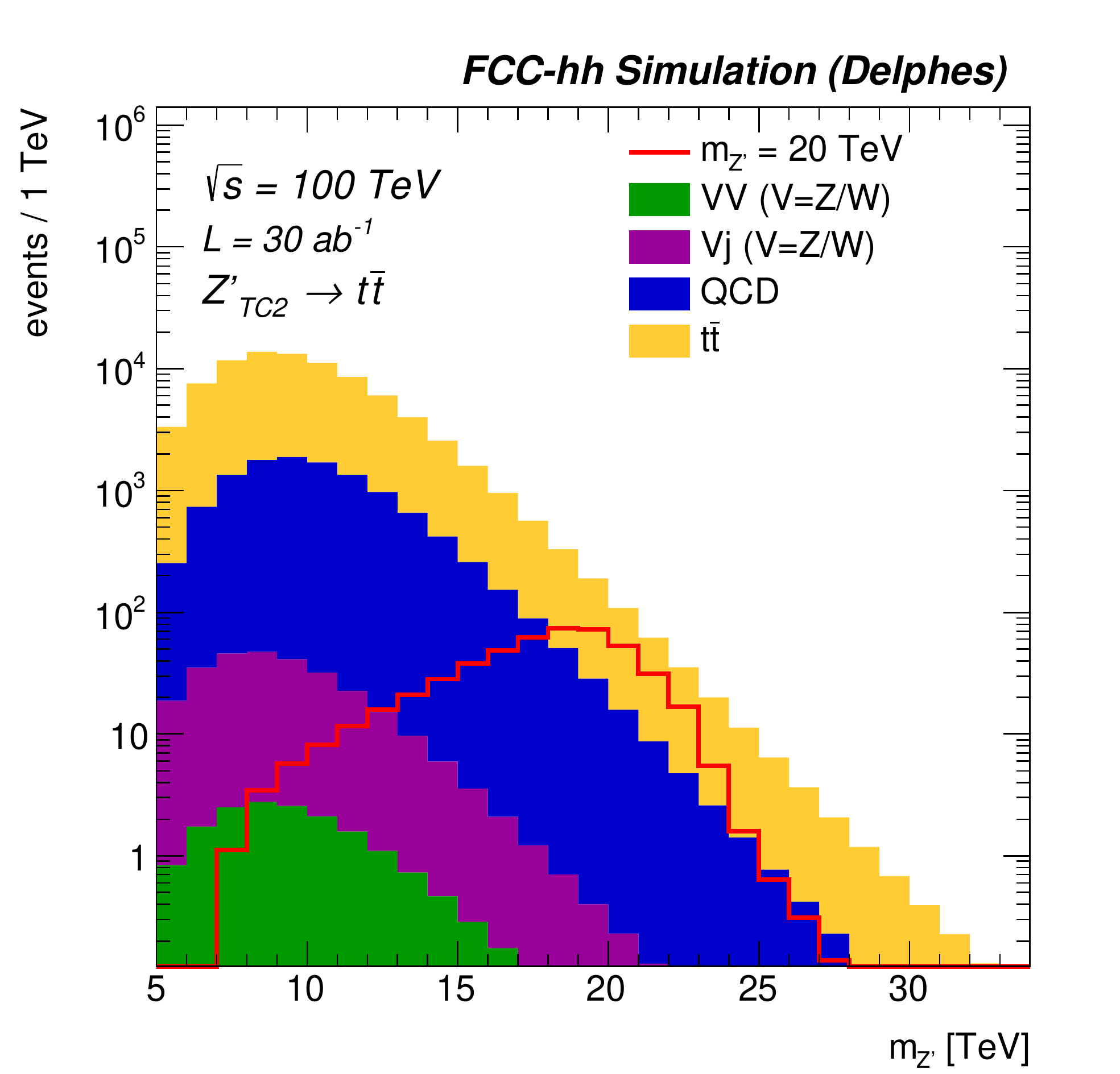}
   \includegraphics[width=0.48\columnwidth]{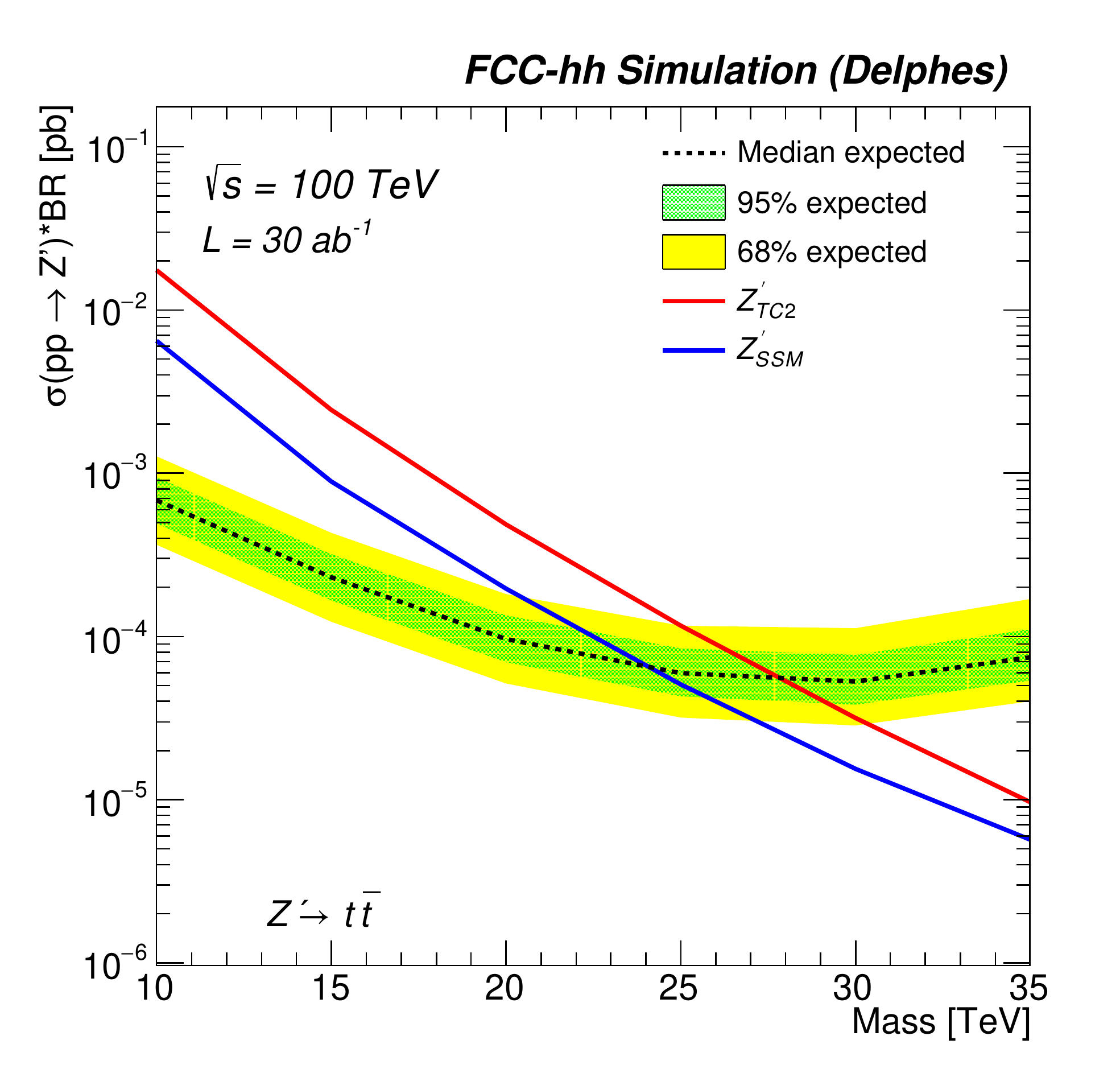}
  \caption{Invariant mass distribution of the two selected top-jets (left) for a 20\,TeV signal (left), 95\% CL limit versus mass (right).}
  \label{figure:hadronicresonances:tt}
\end{figure}

\begin{figure}[!htb]
  \centering
    \includegraphics[width=0.48\columnwidth]{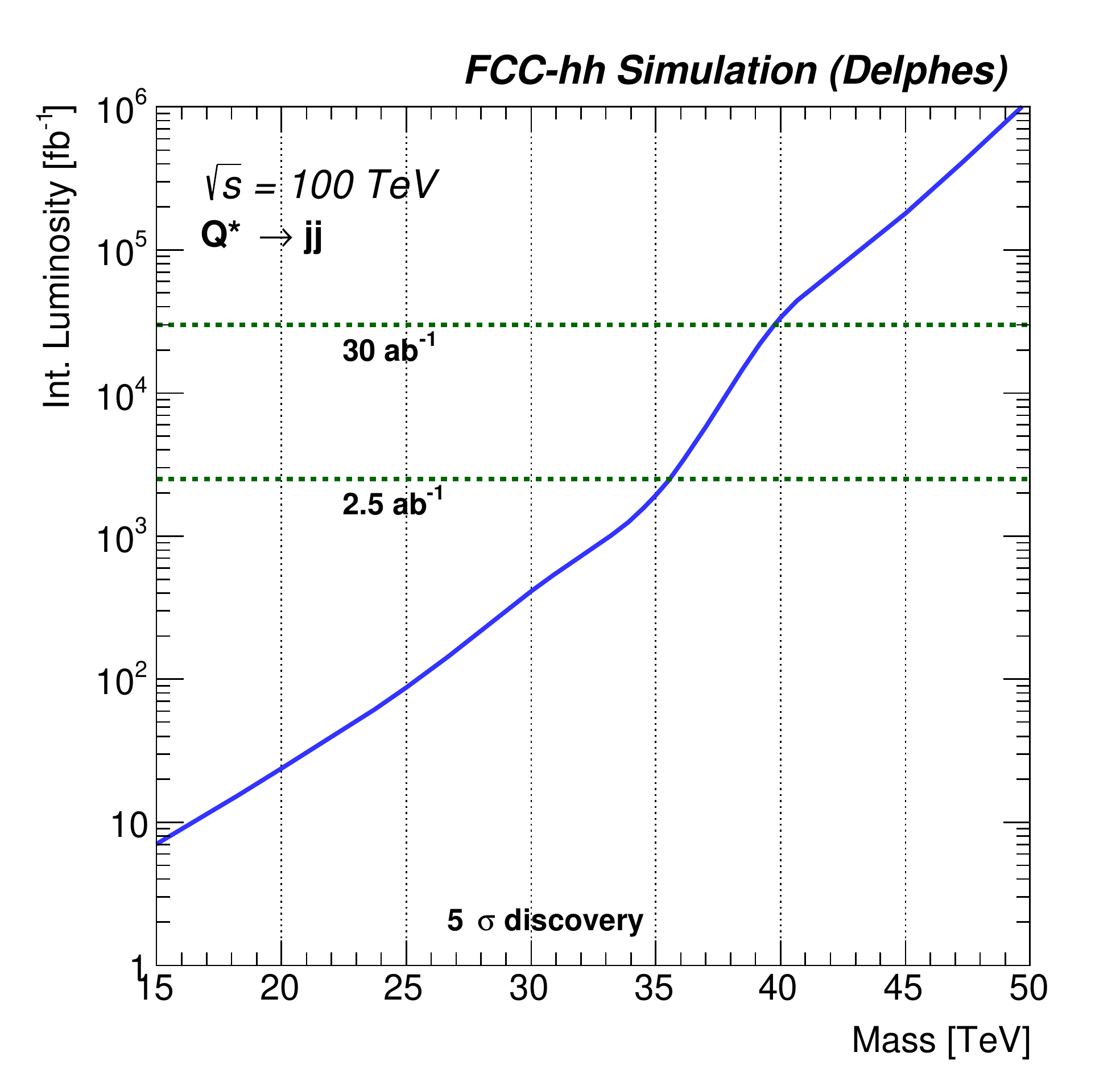}
   \includegraphics[width=0.48\columnwidth]{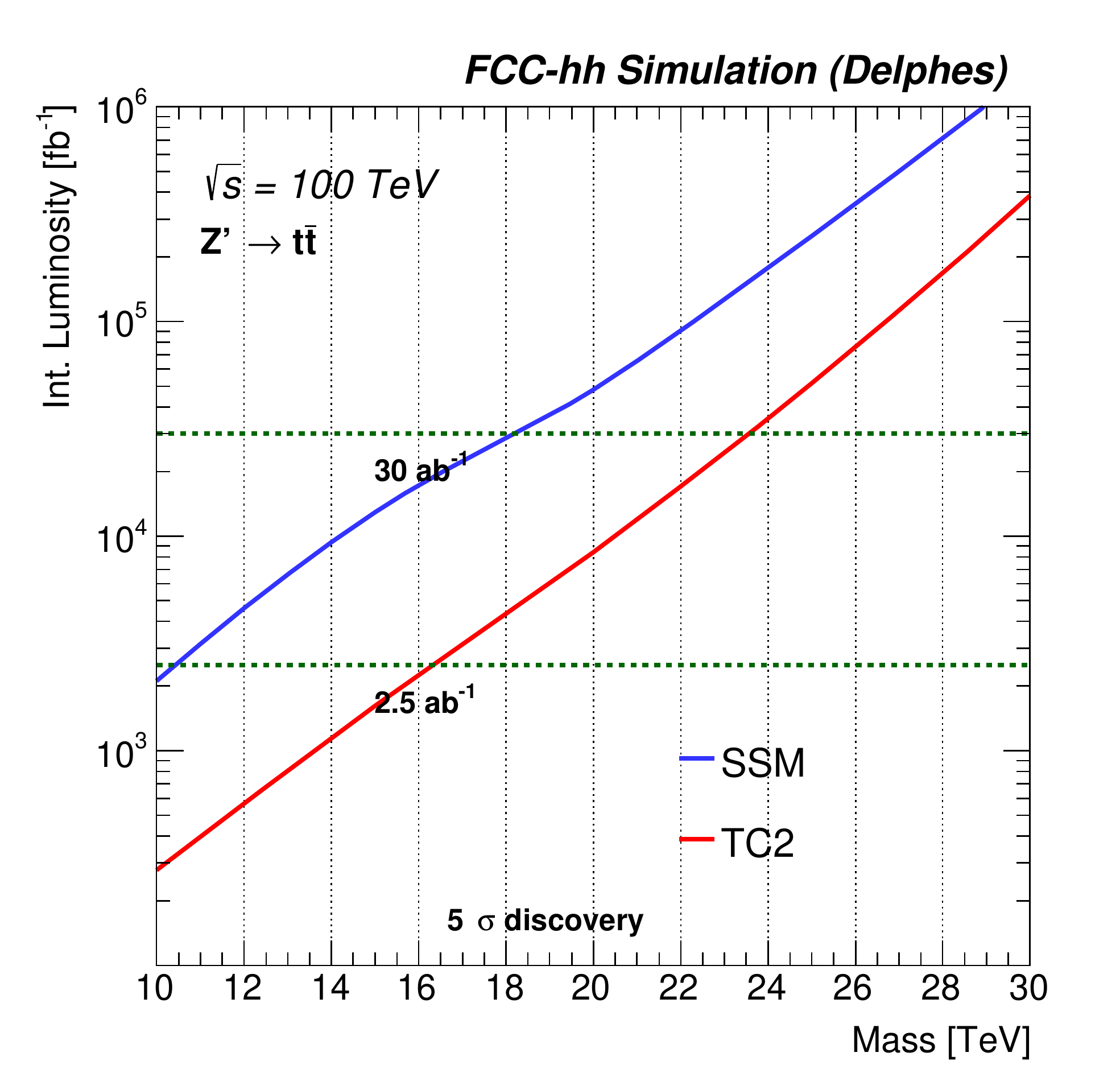}
  \caption{$5\sigma$ discovery reach for the $jj$ (left) and $t\bar{t}$ (right) final states.}
  \label{figure:hadronicresonances:vslumi}
\end{figure}

To resolve the jet sub-structure, track jets are found to perform better compared to particle-flow jets, thus the \rsg\ and \Zptt\ searches make use of track jets. As no lepton veto is applied, there is also some acceptance for leptonic decays and the sensitivity to semi-leptonic or \ttbar\ decays is enhanced by adding the $\metvec$ vector to the closest jet 4-momentum (among the two leading jets).
We require two jets with a $\pt$>3\,TeV and $|\eta|<3$ and $\Delta(\eta_1,\eta_2)<2.4$. Both jets must be tagged as ``top jets''  (using the tagger defined in Section~\ref{subsec:mvatagger}). In addition, the two selected high-\pt\ jets must be tagged as b-jets. Finally, to further reject QCD events, we require for both jets the soft-dropped mass to be larger than 40\,GeV. Figure~\ref{figure:hadronicresonances:tt} (left) shows the di-top invariant mass distribution after the final event selection for a 20\,TeV signal from a Topcolor $Z'$ and backgrounds. Thanks to the BDT discriminant, the largest background contribution is top pair production itself and the QCD contribution is now the second leading one. The right panel shows the 95\% CL exclusion limit obtained with 30\,ab$^{-1}$ of data and the right panel of Fig.~\ref{figure:hadronicresonances:vslumi} shows the integrated luminosity required to reach a $5\sigma$ discovery as a function of the Z$^\prime$ mass. Further developments to improve the mass resolution could be considered to improve the sensitivity, but already with such wide spectrum, exclusions between 25 and 28\,TeV and discoveries between 18 and 24\,TeV are reached depending on the model (\ZpSSM\ or leptophobic \ZpTC). A more extensive study of the \ttbar\ decay at a 100\,TeV collider ignoring the simulation of the detector response can be found in Ref.~\cite{Auerbach:2014xua}. 

\subsubsection{The \texorpdfstring{\ww}{ww} final state}
\label{sec:hadww}
\begin{figure}[!htb]
  \centering
  \includegraphics[width=0.48\columnwidth]{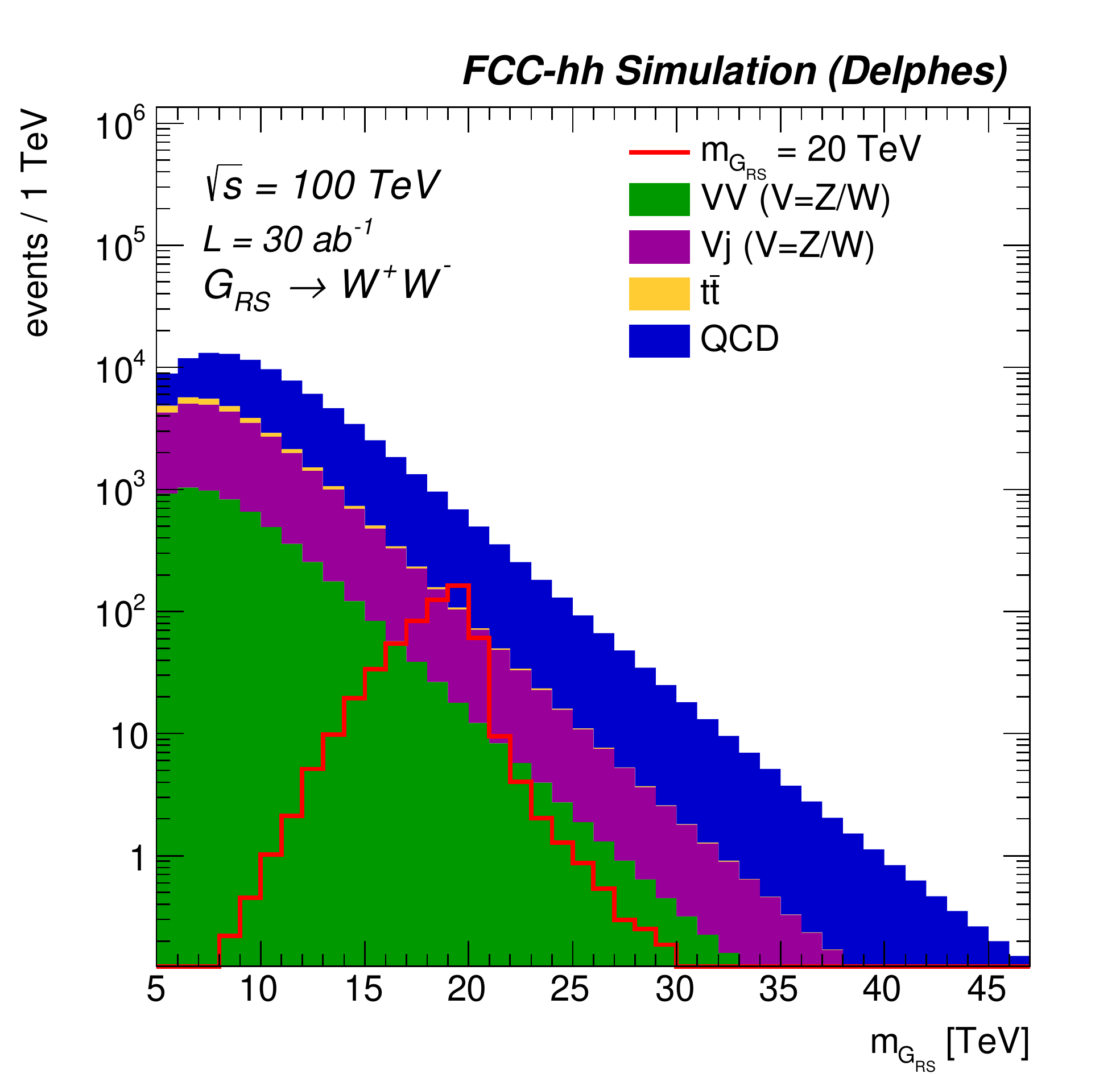}
  \includegraphics[width=0.48\columnwidth]{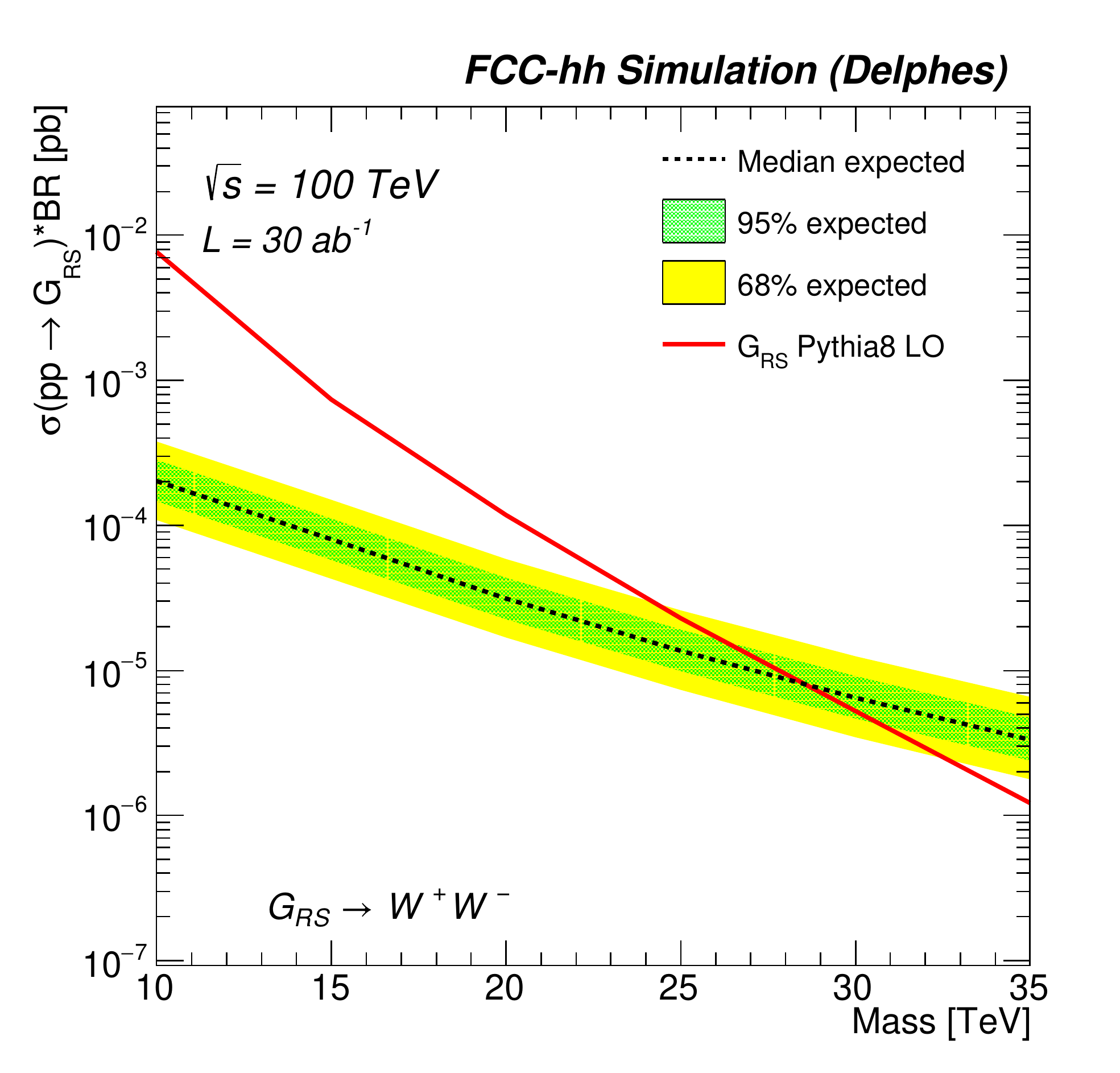}
  \\
  \includegraphics[width=0.48\columnwidth]{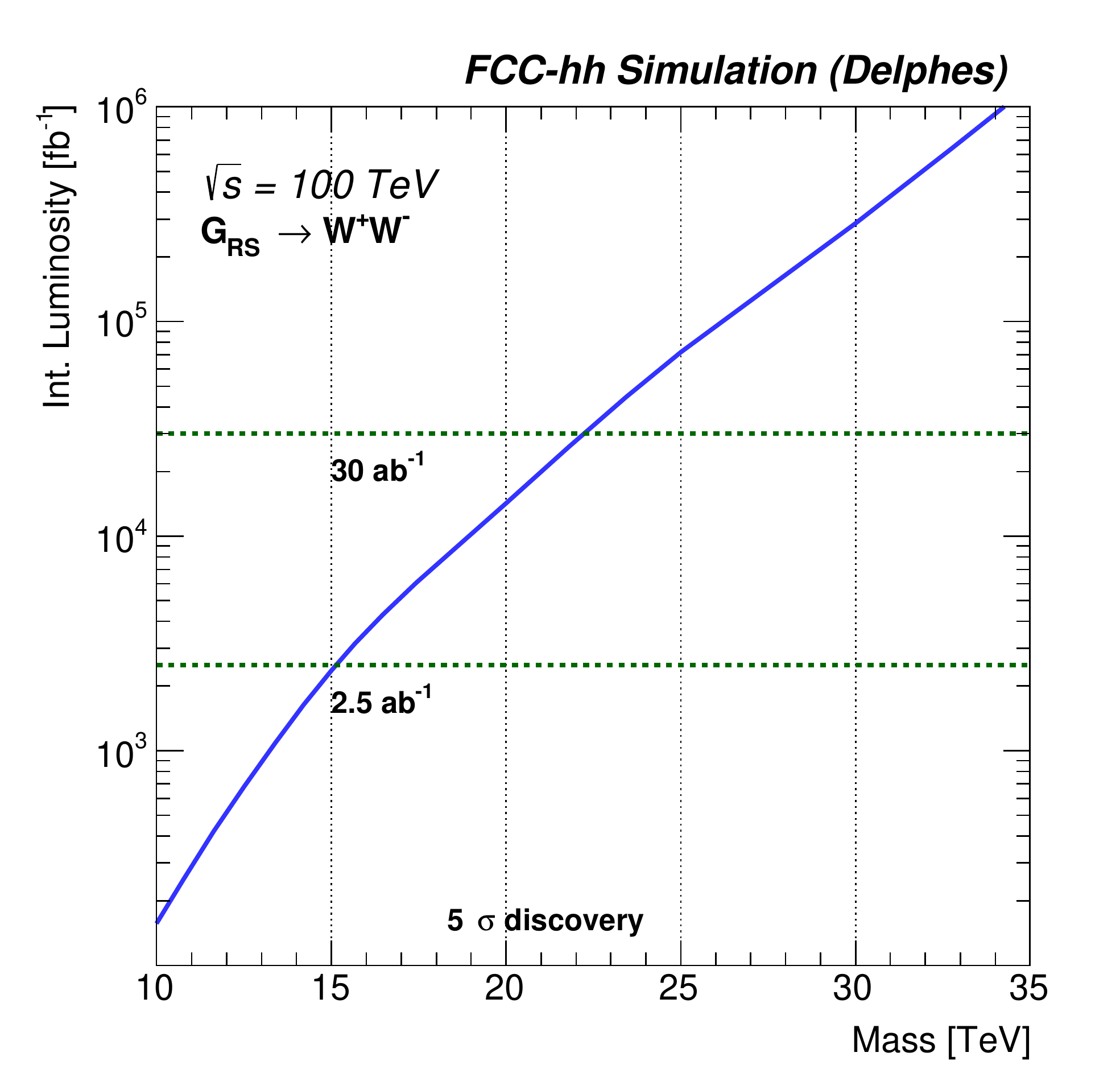}
  \caption{Invariant mass distribution of the two selected W-jets for a 20\,TeV signal (left), 95\% CL limit versus mass (right) and $5\sigma$ discovery reach (bottom).}
  \label{figure:hadronicresonances:ww}
\end{figure}

The event selection in this case consists of two jets with a $\pt$>3\,TeV, $|\eta|<3$ and $\Delta(\eta_1,\eta_2)<2.4$. Both jets must be $W$ tagged (see Section~\ref{subsec:mvatagger}). Again, to further reject QCD events, we require for both jets the soft-dropped mass to be larger than 40\,GeV. Figure~\ref{figure:hadronicresonances:ww} (left) shows the di-boson invariant mass distribution after the final selection for a 20\,TeV signal and background. Given the very good performance of the BDT discriminant, the QCD contribution is greatly reduced. The right panel shows the 95\% CL exclusion limit obtained with 30\,ab$^{-1}$ of data and the bottom panel shows the integrated luminosity required to reach a $5\sigma$ discovery as a function of the Randall-Sundrum graviton mass. Further developments to improve the W-jet/QCD could be considered, to improve the sensitivity as well as combining with leptonic channels, but, already with the current assumptions, the exclusion of 28\,TeV (Fig.~\ref{figure:hadronicresonances:ww} right) and the discovery of a 22\,TeV signal are obtained (Fig.~\ref{figure:hadronicresonances:ww} bottom) .

\section{Comparison with the 27 TeV HE-LHC}
\label{sec:ana27tev}
\begin{figure}[!htb]
  \centering
  \includegraphics[width=0.49\columnwidth]{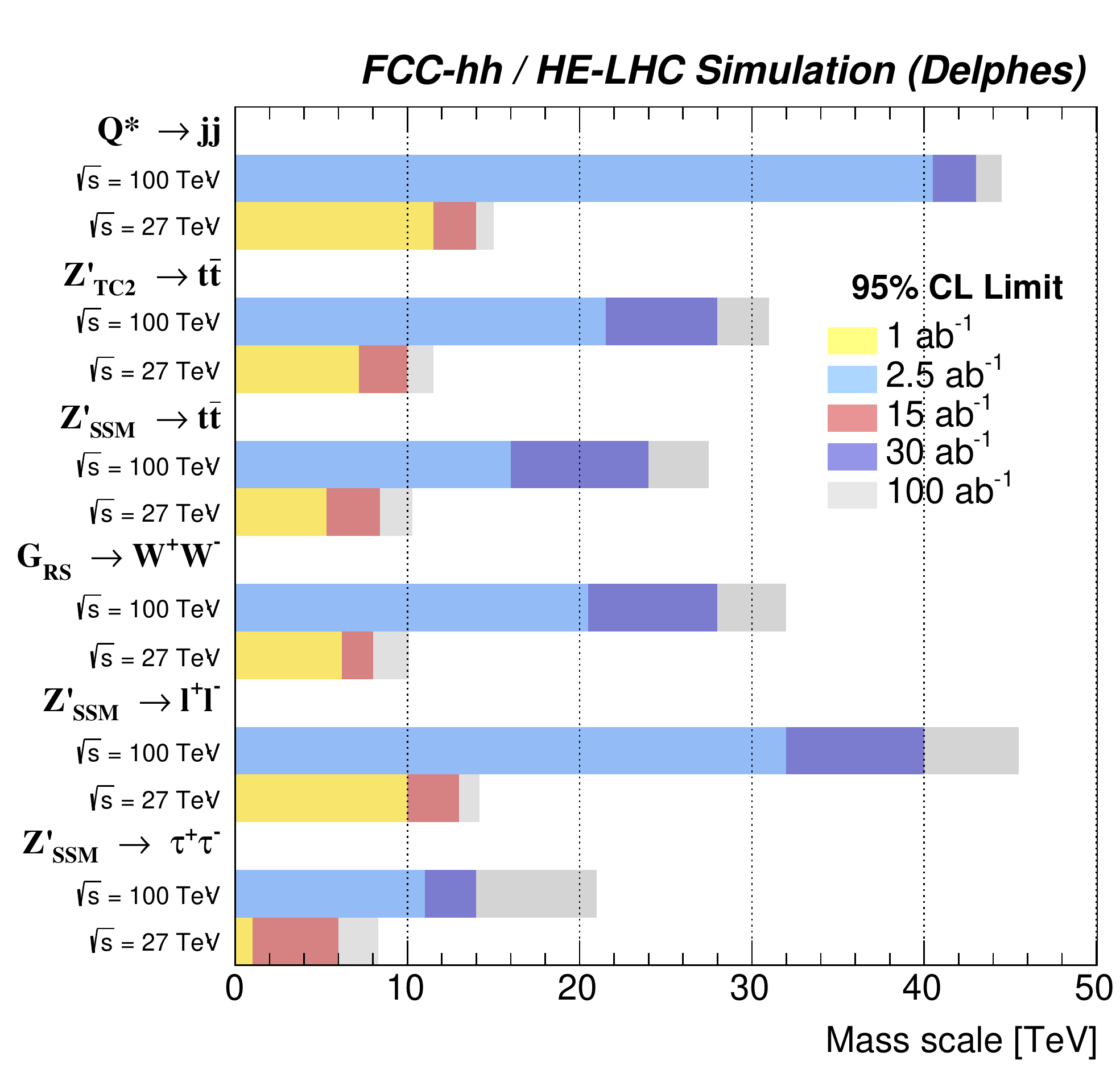}
  \includegraphics[width=0.49\columnwidth]{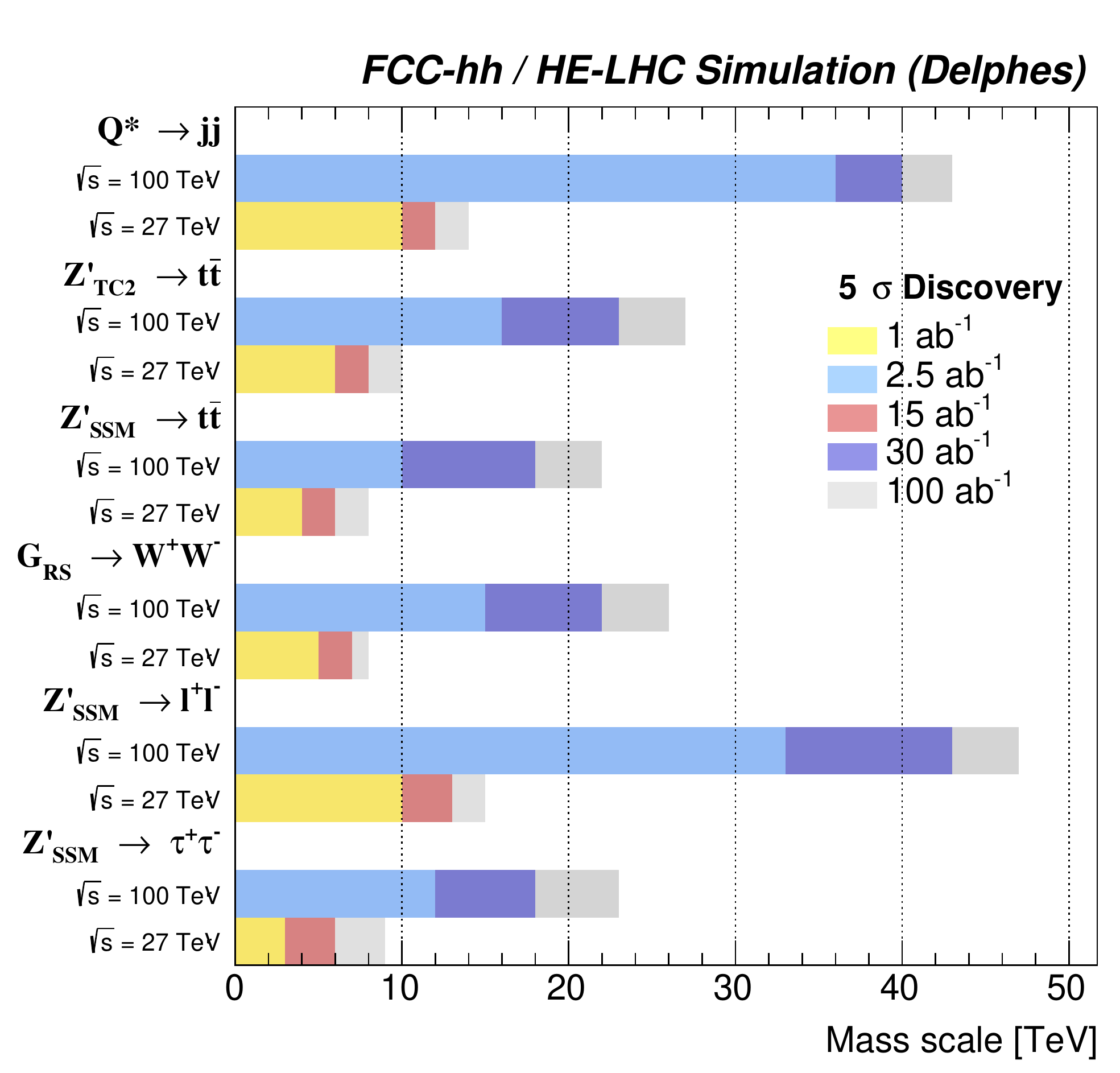}
  \caption{Summary of the 95\% CL limits (left) and $5\sigma$ discovery reach (right) as a function of the resonance mass for different luminosity scenario of FCC-hh and HE-LHC.}
  \label{figure:resonances:summary}
\end{figure}

We briefly present here the results of our study for the 27\,TeV HE-LHC. The detector simulation is based on the hybrid ATLAS/CMS HL-LHC detector parameterisation introduced in Section~\ref{subsec:detparam}. The analysis strategies remain identical to what was already presented in the previous sections, the only difference being the re-optimisation of some selection thresholds, to account for the lower center of mass energy. The changes can be summarised as follows:
\begin{itemize}
\item \Zpee\ and \Zpmumu : lepton $\pt$ threshold lowered from 1 to 0.5\,TeV
\item \Zptata\ : mass dependent selection as shown in Table~\ref{tab:leptonicresonances:tautau27}
\item \rsg, \Zptt, \qjj\ : jet $\pt$ threshold lowered from 3 to 1\,TeV
\end{itemize}
The results are summarised in Fig.~\ref{figure:resonances:summary}, together with a comparison to FCC-hh for the 95\% CL (left) and 5$\sigma$ discovery reach (right). Additional summary plots for FCC-hh and HE-LHC alone can be found in Appendix~\ref{sec:app:sumplots}.
A more extensive study of the di-jet decay at the HE-LHC ignoring the simulation of the detector response can be found in Ref.~\cite{Chekanov:2017pnx}. 

\begin{table}[!htb]
   \centering
\begin{tabular}{c|c|c|c}
   $\Zp$ mass [TeV] &  $\Delta \phi(\tau_1, \tau_2)$&  $\Delta R(\tau_1, \tau_2)$ & $\met$\\
  \hline
  \hline
   $2$ & > 2.4 & > 2.4 and < 3.9 & > 80 GeV\\
   $4$ & > 2.4 & > 2.7 and < 4.4 & > 80 GeV\\
   $6$ & > 2.4 & > 2.9 and < 4.4 & > 80 GeV\\
   $8$ & > 2.6 & > 2.9 and < 4.6 & > 80 GeV\\
  $10$ & > 2.8 & > 2.9 and < 4.1 & > 60 GeV\\
  \end{tabular}
  \caption{Mass dependent cuts optimised to maximise the sensitivity for the \Zptata\ resonance search.}
  \label{tab:leptonicresonances:tautau27}
\end{table}

\section{Characterisation of a \texorpdfstring{\Zp}{zp} discovery}
\label{sec:zprimedisc}

\subsection{Context of the study}
We consider in this section a scenario in which a heavy dilepton resonance is observed by the end of HL-LHC run. In this case, considering that current limits are already pushing to quite high values the possible mass range, a collider with higher energy
in the \com would be needed to study the resonance properties, since too few events will be available at \sqrtslhc. In this section we present the discrimination potential, among six $Z'$ models, of the 27~TeV HE-LHC, with an assumed integrated luminosity of \intlumihelhc. Under the assumption that these $Z'$'s decay only to SM particles, we show that there are sufficient observables to perform this model differentiation in most cases.

\subsection{Bounds from HL-LHC}
\begin{figure}[htbp]
  \centering
    \includegraphics[trim={2cm 3cm 3cm 2.5cm},clip,width=0.48\columnwidth]{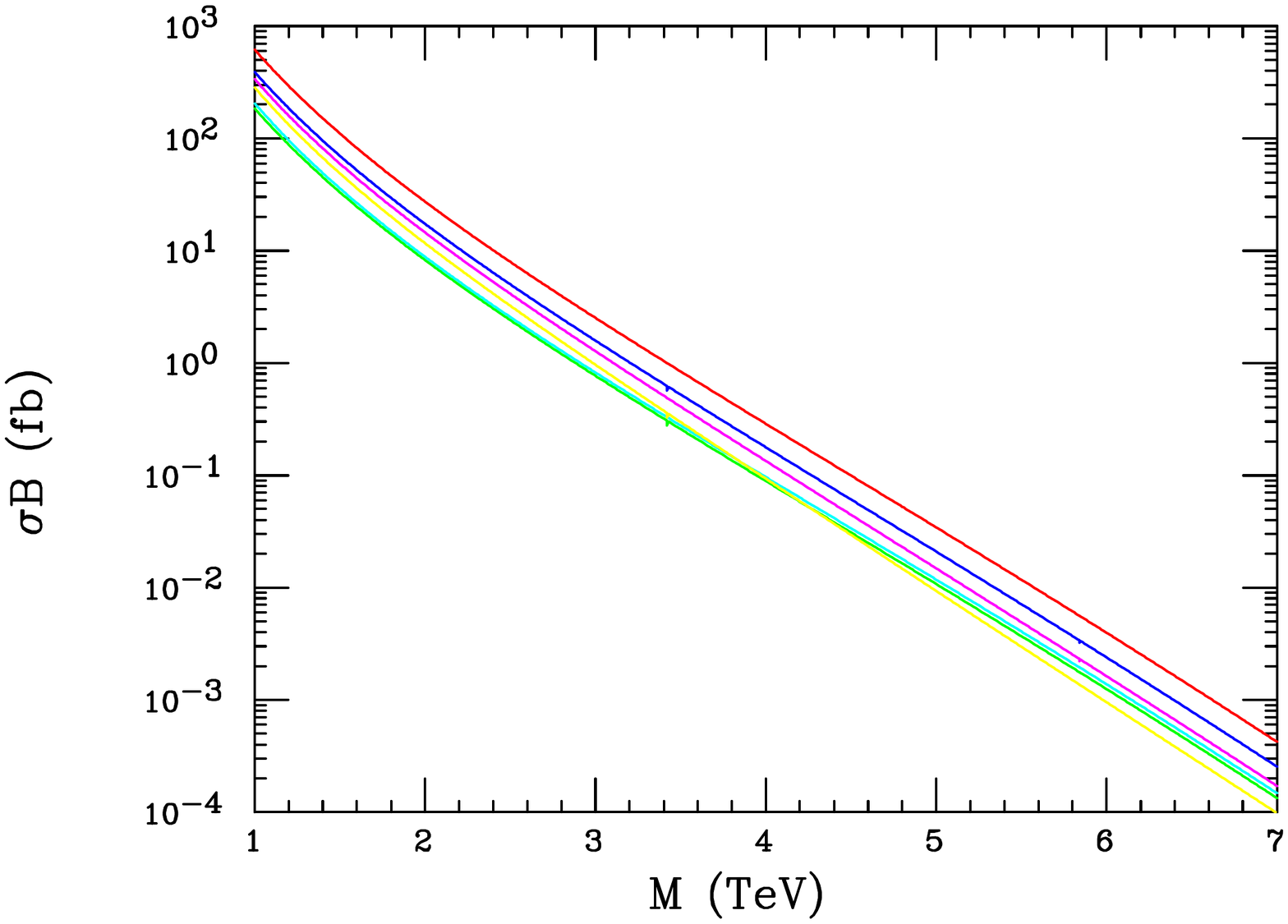}
    \includegraphics[trim={2cm 3cm 3cm 2.5cm},clip,width=0.48\columnwidth]{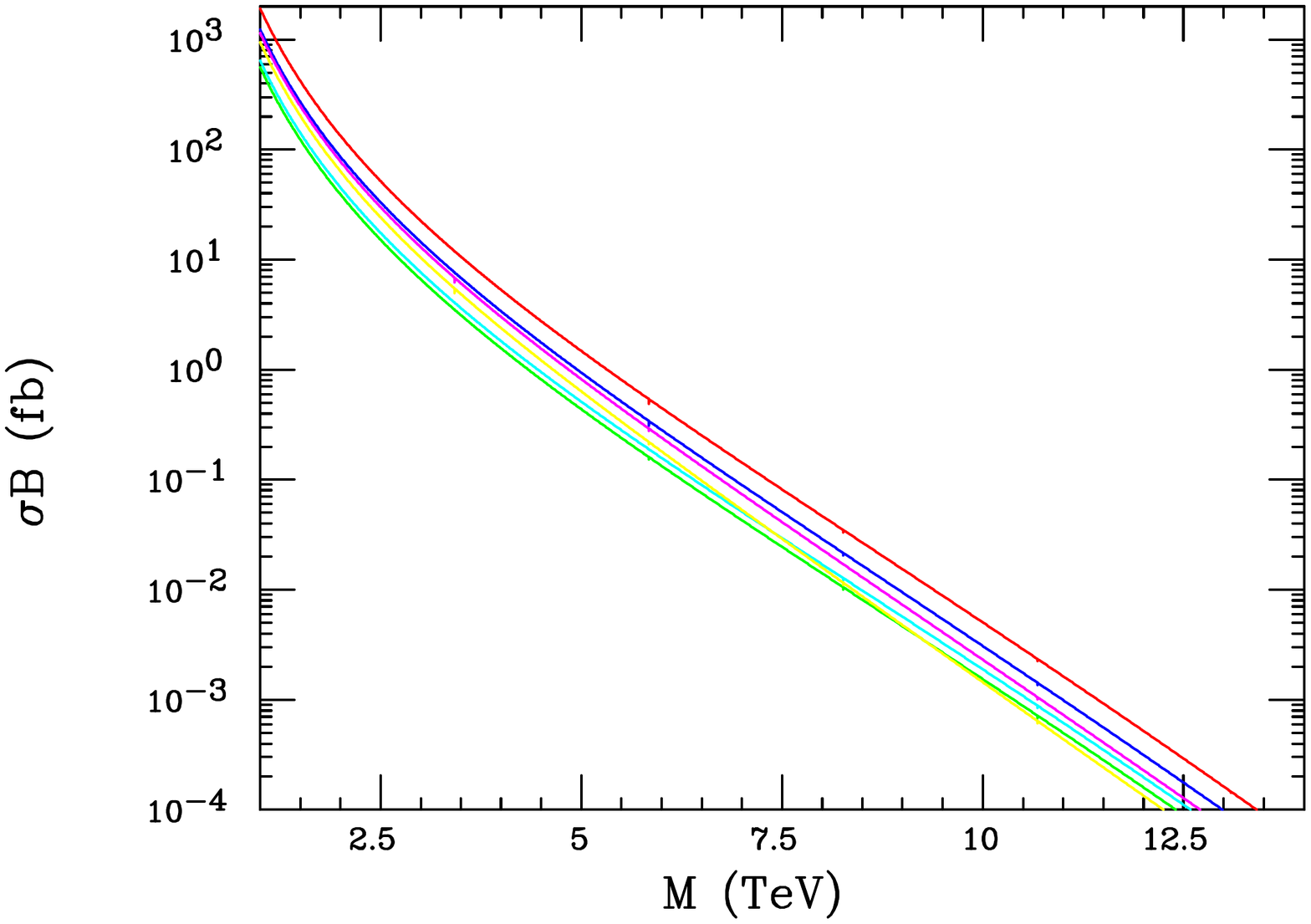}
    \caption{Left: $\sigma B_l$ in the NWA for the $Z'$ production at the $\sqrt s=14$\,TeV LHC as functions of the $Z'$ mass: SSM(red), LRM (blue), $\psi$(green), $\chi$(magenta),
$\eta$(cyan), I(yellow). (Right) $\sigma B_l$ of $Z'$ in models described in (left) at $\sqrt s=27$\,TeV.}
\label{fig:pheno:toy}
\end{figure}

As a starting point we need to estimate what are, for $\sqrt s=14$\,TeV, the typical exclusion/discovery reaches for standard reference $Z'$ models, assuming \intlumihllhc\ and employing only the $e^+e^-$ and $\mu^+\mu^-$ channels. To address this and the other questions below we will use the same set of $Z'$ models as employed
in Ref.~\cite{Rizzo:2014xma} and mostly in Ref.~\cite{Han:2013mra}. We employ the MMHT2014 NNLO PDF set~\cite{Harland-Lang:2014zoa}
throughout, with an appropriate constant $K$-factor (=1.27) to account for higher order QCD corrections. The production cross section times leptonic branching fraction is shown in Fig.~\ref{fig:pheno:toy} (left) for these models at \sqrtslhc\ in the narrow width approximation (NWA). We assume here that these $Z'$ states only decay to SM particles.

Using the present ATLAS and CMS results at 13\,TeV,~\cite{Aaboud:2017buh} and~\cite{Sirunyan:2018exx}, it is straightforward to estimate by extrapolation the
exclusion reach at \sqrtslhc\ using the combined $ee+\mu\mu$ final states. This is given in the first column of Table~\ref{tab:pheno:spec}. For discovery, only the $ee$ channel is used, due to the poor $\mu\mu$-pair invariant mass resolution near $M_{Z'}=6$\,TeV. Estimates of the $3\sigma$ evidence and $5\sigma$
discovery limits are also given in Table~\ref{tab:pheno:spec}. This naive extrapolation can be compared to the ATLAS HL-LHC prospect analysis in Ref.~\cite{CidVidal:2018eel} and is found to be agreement. 
Based on these results, we will assume in our study for the HE-LHC that we are dealing with a $Z'$ of mass 6\,TeV. Figure~\ref{fig:pheno:toy} (right) shows the NWA cross sections for the same set of models, at \sqrtshelhc. We note that very large statistical samples will be available, with \intlumihelhc, for $M_{Z'}=6$\,TeV and in both dilepton channels.

\begin{table}
\centering
\begin{tabular}{c|c|c|c}

  Model &   95$\%$ \cl  &  $3\sigma$  & $5\sigma$   \\
\hline
\hline
SSM    &     6.62     &  6.09    &  5.62     \\
LRM    &   6.39     & 5.85     & 5.39  \\
$\psi$    &  6.10   & 5.55   & 5.07  \\
$\chi$   &  6.22    & 5.68    & 5.26   \\
$\eta$   &  6.15     &  5.59  &  5.16   \\
~I        & 5.98   &  5.45   &  5.05  \\
\end{tabular}
\caption{Mass reach for several $Z'$ models at \sqrtslhc\ with \intlumihllhc. }
\label{tab:pheno:spec}
\end{table}
%

\subsection{Definition of the discriminating variables}
\label{par:vardef}

The various $Z'$ models can be disentangled with the help of 3 inclusive observables: the production cross sections for different leptonic and hadronic final states, the leptonic forward-backward asymmetry $A_{FB}$ and the rapidity ratio $r_y$. The variable $A_{FB}$ can be seen as an estimate of the charge asymmetry
\begin{equation}
A_{FB} = A_C =  \frac{\sigma(\Delta|y| > 0) - \sigma(\Delta|y| < 0)}{\sigma(\Delta|y| > 0) + \sigma(\Delta|y| < 0)},
\end{equation}
where $\Delta|y| = |y_l| - |y_{\bar{l}}|$. This definition is equivalent to
\begin{equation}
A_{FB} = \frac{\sigma_F - \sigma_B}{\sigma_F + \sigma_B},
\end{equation}
with $\sigma_F = \sigma (\text{cos} ~\theta^{*}_{cs})>0$ and $\sigma_B = \sigma (\text{cos}~\theta^{*}_{cs})<0$ where $\theta^*_{cs}$ is the Collins-Soper frame angle. The variable $r_y$ is defined as the ratio of central over forward events:
\begin{equation}
r_y = \frac{\sigma(|y_{Z'}| < y_1)}{\sigma(y_1 < |y_{Z'}| <y_2)},
\end{equation}
where $y_1=0.5$ and $y_2=2.5$.


The results in this section have been obtained assuming the \delphes~\cite{deFavereau:2013fsa} parametrisation of the HE-LHC detector~\cite{delphes_card_helhc}. In such a detector, muons at $\eta \approx 0$ are assumed to be reconstructed with a resolution $\sigma(p)/p \approx 7\%$ for $\pt~=~3$\,TeV.

\subsubsection{Leptonic final states}
\label{par:lepana}

The potential for discriminating various $Z'$ models is first investigated using the leptonic $ee$ and $\mu\mu$ final states only. The signal samples for the 6 models have been generated with \py~\cite{Sjostrand:2014zea} as described in Section~\ref{subsec:mcprod} with the only difference being that the interference between the signal and Drell-Yan is included. The $Z'$ decays assume lepton flavour universality, with branching ratio $B_l$. For a description of the event selection and a discussion of the discovery potential in leptonic final states for the list of $Z'$ models being discussed here, the reader should refer to Section~\ref{sec:lep} and~\ref{sec:ana27tev}. We simply point out here that with \intlumihelhc\,, all $Z'$ models with $m_{Z'}~\lesssim~10$\,TeV can be excluded at \sqrtshelhc.

\begin{figure}[!htb]
  \centering
   \includegraphics[width=0.48\columnwidth]{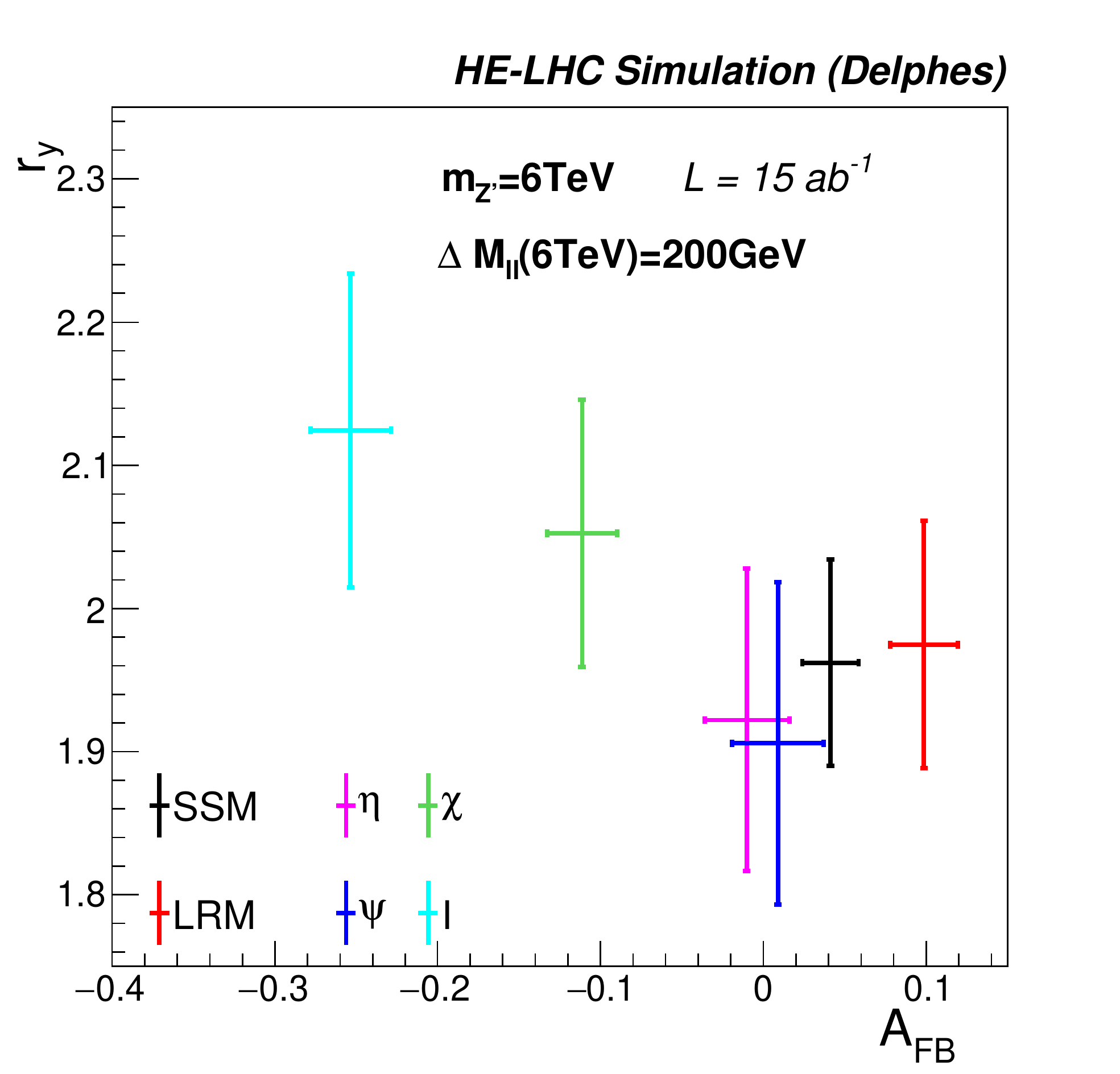}
      \includegraphics[width=0.48\columnwidth]{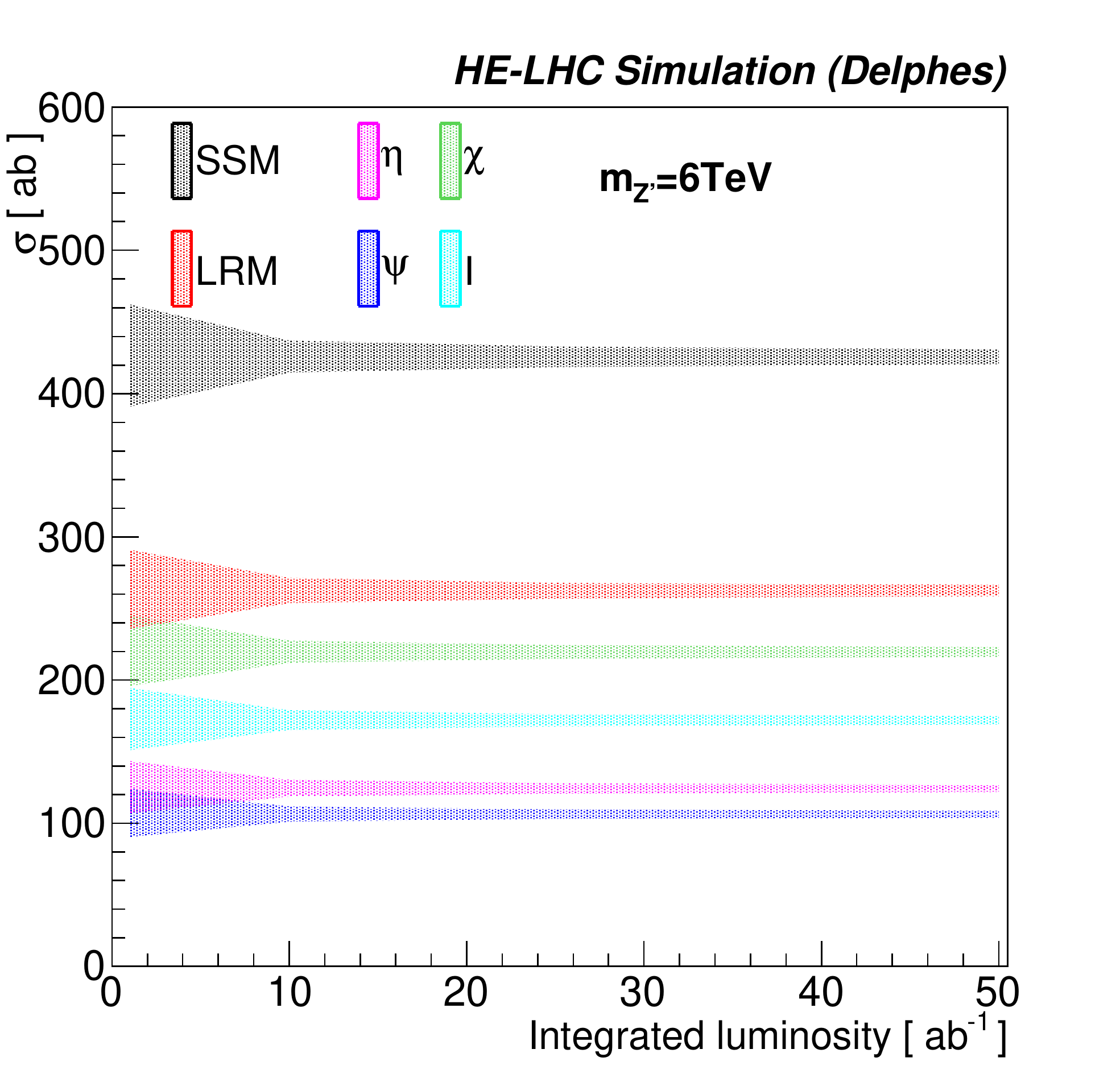}
    \includegraphics[width=0.48\columnwidth]{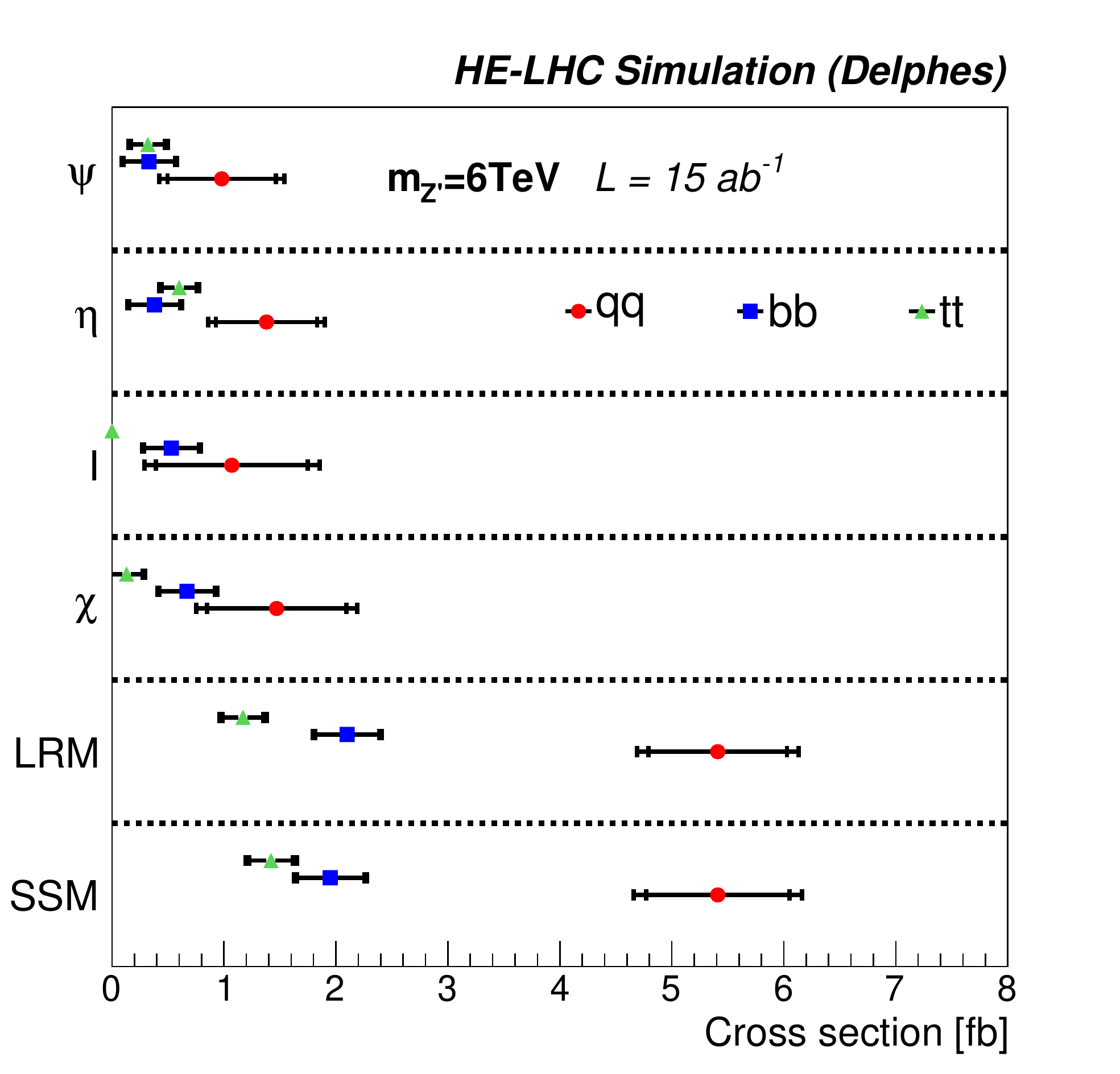}
   
  \caption{Left: Scatter plot of $r_y$ versus $A_{FB}$ with a 200~GeV mass window. The full interference is included. Right: Fitted signal cross-section together with its corresponding error versus integrated luminosity. Bottom: Fitted cross-section of the three hadronic analyses. Statistical and full uncertainties are shown on each point.}
  \label{fig:ana:res}
\end{figure}

Figure~\ref{fig:ana:res} (left) shows the correlated predictions for the $A_{FB}$ and the rapidity ratio $r_y$ observables defined previously, for these six models given the above assumptions. Although the interference with the SM background was included in the simulation, its effect is unimportant due to the narrowness of the mass window around the resonance that was employed. Furthermore, the influence of the background uncertainty on the results has been found to have little to no impact on the model discrimination potential. Therefore the displayed errors on $A_{FB}$ and $r_y$ are of statistical origin only. The results show that apart from a possible near degeneracy in models $\psi$ and $\eta$, a reasonable $Z'$ model separation can indeed be achieved.

Using a profile likelihood technique, the signal strength $\mu$, or equivalently, $\sigma \times B_l$, can be fitted together with its corresponding error using the the di-lepton invariant mass shape. The quantity $\sigma \times B_l$ and its total estimated uncertainty is shown in Fig.~\ref{fig:ana:res} (right) as a function of the integrated luminosity. The $\sigma \times B_l$ measurement seems to be able to resolve the degeneracy between the $\psi$ and $\eta$ models with \intlumihelhc. It should be noted however that since the cross-section can easily be modified by an overall rescaling of the couplings or via the existence of decays into non-SM states, further handles will be needed for a convincing discrimination.

\subsubsection{Hadronic final states}
\label{par:hadana}

Model discrimination can be improved by including an analysis involving three additional hadronic final states: $t\bar{t}$, $b\bar{b}$ and $q\bar{q}$, where $q=u,d,c,s$. The sample production and event selection for the $t\bar{t}$, $q\bar{q}$ final states have been described in Section~\ref{sec:hadronic}. We simply remind the reader that the analysis involves requiring the presence of two central high-$\pt$ jets. In order to ensure complete orthogonality between the various final states, jets are required to be tagged as follows. In the $Z' \rightarrow t\bar{t}$ analysis both jets should be \emph{top-tagged}. For the $Z' \rightarrow b\bar{b}$ final state both jets are required to be \emph{b-tagged} and we veto events containing at least one top-tagged jet. Finally, in the $Z' \rightarrow q\bar{q}$ analysis, we veto events that contain at least one b-tagged or top-tagged jet.

Figure~\ref{fig:ana:res} (bottom) summarises the discrimination potential in terms of fitted cross-section of the different models considering the three aforementioned hadronic decays, $t\bar{t}$,  $b\bar{b}$ and $q\bar{q}$. A good overall discrimination among the various models can be achieved using all possible final states. For example, the SSM and $\psi$ models, which have very close predictions for $r_y$ and $A_{FB}$, have measurably different fractions of $t\bar{t}$ or $b\bar{b}$ final states. We note however that the degeneracy between $\eta$ and $\psi$ can only be partially resolved at $\approx~1~\sigma$ by exploiting the difference in $t\bar{t}$ yield. We note that ratios of these individual production cross sections are insensitive to the possible existence of other non-SM decay modes of the $Z'$. 


\section{Conclusions}
\label{sec:conc}
This paper had three main goals: (i) to determine the discovery reach of the FCC-hh and HE-LHC at the highest masses, using as benchmarks several BSM models of $s$-channel resonance production, (ii) to define performance targets for the detectors, and (iii) to study the power of the HE-LHC to discriminate among different models of resonances that could be just visible at HL-LHC. We confirmed the expectation that the discovery reach scales approximately as the increase in the beam energy: this is not a trivial finding, since the energy measurement and the reconstruction of the multi-TeV decay final states (dileptons or different types of dijets) is not guaranteed, and requires important improvements with respect to the performance of the LHC detectors (e.g. higher calorimeter granularity for the reconstruction and identification of jets, or better momentum resolution for muons). That these improvements are potentially within the reach of foreseeable technology, as indicated by the preliminary detector design proposals for FCC-hh~\cite{Benedikt:2651300}, indicates that the FCC-hh physics potential can be fully exploited. 

We also studied the discrimination potential of six $Z'$ models at HE-LHC. The exercise was performed assuming the evidence of an excess observed at \sqrtslhc\ at a mass $m_{Z'}\approx~6$\,TeV. Overall it was found that the increased production cross section and the corresponding statistical increase from HL-LHC to HE-LHC are sufficient to analyze an extended set of observables, whose global behaviour provides important information to distinguish among most models. Further studies, using for example 3-body decay modes or associated Z' production (with jets or with SM gauge bosons), could be considered to provide additional handles characterizing the resonance properties.

\acknowledgments
We thank Ben Allanach for helpful discussions on the interpretation of the $R_{K^{(*)}}$ anomalies and for providing the ``naive'' model to be tested in the context of the FCC-hh detector studies. The work of TGR was supported by the Department of Energy, Contract DE-AC02-76SF00515.

\appendix
\section{Discussion of the detector performance}
\label{sec:app:detperf}

\begin{figure}[!htb]
  \centering
    \includegraphics[width=0.48\columnwidth]{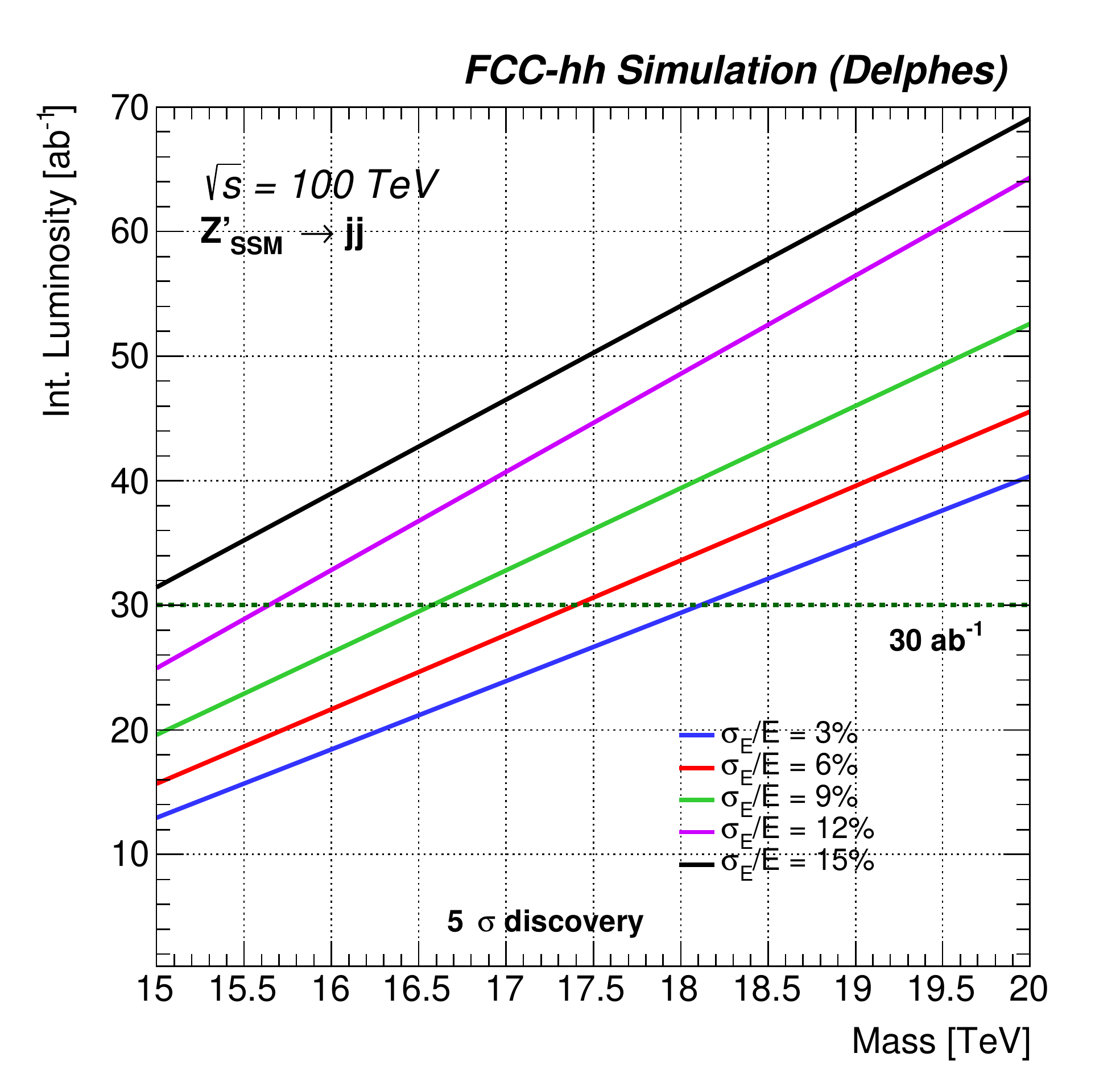}
  \includegraphics[width=0.48\columnwidth]{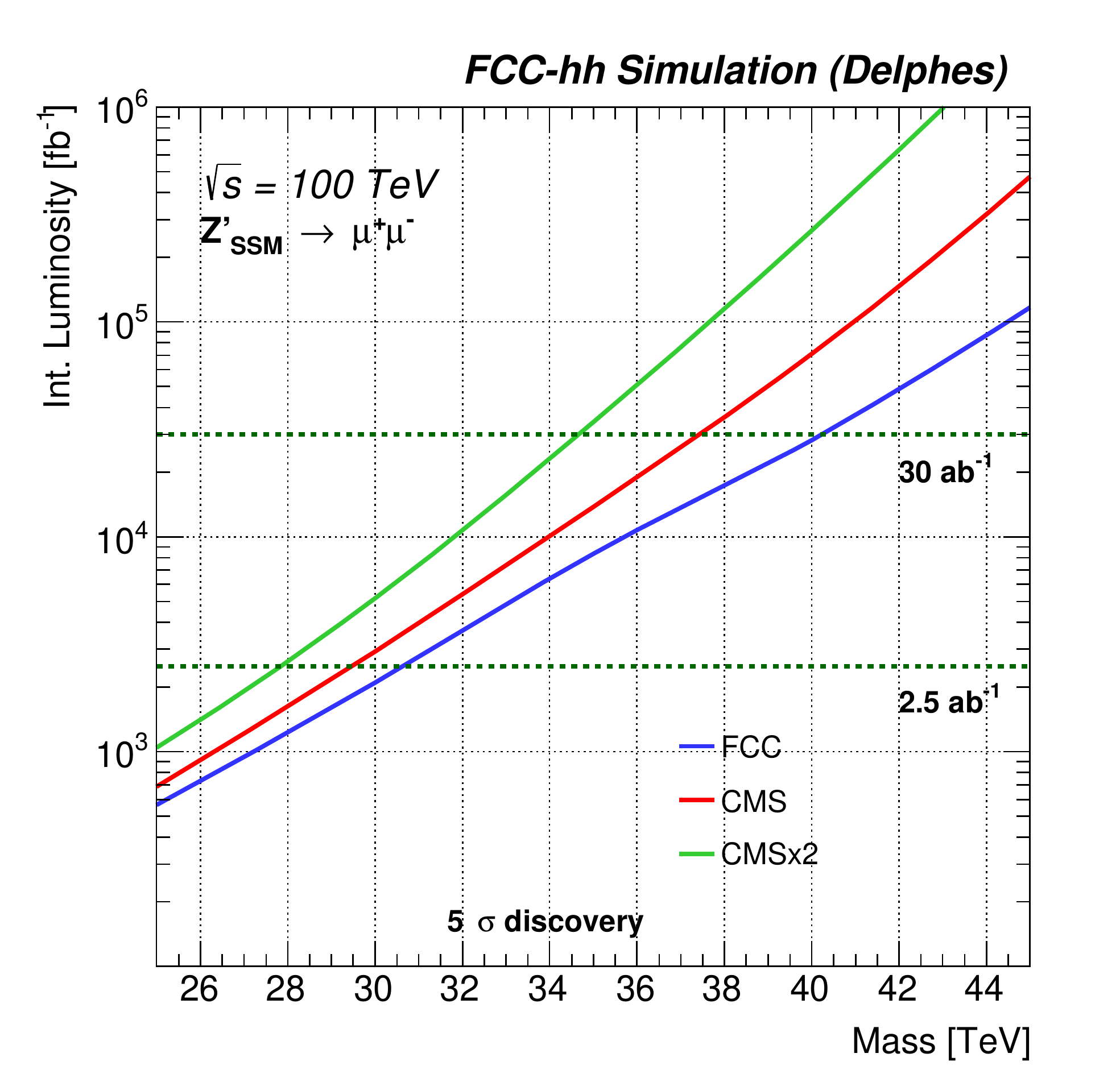}
  \includegraphics[width=0.48\columnwidth]{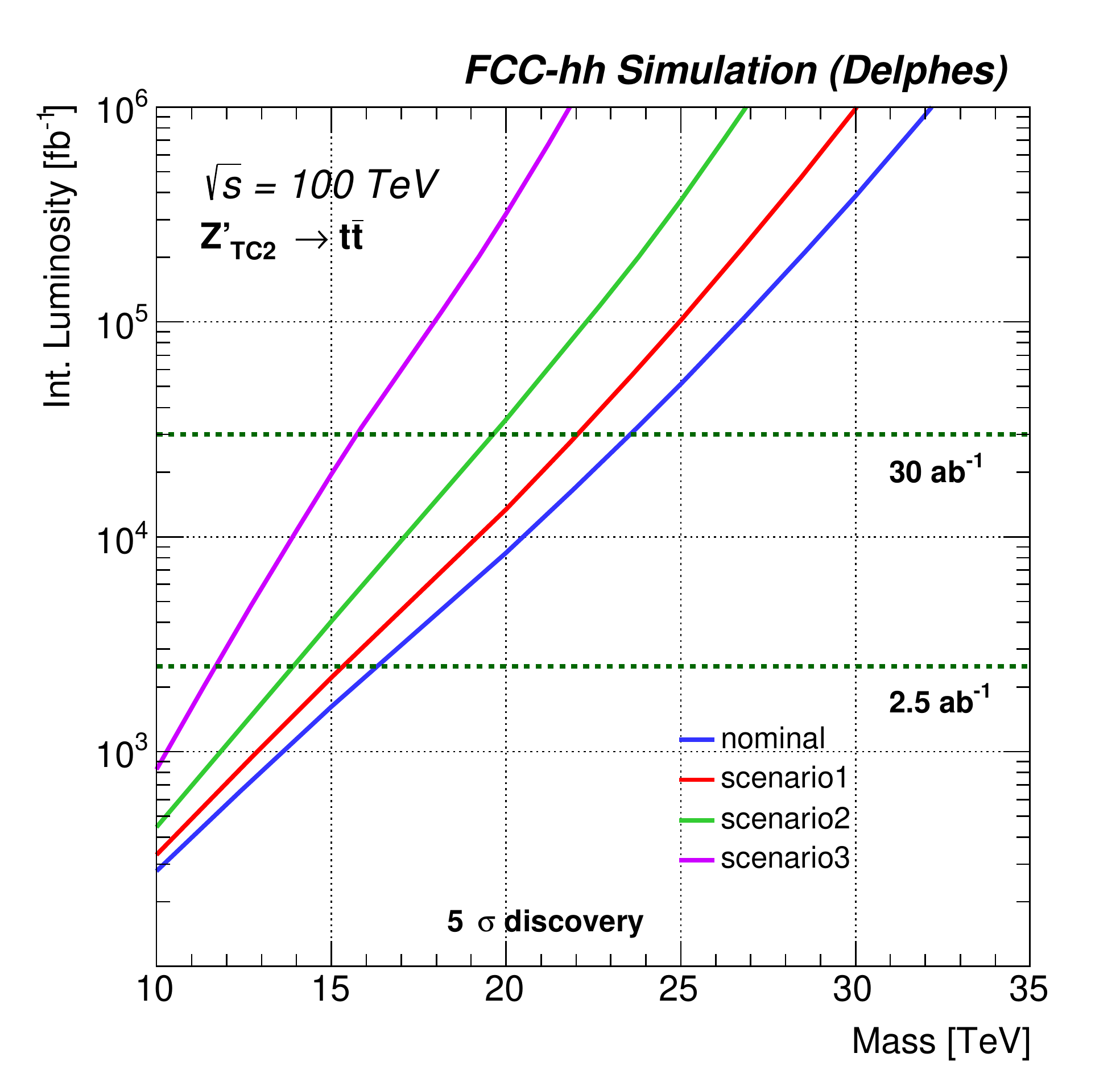}
  \caption{Luminosity versus mass for a $5\sigma$ discovery for different calorimeter resolution (left), muon resolutions (right) and b-tagging scenarios (bottom). }
  \label{figure:detperf}
\end{figure}

In the calorimeters, the energy resolution at high energy is determined by the constant term. The value of the constant term is different for ECAL and HCAL calorimeters. It is ultimately determined by the choice of the calorimeter technology and the design. Large constant terms  typically originate from inhomogenities among different detector elements and energy leakages due to sub-optimal shower containment. The calorimeters of the FCC-hh detector must therefore be capable of containing EM and hadronic showers in the multi-TeV regime in order to achieve small constant terms.
Comparing with the LHC experiments, we require a performance of $\sigma_E/E~ \approx 0.3 \%$ and $\sigma_E/E~ \approx 3\%$ for the ECAL and HCAL, respectively. As shown in Fig.~\ref{figure:detperf} (left), the effect induced by the magnitude of the hadronic calorimeter constant term on the expected discovery reach for heavy \ZpSSM\ resonances decaying hadronically is sizable. We note that, despite the fraction of electromagnetic energy from $\pi^0$'s large in jets, the sensitivity is entirely driven by the hadronic calorimeter resolution given its worse intrinsic resolution.

Muons cannot be reconstructed with calorimetric methods~\footnote{~Calorimetric information can however help for muon identification. For example a 20\,TeV muon deposits through radiative energy loss on average $\Delta E=$~200 GeV in 3 meters of iron, corresponding to 1\% of the initial muon energy.}. Since the muon momentum is obtained through a fit of the trajectory that uses as input a combination of track and muon spectrometer hits, the muon momentum resolution degrades with increasing momentum, as $\frac{\sigma_p}{p}= a \oplus b~p$ where $a$ is the constant term determined by the amount of material responsible for multiple scattering in the tracking volume. As with jets, electrons and photons, a good muon momentum resolution at multi-TeV energy is crucial for maintaining a high sensitivity in searches for heavy new states that might decay to muons. The reach for a \Zpmumu\ resonance obtained with various assumptions on the muon resolution is illustrated in Fig.~\ref{figure:detperf} (right). The best sensitivity is achieved with an assumed $\sigma_p/p~ \approx 5\%$ at $\pt = 20$~TeV corresponding to our target for the FCC-hh detector, as opposed to the projected CMS resolution of $\sigma_p/p~\approx 40\%$. In order to reconstruct and measure accurately the momentum of $\pt=20~$TeV a large lever arm is needed and excellent spatial resolution and precise alignment of the tracking plus muon systems is also needed. The specifics of the design that allows to reach such required performance are discussed in Ref.~\cite{Benedikt:2651300}.

New heavy states could decay to multi-TeV $c$ and $b$-quarks. FCC-hh detectors must therefore be capable of efficiently identifying multi-TeV long-lived hadrons. A $\pt=5$~TeV b-hadron is qualitatively very different from $\pt=100$~GeV b-hadron. The latter decays on average within the vertex detector acceptance and can be identified by means of displaced vertex reconstruction. Conversely, the former decays on average at a distance $\gamma c \tau = 50$~cm, well outside the pixel detector volume. Reconstructing such highly displaced b-jets will require a paradigm shift in heavy flavour reconstruction. The success of algorithms exploring large hit multiplicity discontinuities among subsequent tracking layer heavily relies on excellent granularity of the tracking system, in both longitudinal and transverse directions. High efficiencies ($\epsilon_b > 60\%$) for corresponding low mis-identification probability ($\epsilon_{u,d,s} < 1\%$) from light jets have to be achieved up to $\pt=5$~TeV. For example, searches for heavy resonances decaying to hadronic $\ttbar$ pairs heavily rely on efficient b-tagging performance at such energies. The discovery reach for a specific $Z'$ model assuming several scenarios for b-jet identification at high energies is shown in Fig.~\ref{figure:detperf} bottom. Various scenarios of b-tagging efficiencies at very large \pt\ are considered. The nominal efficiency is given in Table~\ref{tab:effs}, and scenarios 1,2 and 3 correspond to reduction of the slope respectively by a factor 25\%, 33\% and 50\%. As expected the discovery reach strongly depends on the b-tagging performances.

\section{Tagging rate function}
\label{sec:app:trf}
Given a jet with specific values of $\eta$, $\pt$ and with flavour $f$, its tagging probability can be denoted as:
\begin{equation*}
	\varepsilon \left(f,|\eta|,\pt\right)
\end{equation*}
\newline
For a given event with $N$ jets, its probability of containing exactly one $b$-tag jet can be computed as:
\begin{equation*}
	P_{=1} = \sum\limits_{i=1}^N \left( \varepsilon_{i} \prod\limits_{i \neq j} \left( 1 - \varepsilon_{j} \right) \right)
\end{equation*}
\newline
In the same way, it can be used to compute the probability for inclusive $b$-tag selections:
\begin{align*}
	P_{=0} &= \prod\limits_{i=1}^N \left( 1 - \varepsilon_{j} \right) \\
	P_{\geq 1} &= 1 - P_{=0}
\end{align*}
\newline
It was verified that the TRF methods agree well with the direct tagging.

\section{Background fit}
\label{sec:app:bgfit}
\begin{figure}[!htb]\centering
\includegraphics[width=0.49\columnwidth]{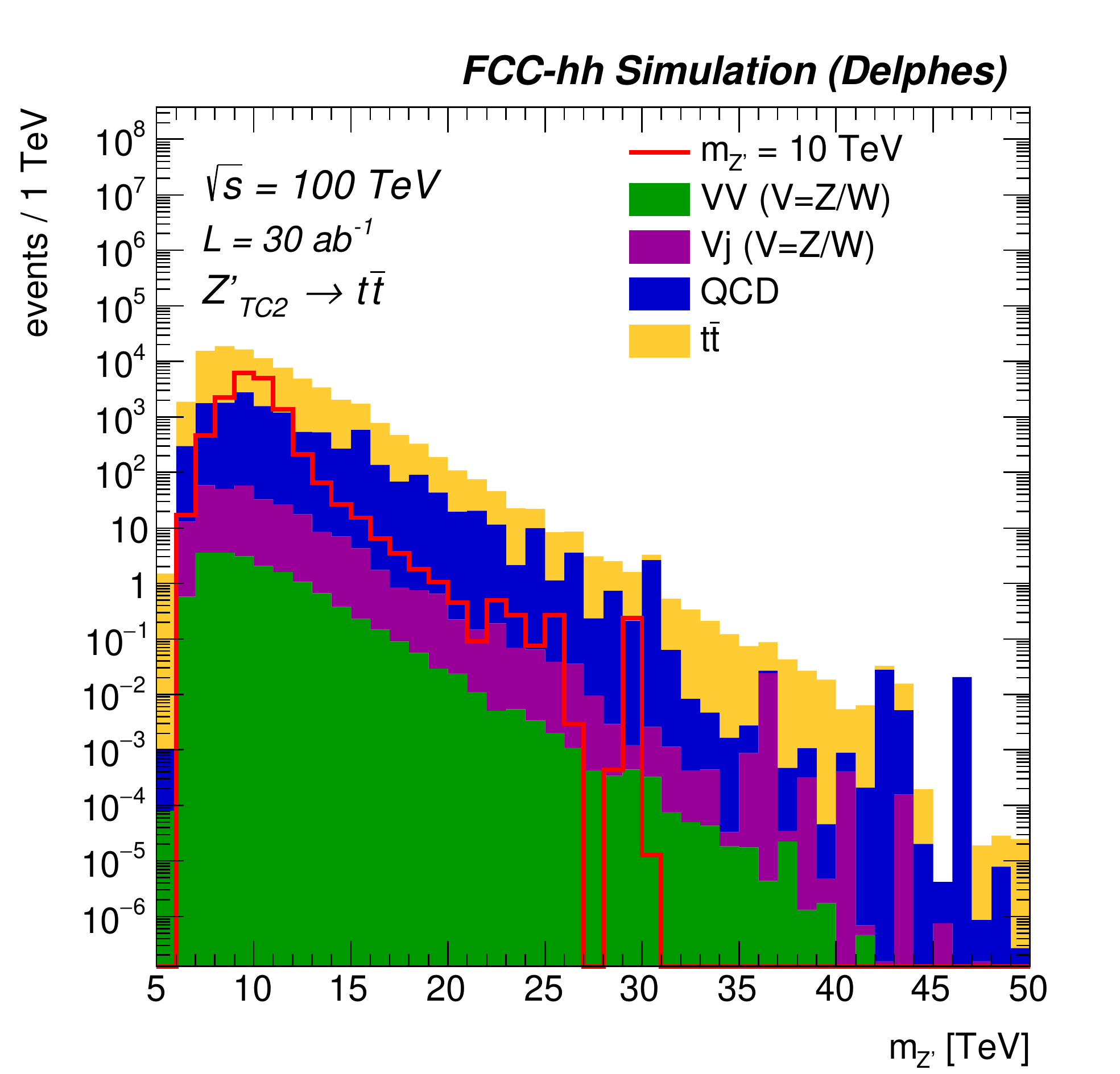}
\includegraphics[width=0.49\columnwidth]{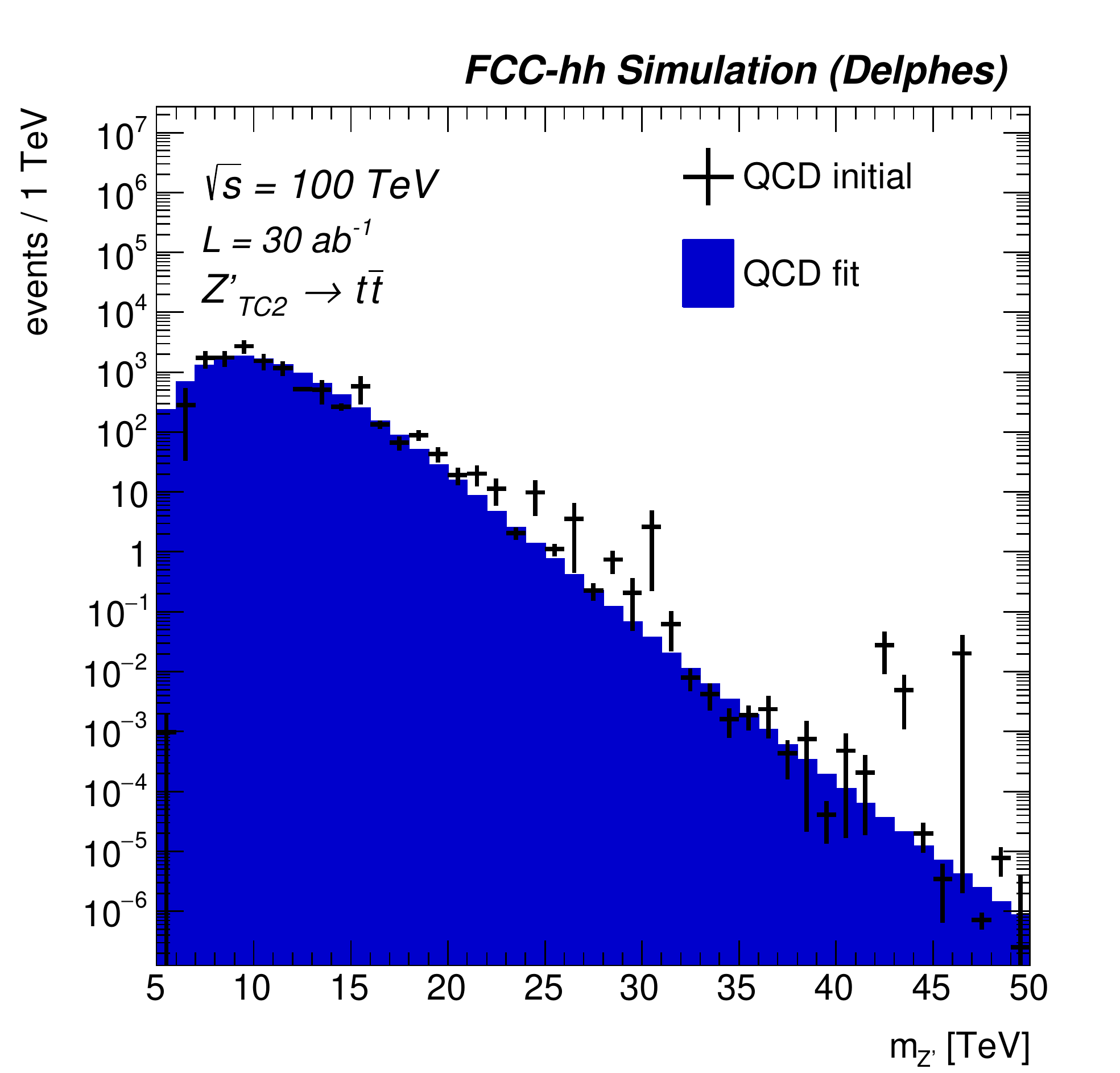}
\caption{QCD Background invariant mass spectra prior to fit.}
\label{fig:hadronicresonances_nofit}
\end{figure}

The function used to fit the background shapes when Monte Carlo statistics are insufficient can be expressed as:
\begin{equation}
\label{eq:fitfunc}
f(z)=p_1(1-z)^{p_2}z^{p_3}z^{p_{4}logz}
\end{equation}
where $z=m_{jj}/\sqrt{s}$, with $m_{jj}$ the invariant mass of the two highest energetic objects and $\sqrt{s}$ the center of mass energy. Fitting the invariant mass distribution with the function~\ref{eq:fitfunc} allows to obtain a smooth shape, while the overall normalisation is taken prior to the fit. Figure~\ref{fig:hadronicresonances_nofit} (left)  shows the \Zptt\ invariant mass distribution after the final selection for the various backgrounds and a 10\,TeV signal. Large statistical fluctuations can be observed especially for the QCD background that is heavily suppressed thanks to the multivariate object tagger. The right panel represents the same QCD invariant mass distribution before the fit (dots) and after the fit (plain). Good agreement is observed and the fitted distribution normalised to the pre-fit yields is used to obtain the results in the statistical analysis.

\section{Multivariate object tagger}
\label{sec:app:mva}
\begin{figure}[!htbp]\centering
\includegraphics[width=0.48\textwidth]{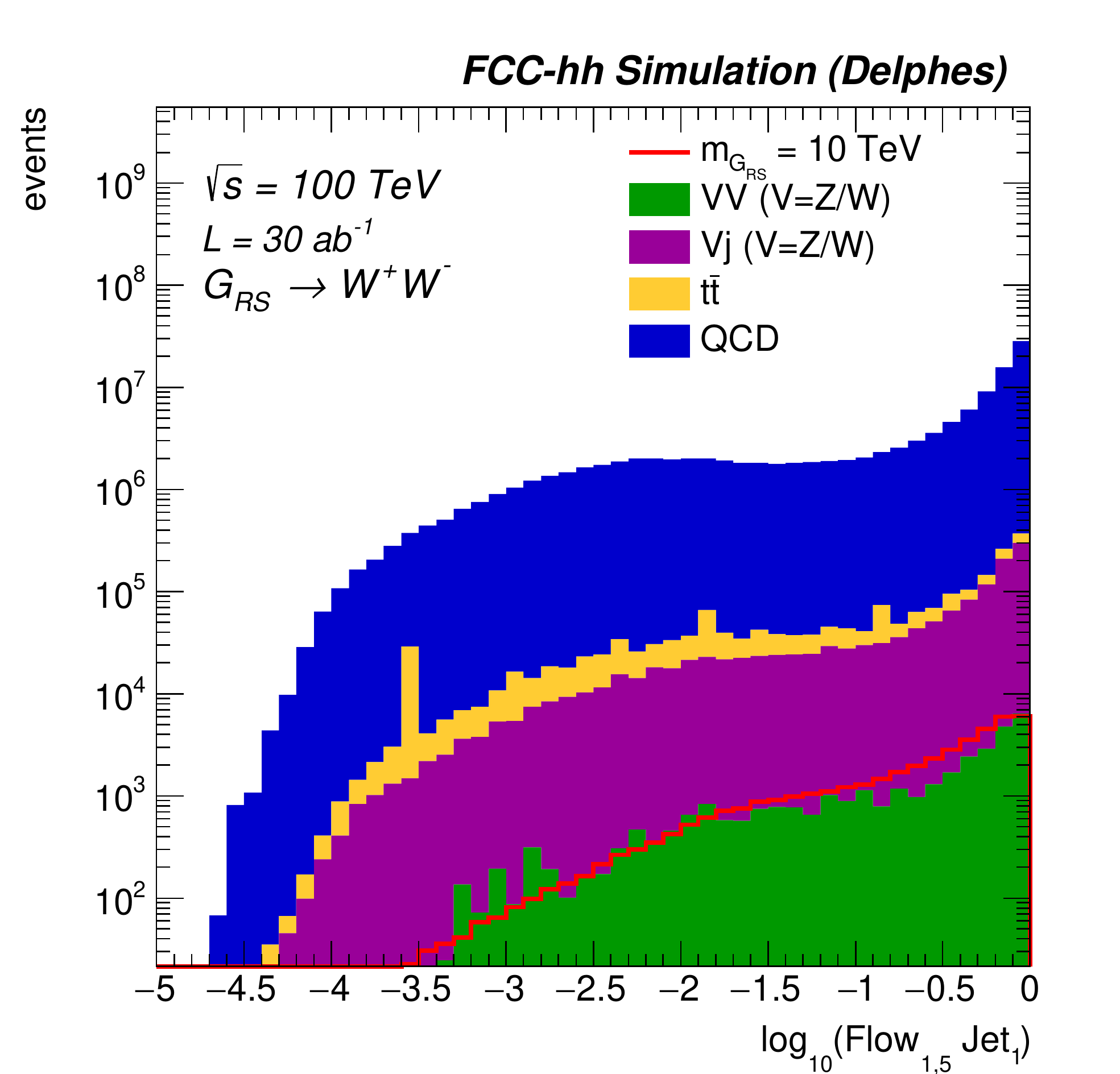}
\includegraphics[width=0.48\textwidth]{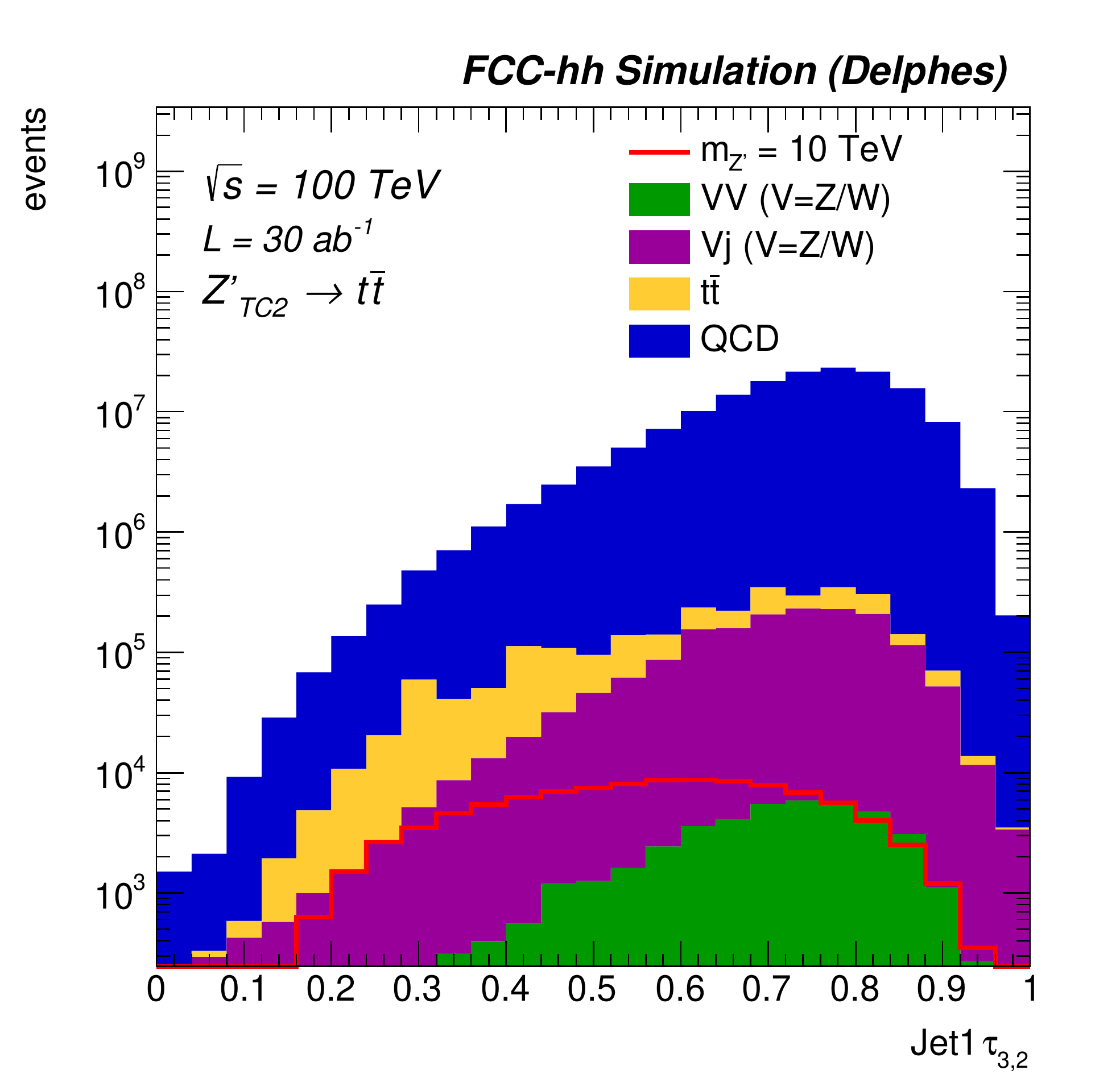}
\includegraphics[width=0.48\textwidth]{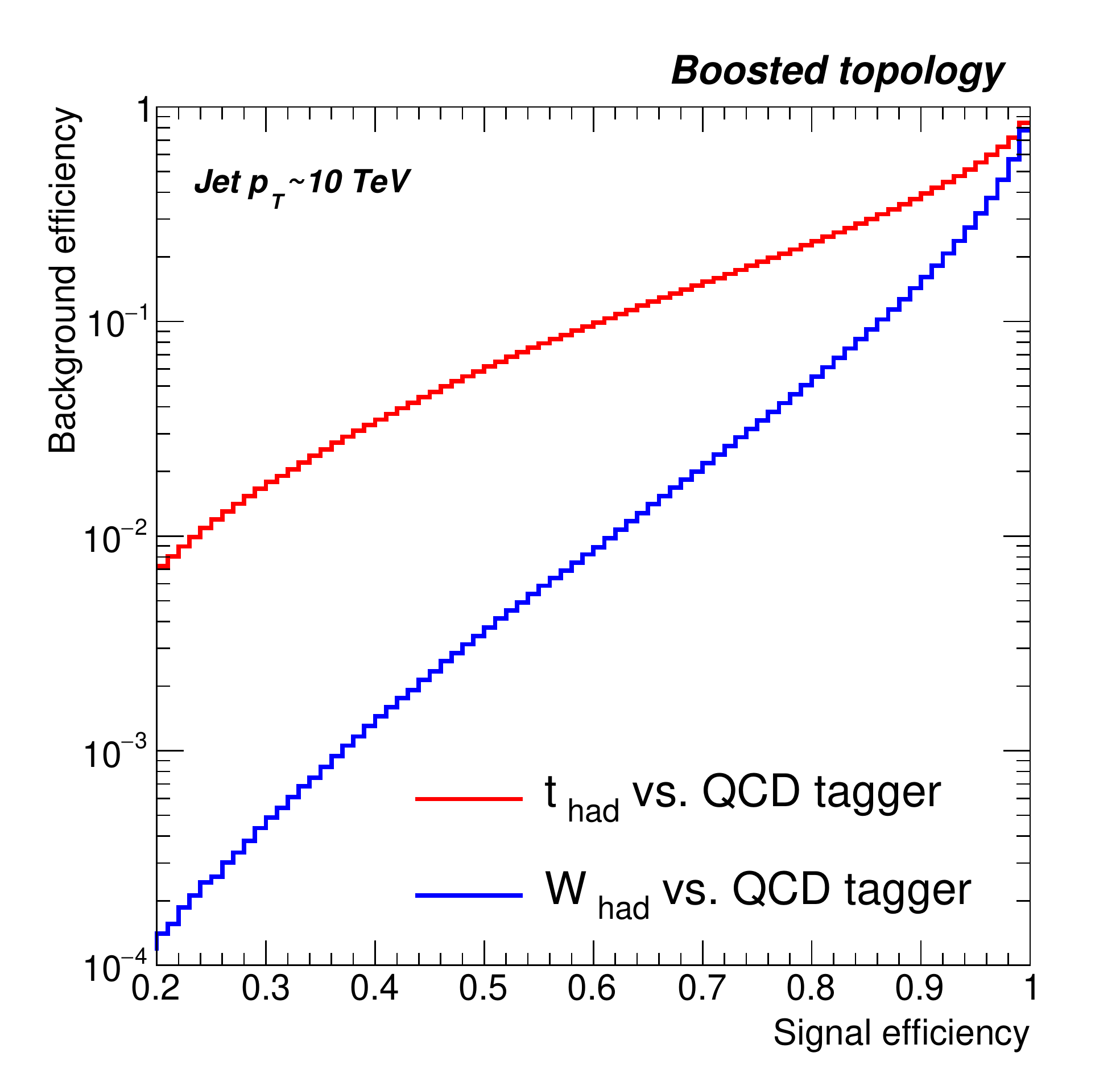}
\caption{Left: Energy-flow $E_{F}(n=1,\alpha=0.05)$ observable for the leading jet $\pt$ of the \rsg\ analysis at the pre-selection level (two high $\pt$ jets). Right: $\tau_{32}$ observable for the leading jet $\pt$ of the \Zptt\ analysis at the pre-selection level (two high $\pt$ jets). Bottom: Light jet rejection versus tagging efficiency for the $W$-tagger (blue) and top-tagger (red)}
\label{fig:TMVA_final_result}
\end{figure}

The training samples are built from \Zp\ samples with $m_{Z'}=20$~TeV using jets that do not contain leptons. The $E_{F}(n=1,\alpha=0.05)$ observable used as input to the $W$-tagger is shown in Fig.~\ref{fig:TMVA_final_result} (left) and the $\tau_{32}$ observable used as input to the top-tagger in Fig.~\ref{fig:TMVA_final_result} (right). The evolution of the light (u,d, and s quark) flavour jet efficiency (mis-tag rate) versus the $W$ and top tagging efficiencies for both taggers is shown in Fig.~\ref{fig:TMVA_final_result} (bottom). Several cross-checks have been performed to further validate the multivariate training procedure. By removing highly correlated variables it has been checked that the same performance is achieved. In addition, the BDT response has been tested with different signal masses. For the selection used in the analysis (BDT score greater than 0.15), the shape of the BDT does not significantly impact the signal efficiency. The list of input variables used to train the BDT, ordered by the training weight, can be found in Table~\ref{tab:TMVA_summary}.

\begin{table}[!htb]\centering
\begin{tabular}{| l | c | l | c |}
\hline
  \multicolumn{2}{|c|}{$W$ tagger}  & \multicolumn{2}{c|}{top tagger} \\
  \hline
 variable & weight & variable & weight \\
\hline
 $\tau_3$ (track jet, R=0.2)      & 0.12      & $\tau_1$ (track jet, R=0.2) & 0.21  \\
 $\mSD$  (track jet, R=0.2)      & 0.11      & $\mSD$  (track jet, R=0.2) & 0.17 \\
 $\tau_{31}$  (track jet, R=0.2) & 0.10     & $\tau_{31}$  (track jet, R=0.2)  & 0.11 \\
 $E_{F}(n=5,\alpha=0.05)$                               & 0.09     &  $\tau_2$ (track jet, R=0.2) & 0.10 \\
 $E_{F}(n=4,\alpha=0.05)$                               & 0.09     & $\tau_3$ (track jet, R=0.2) & 0.09 \\
 $E_{F}(n=1,\alpha=0.05)$                               & 0.08     & $\mSD$  (track jet, R=0.8)& 0.09 \\
 $E_{F}(n=2,\alpha=0.05)$                               & 0.07     &  $\mSD$  (track jet, R=0.4) & 0.09 \\
 $E_{F}(n=3,\alpha=0.05)$                               & 0.06     & $\tau_{32}$  (track jet, R=0.2) & 0.08 \\
 $\tau_{21}$  (track jet, R=0.2)& 0.06   & $\tau_{21}$  (track jet, R=0.2) & 0.06 \\
 $\mSD$  (track jet, R=0.8) & 0.06 &  &\\
 $\mSD$  (track jet, R=0.4) & 0.06 & & \\
 $\tau_1$ (track jet, R=0.2) & 0.05      &  &\\
 $\tau_2$ (track jet, R=0.2) & 0.04      &  &\\
 $\tau_{32}$  (track jet, R=0.2) & 0.02    &  &\\
\hline
\end{tabular}
\caption{Summary of the input variables to the BDT and their relative weight for both $W$ and top taggers.}
\label{tab:TMVA_summary}
\end{table}

\section{Summary plots}
\label{sec:app:sumplots}
\begin{figure}[!htb]
  \centering
  \includegraphics[width=0.49\columnwidth]{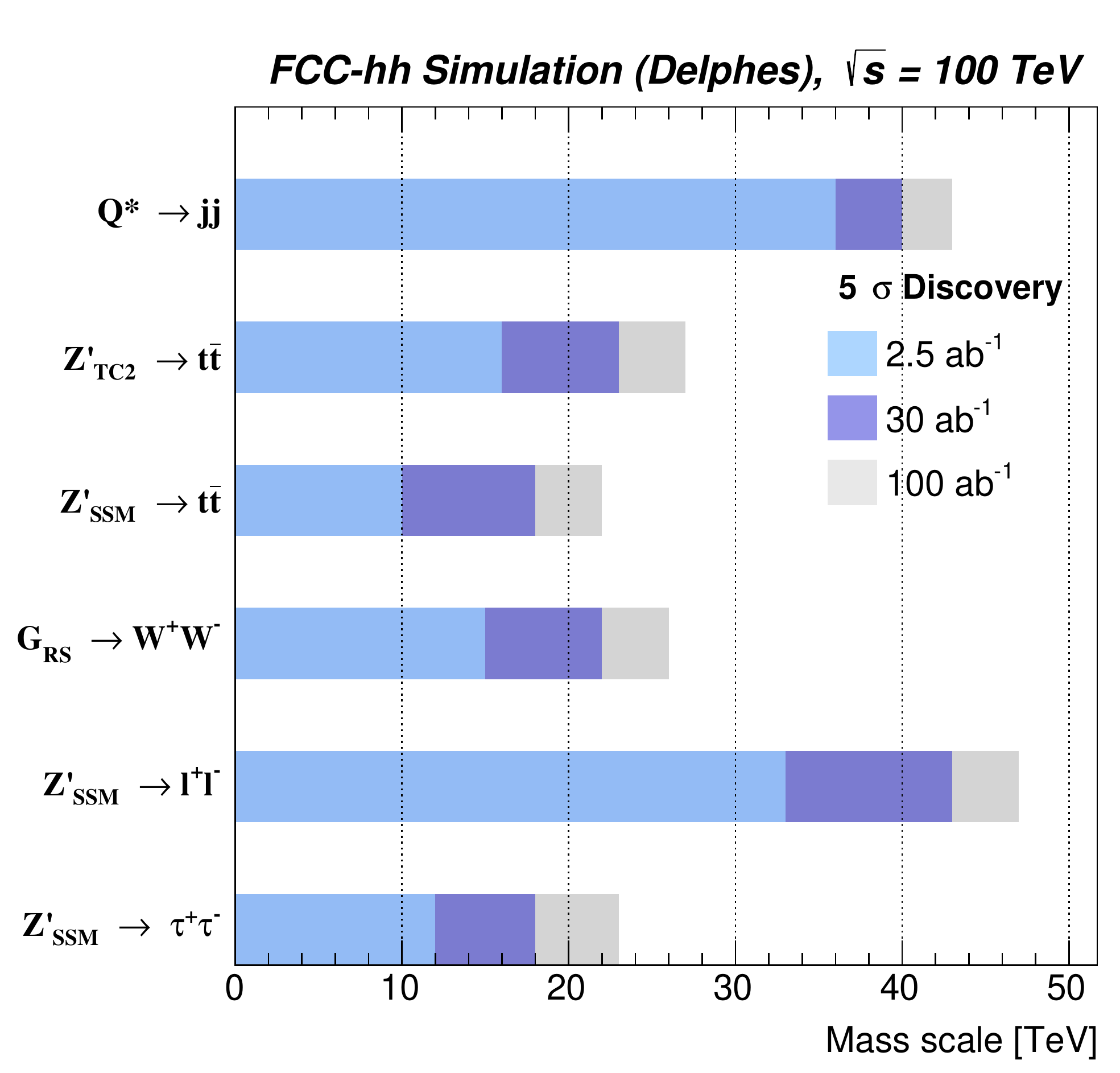}
  \includegraphics[width=0.49\columnwidth]{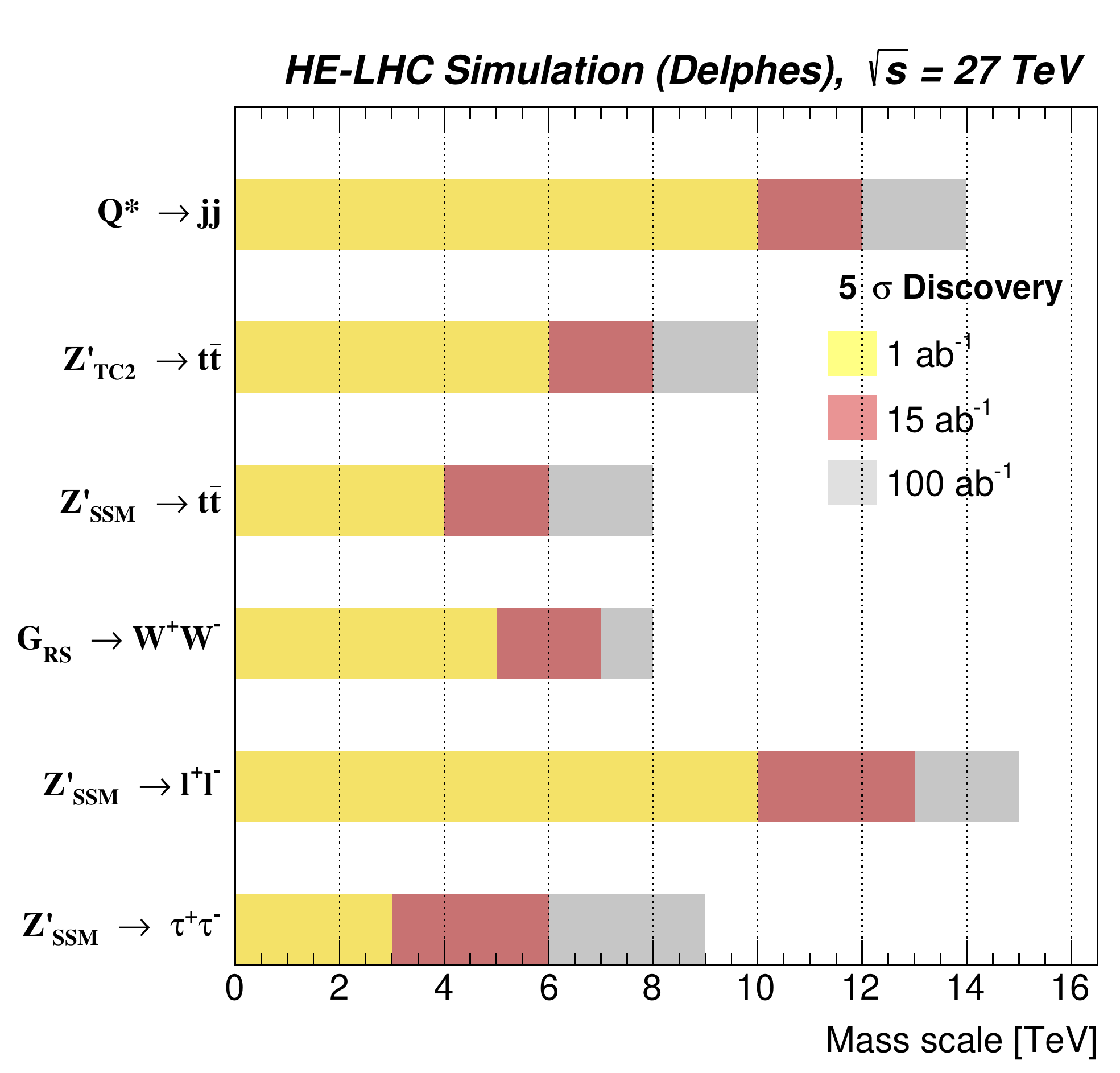}
   \includegraphics[width=0.49\columnwidth]{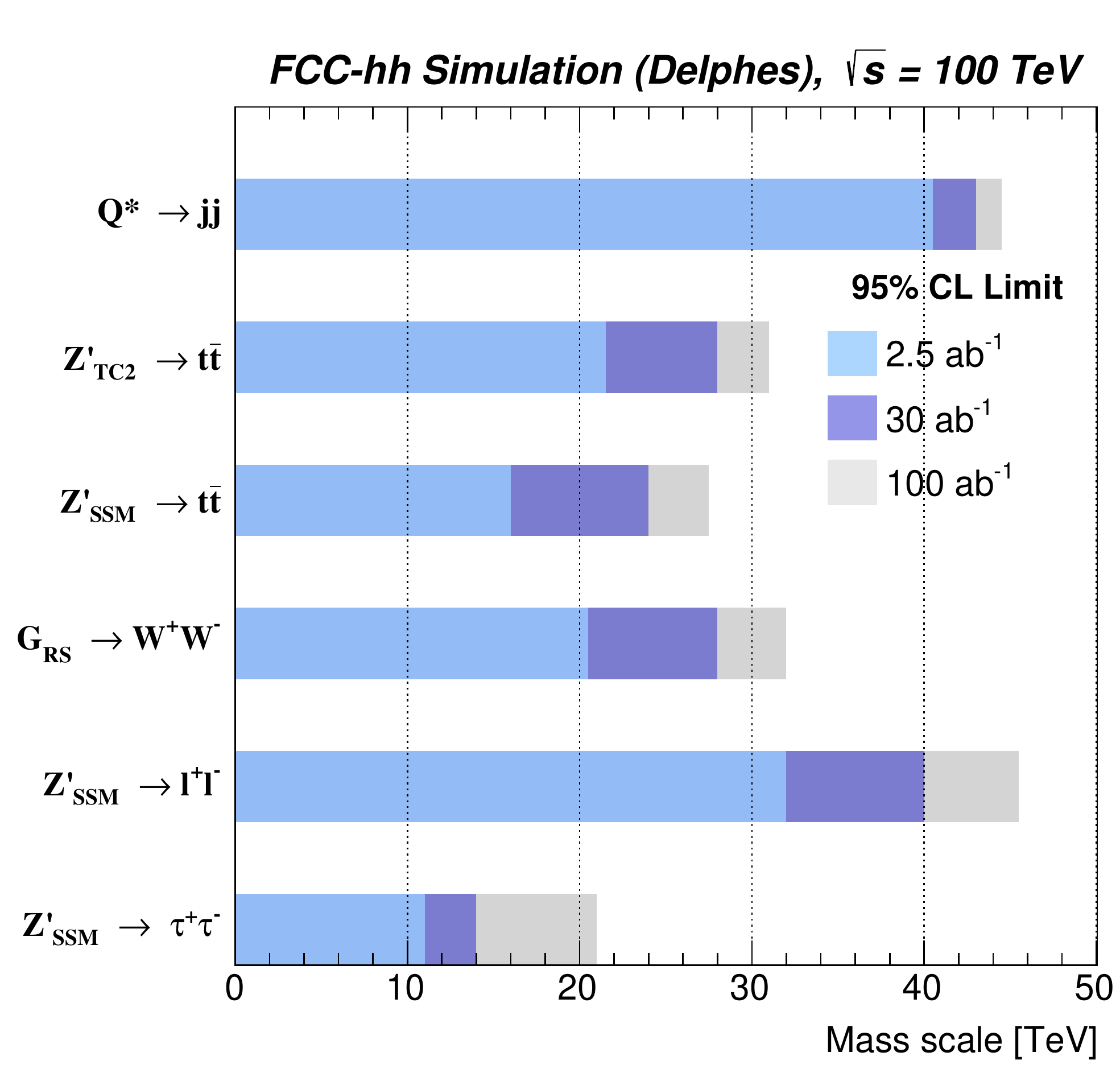}
  \includegraphics[width=0.49\columnwidth]{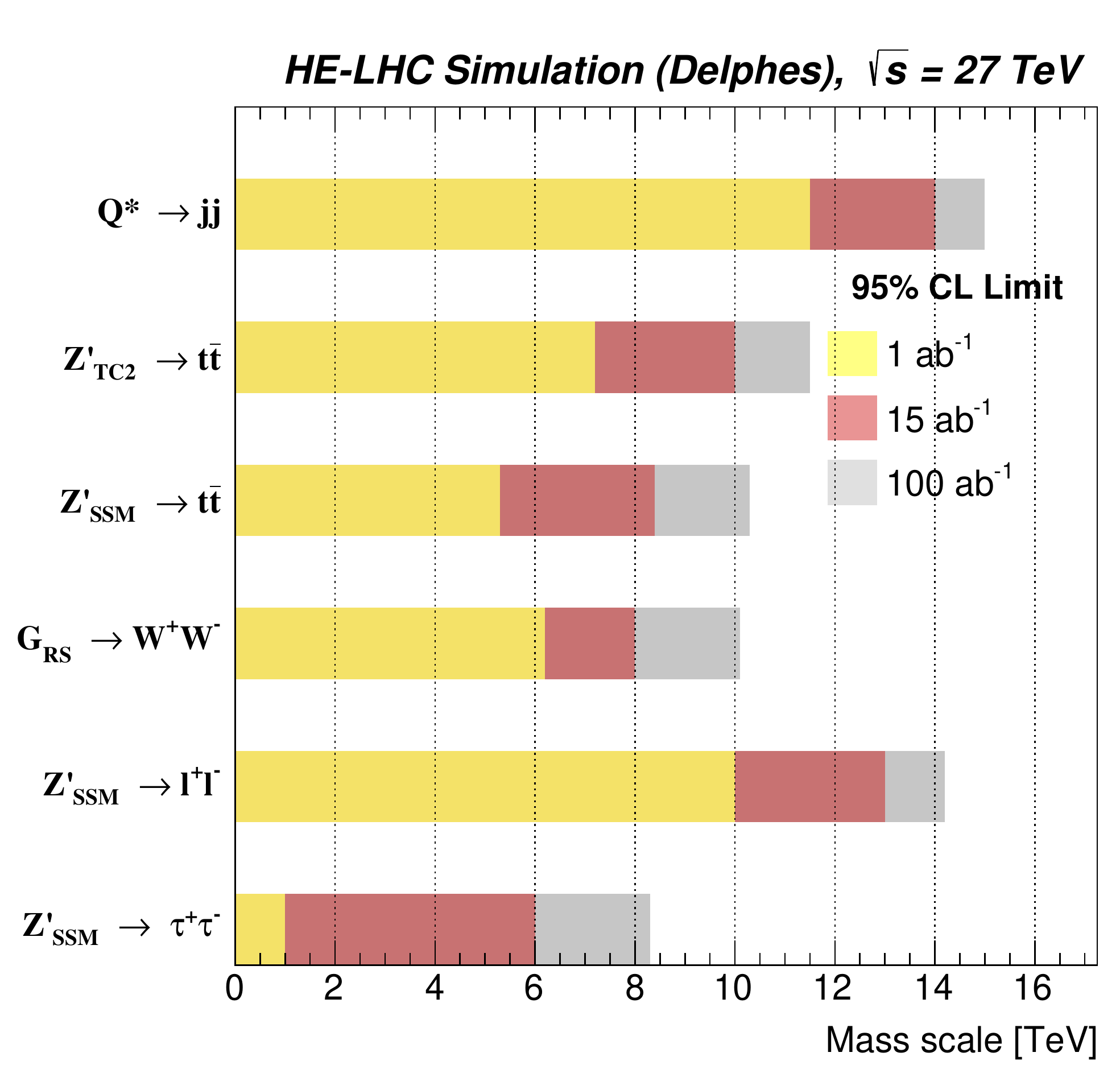}
  \caption{Summary of a $5\sigma$ discovery reach (top) and 95\% CL limits (bottom) as a function of the resonance mass for different luminosity scenario of FCC-hh (left) and HE-LHC (right).}
  \label{figure:resonances100:summary}
\end{figure}

The discovery potential (top) and 95\% CL limits (bottom) for the heavy resonances presented in this document are summarised in Fig.~\ref{figure:resonances100:summary} for FCC-hh (left) and HE-LHC (right).


\bibliographystyle{JHEP}
\clearpage
\bibliography{paper}



\end{document}